\documentclass[11pt]{article}
\usepackage[T1]{fontenc}
\usepackage{lmodern}
\usepackage[letterpaper,margin=1in]{geometry}
\usepackage{amsmath,amssymb,amsthm}
\usepackage{multirow,graphicx,caption,subcaption,url,enumitem,rotating}
\usepackage[dvipsnames]{xcolor}
\usepackage{imakeidx,booktabs,longtable,array,pdflscape}
\usepackage{microtype,setspace}
\usepackage[authoryear,round]{natbib}
\usepackage[export]{adjustbox}
\usepackage{etoolbox,xparse}
\usepackage{hyperref}
\hypersetup{hidelinks,
 pdftitle={Semiparametric causal mediation analysis in cluster-randomized trials with post-treatment confounding},
 pdfauthor={Daeyoung Ham, Chao Cheng, Fan Li}}
\makeindex
\let\myitem\relax

\let\indep\relax

\let\Var\relax

\let\EE\relax
\let\PP\relax

\let\RR\relax

\let\NN\relax

\let\mc\relax

\makeatletter
\newcommand{\myitem}[1]{%
\item[#1]\protected@edef\@currentlabel{#1}%
}
\makeatother

\newcommand{\indep}{\rotatebox[origin=c]{90}{$\models$}}

\DeclareMathOperator*{\Var}{Var}

\newcommand{\EE}{{\mathbb E}}
\newcommand{\PP}{{\mathbb P}}

\newcommand{\RR}{\mathbb{R}}

\newcommand{\NN}{\mathbb{N}}

\newcommand{\mc}[1]{\mathcal {#1}}

\theoremstyle{plain}
\newtheorem{theorem}{Theorem}
\newtheorem{proposition}{Proposition}

\newtheorem{lemma}{Lemma}

\theoremstyle{definition}

\newtheorem{assumption}{Assumption}

\theoremstyle{remark}

\allowdisplaybreaks
\newif\ifReviewRemarks
\newif\ifReviewColors
\ReviewRemarkstrue
\ReviewColorstrue

\definecolor{CodexRevision}{RGB}{0,105,135}

\let\UnboundedGraphics\includegraphics
\RenewDocumentCommand{\includegraphics}{s O{} m}{%
  \adjustbox{max height=.75\textheight}{\UnboundedGraphics[#2]{#3}}}

\title{Semiparametric causal mediation analysis in cluster-randomized trials with post-treatment confounding}
\author{\begin{tabular}{c}
Daeyoung Ham$^{1}$, Chao Cheng$^{2}$, and Fan Li$^{3,4}$\\[8pt]
\small $^{1}$Department of Statistics and Data Science, University of Texas at San Antonio,\\
\small San Antonio, TX, USA\\[3pt]
\small $^{2}$Department of Statistics and Data Science, Washington University in St. Louis,\\
\small St. Louis, MO, USA\\[3pt]
\small $^{3}$Department of Biostatistics, Yale School of Public Health, New Haven, CT, USA\\[3pt]
\small $^{4}$Center for Methods in Implementation and Prevention Science,\\
\small Yale School of Public Health, New Haven, CT, USA\\[8pt]
\small \href{mailto:daeyoung.ham@utsa.edu}{daeyoung.ham@utsa.edu};
\href{mailto:chaoc@wustl.edu}{chaoc@wustl.edu};
\href{mailto:fan.f.li@yale.edu}{fan.f.li@yale.edu}
\end{tabular}}
\date{}
\begin{document}
\maketitle
\begin{abstract}

Causal mediation analysis of cluster-randomized trials can reveal how interventions affect outcomes through individual and spillover pathways. Most existing methods target natural mediation effects under assumptions that exclude treatment-induced mediator--outcome confounding. We develop a semiparametric framework for interventional mediation effects that accommodates such confounding and within-cluster interference without requiring cross-world independence assumptions. We identify cluster-average and individual-average direct and indirect effects and decompose the indirect effect into individual and spillover components, distinguishing pathways through an individual's own mediator and those of other cluster members. 
These contrasts evaluate hypothetical interventions on the joint mediator distribution and can differ from treatment effects carried through individual causal pathways represented by natural effects, thereby providing an intervention-oriented perspective that complements natural effects.
We derive efficient influence functions to construct estimators using cluster-level cross-fitting, characterize their multiple-robustness properties, and establish conditions for asymptotic normality and semiparametric efficiency. To facilitate implementation, we develop a lower-dimensional reparameterization that reduces the need for multivariate density estimation and integration and allows nuisance functions to be estimated using parametric models or machine learning. Simulation studies evaluate finite-sample performance, and an application to a cluster-randomized trial of cognitive behavioral therapy for chronic pain illustrates the proposed methods.

\end{abstract}
\noindent\textbf{Keywords:} Cluster-randomized trials; Efficient influence functions; Interventional mediation; Post-treatment confounding; Within-cluster interference.
% PRESERVED MANUSCRIPT START
\section{Introduction}
Cluster-randomized trials, in which groups of individuals rather than individual participants are randomized to treatment conditions, are widely used to evaluate interventions in health, biomedical, and social science research. The primary analysis of a cluster-randomized trial typically assesses whether an intervention is effective by estimating its total effect on a final outcome \citep{hayes2017cluster,li2025model}. The total effect alone, however, does not reveal the mechanisms through which the intervention operates. Causal mediation analysis complements this overall assessment by examining pathways through post-treatment intermediate variables, referred to as mediators \citep{imai2010general}. By distinguishing \textit{indirect effects} operating through the mediator from \textit{direct effects} operating through other pathways, causal mediation analysis can deepen our understanding of how an intervention works and inform the design and refinement of future interventions.

A growing literature has investigated causal mechanisms in cluster-randomized trials. Relative to simpler individually randomized trials, mediation analysis in cluster-randomized trials is further complicated by within-cluster interference, under which one individual's mediator may affect not only that individual's own outcome but also the outcomes of other members of the same cluster. In the single-mediator setting, \citet{vanderweele2013mediation} decomposed the natural indirect effect into an individual mediation effect, capturing pathways through an individual's own mediator, and a spillover mediation effect, capturing pathways through the mediators of other individuals in the same cluster. \citet{cheng2026semiparametric} established semiparametric efficiency theory for inferring these effect estimands and developed efficient estimators that accommodate data-adaptive machine-learning methods for nuisance modeling. Subsequent work has extended this literature to settings with multiple mediators. For two unordered mediators, \citet{ohnishi2026bayesian} decomposed the mediator-specific exit indirect effect \citep{xia2022decomposition} into exit individual and exit spillover mediation effects, thereby providing a finer characterization of individual and spillover pathways, and developed a Bayesian nonparametric approach for estimation and inference. \citet{tong2026multiple} generalized these exit mediation effects to an arbitrary number of mediators without requiring a known causal ordering among them and constructed efficient debiased machine-learning estimators. Together, these developments have substantially expanded the range of causal mechanisms that can be studied in cluster-randomized trials. Existing methods, however, mostly focus on settings without treatment-induced confounding of the mediator--outcome relationship.

Treatment-induced mediator-outcome confounding arises when the  treatment affects an intermediate post-treatment variable that, in turn, influences both the mediator and the outcome. In this setting, natural mediation effects are generally not identified under sequential ignorability assumptions, and identifying them requires additional unverifiable restrictions on counterfactual quantities \citep{avin2005identifiability,tchetgen2014identification,xia2023identification}. One approach to this problem is to target the natural mediation effects and assess their sensitivity to the unidentified cross-world dependence between the potential mediator and the potential outcome. For example, \citet{ohnishi2025identification} developed a Bayesian sensitivity analysis framework for natural mediation effects in cluster-randomized trials subject to post-treatment confounding, using a Gaussian copula model to parameterize the residual dependence between counterfactual mediator and outcome under different treatment conditions. As a complementary approach, we target interventional analogues of natural mediation effects, which replace individual-specific natural mediator values with stochastic draws from treatment-specific mediator distributions \citep{petersen2006estimation,vanderweele2014effect,vansteelandt2017interventional}. 
In the independent data setting, these effects can be identified in the presence of observed treatment-induced mediator-outcome confounding under appropriate exchangeability and positivity conditions, without specifying the unidentified cross-world dependence required for natural effects \citep{diaz2021efficient,hejazi2023nonparametric}. 
In cluster-randomized trials, an individual's outcome may depend on both their own mediator and those of other cluster members, motivating interventions on the joint mediator distribution to assess how changes in marginal distributions and their dependence affect the outcomes.
The resulting contrasts quantifies the potential value of hypothetical mediator interventions by comparing mean outcomes under treatment and control mediator distributions while holding treatment fixed, and can differ from the treatment effects carried through individual causal pathways that natural effects explain \citep{miles2023causal}.
Interventional effects provide a complementary, intervention-oriented perspective on the role of mediators.
% Beyond this identification advantage, they address a substantive question on how would outcomes change if, while holding treatment fixed, the mediator distribution were shifted from that under control to that under treatment. This perspective can inform the value of interventions targeting the mediator and, in cluster-randomized trials, motivate distinguishing the consequences of changing an individual's own mediator from those of changing the mediators of other cluster members. Interventional effects therefore provide a complementary, intervention-oriented perspective on the role of mediators \citep{miles2023causal}.

We seek to develop novel semiparametric methods for interventional mediation analysis in cluster-randomized trials with treatment-induced mediator-outcome confounding. Our contributions are threefold. First, we define interventional direct and indirect effects by drawing the within-cluster mediator vector from its treatment-specific distribution conditional on baseline information, and further decompose the interventional indirect effect into individual and spillover components operating through an individual's own mediator and the mediators of other cluster members, respectively. The proposed estimands allow arbitrary within-cluster interference and accommodate both cluster-average and individual-average target populations through general cluster-level weighting \citep{kahan2023estimands}. To our knowledge, this is the first effort to address interventional effect estimands in the presence of mediator spillover. Second, we derive the efficient influence functions and construct cluster-level cross-fitted estimators that accommodate flexible nuisance estimation \citep{chernozhukov2018double}. Exact bias decompositions characterize their multiple robustness properties. 
The estimators are consistent under the required multiple robustness conditions for the nuisance functions. 
And, when all nuisance estimators are consistent and the required product rate conditions hold, the estimators are asymptotically normal and semiparametrically efficient.
Third, to facilitate practical implementation, we further develop a lower-dimensional reparameterization of the efficient influence function that replaces high-dimensional post-treatment confounder densities and integration over the full mediator vector with density ratios, treatment classifiers, univariate mediator densities, and conditional mean regressions. In particular, we adapt the augmented-classification approach of \citet{rudolph2024practical} to estimate the key density ratio without directly modeling the corresponding high-dimensional conditional densities. We establish corresponding robustness and efficiency results under compatibility conditions on the auxiliary regressions and examine both implementations in simulation studies. We illustrate the proposed methods in the PPACT cluster-randomized trial of cognitive behavioral therapy for chronic pain.

\section{Notation, estimands and assumptions}\label{sec: notation and DGP}
Consider a cluster-randomized trial with $K$ clusters, indexed by $i\in\{1,\ldots,K\}$. Let $N_i$ denote the size of cluster $i$, and let $A_i\in\{0,1\}$ denote its treatment assignment, with $A_i=1$ indicating treatment and $A_i=0$ indicating control. All individuals in a cluster receive the same assignment. Baseline information consists of cluster-level covariates $V_i\in\RR^{d_{\bf V}\times 1}$ and individual-level covariates $X_{ij}\in\RR^{d_{\bf X}\times 1}$, for $j\in\{1,\ldots,N_i\}$. We also write $C_i=\{V_i,X_i\}$ for the full baseline covariate information in the cluster, with $X_i=[X_{i1},\ldots,X_{iN_i}]^T\in\RR^{N_i\times d_{\bf X}}$. 
After assignment, we observe post-treatment confounders $Z_{ij}\in\RR^{d_{\bf Z}\times 1}$, followed by a mediator $M_{ij}\in\RR$ and an outcome $Y_{ij}\in\RR$. $Z_{ij}$ may be affected by treatment and may, in turn, affect both the mediators and the outcomes. 
Denote $Z_i=[Z_{i1},\ldots,Z_{iN_i}]^T\in\RR^{N_i\times d_{\bf Z}}$, $M_i=[M_{i1},\ldots,M_{iN_i}]^T\in\RR^{N_i\times 1}$, and $Y_i=[Y_{i1},\ldots,Y_{iN_i}]^T\in\RR^{N_i\times 1}$. For an individual $j$, $M_{i(-j)}\in\RR^{(N_i-1)\times 1}$ denotes the mediators of all other individuals in the same cluster. The observed data for cluster $i$ are $O_i=\{N_i,C_i,A_i,Z_i,M_i,Y_i\}$. We suppress the cluster index ``$i$" when referring to a generic cluster, using a ``$\bullet$" in its place for individual-level variables; for example, we write $M$ and $M_{\bullet j}$ for $M_i$ and $M_{ij}$.

Let $Z_{\bullet j}(a)$ denote the potential post-treatment confounders for individual $j$ under cluster treatment $a\in\{0,1\}$, and let $M_{\bullet j}(a,z)$ denote the potential mediator when treatment is set to $a$ and the full cluster confounder vector is set to $z=\{z_{\bullet j},z_{\bullet(-j)}\}$. Similarly, $Y_{\bullet j}(a,z,m)=Y_{\bullet j}(a,z_{\bullet j},z_{\bullet(-j)},m_{\bullet j},m_{\bullet(-j)})$ is the potential outcome when treatment, confounders, and mediators are set to $a$, $z$, and $m=\{m_{\bullet j},m_{\bullet(-j)}\}$, respectively. Dependence on the full vectors $z$ and $m$ permits an individual's outcome to depend on both their own and other cluster members' post-treatment variables. 
The potential variables are indexed only by interventions within the same cluster, thereby excluding interference across clusters. 

We assume composition: $Y_{\bullet j}(a,m)=Y_{\bullet j}(a,Z(a),m)$ and $M_{\bullet j}(a)=M_{\bullet j}(a,Z(a))$, where $Z(a)$ and $M(a)$ collect the corresponding potential variables over cluster members. Thus, an intervention on the mediators leaves the post-treatment confounders at their natural values under treatment $a$. We also assume $Y_{\bullet j}(a) = Y_{\bullet j}(a,Z(a),M(a)) = Y_{\bullet j}(a,Z(a),M(a,Z(a)))$, i.e., the potential outcome under treatment alone equals the potential outcome obtained by setting both post-treatment vectors to their natural values under that treatment.

To define interventional mediation effects, let $G(a)$ be a random draw from the conditional distribution of $M(a)$ given $(C,N)$. Conditional on $(C,N)$, we construct $G(0)$ and $G(1)$ to be independent of one another and of all potential variables. Each $G(a)$ is drawn as a \emph{joint cluster mediator vector}, allowing the dependence among its components under treatment $a$.
This construction extends stochastic mediator interventions for intermediate confounding to clustered data setting \citep{diaz2021efficient,rudolph2024practical}. Specifically, we define the {\it interventional direct effect} (${\rm IDE}$) and {\it interventional indirect effect} (${\rm IIE}$) by
$$
{\rm IDE}=h(\theta(1,0),\theta(0,0)), \quad {\rm IIE}=h(\theta(1,1),\theta(1,0)).
$$where $h$ is a known contrast function specifying the effect scale, and
\begin{equation}
\theta(a,a')=\frac{\EE\left[\frac{W}{N}\sum_{j=1}^N Y_{\bullet j}(a, G_{\bullet j}(a'),G_{\bullet (-j)}(a'))\right]}{\EE[W]},\qquad~{\rm for~}a,a'\in\{0,1\}.
\label{def:theta}
\end{equation}
The weight $W$ is a user-specified function of $\{N,V\}$. As illustrated in \citet{li2025model}, $W$ specifies how cluster-specific mean outcomes contribute to the target population mean. In particular, $W=1$ gives equal weight to clusters, whereas $W=N$ gives weight proportional to cluster size, yielding cluster-average and individual-average interventional effect estimands, respectively. Possible scale functions include $h(x,y)=x/y$ and $h(x,y)=(x/(1-x))/(y/(1-y))$, indicating risk ratio and odds ratio scales. In this paper, we focus on the mean difference scale with $h(x,y)=x-y$.

In $\theta(a,a')$, treatment is set to $a$, the confounders take their natural values $Z(a)$, and the entire mediator vector is independently drawn from its distribution under $a'$ among clusters with the same $(C,N)$. 
The IDE therefore changes treatment while holding the mediator intervention at the control distribution; it allows treatment to affect the outcome through the post-treatment confounders with the mediator draw held fixed in distribution. 
The IIE holds treatment at $1$ and changes the joint mediator distribution from control to treatment. 
These effects sum to $\theta(1,1)-\theta(0,0)$. 
This sum need not equal the total effect of treatment alone. That is, even when $a=a'$, the stochastic intervention breaks the natural dependence between $Z(a)$ and $M(a)$ conditional on $(C,N)$. 
This distinction between interventional and total treatment effects is discussed by \citet{diaz2021efficient} and \citet{rudolph2024practical}.

To distinguish changes involving an individual's own mediator from those involving the mediators of other cluster members, we further define the {\it interventional individual mediation effect} (${\rm IIME}$) and the {\it interventional spillover mediation effect} (${\rm ISME}$) as
$$
{\rm IIME}=h(\tau,\theta(1,0)), \quad {\rm ISME}=h(\theta(1,1),\tau),
$$
where
\begin{equation}
\tau=\frac{\EE\left[\frac{W}{N}\sum_{j=1}^N Y_{\bullet j}(1,G_{\bullet j}(1),G_{\bullet (-j)}(0))\right]}{\EE[W]}.
\label{def:tau}
\end{equation}
For each individual $j$, the intervention in $\tau$ draws their own mediator from the treatment distribution and the vector of other members' mediators from the control distribution, independently conditional on $(C,N)$. 
The parameter averages outcomes under these individual-specific interventions; it does not require a single mediator assignment to implement all of them simultaneously.
Note that the IIME and ISME sum exactly to the IIE. Their interpretation also depends on the joint mediator distributions under treatment and control. 
Moving from $\theta(1,0)$ to $\tau$ changes the own-mediator marginal distribution and removes its dependence on the peer mediator vector under control. Moving from $\tau$ to $\theta(1,1)$ changes the peer marginal distribution and restores the dependence between own and peer mediators under treatment. Consequently, these contrasts can reflect changes in mediator dependence as well as changes in the respective marginal distributions.

\begin{assumption}
We require the following identifying assumptions:
\begin{enumerate}
\myitem{(A1)}\label{assm:consistency} If $A_i=a$, then
$Z_i=Z_i(a)$.  If, in addition, $Z_i=z$, then
$M_i=M_i(a,z)$.  If $A_i=a$, $Z_i=z$, and $M_i=m$, then
$Y_{ij}=Y_{ij}(a,z,m)$ for every $j\in\{1,\ldots,N_i\}$.
\myitem{(A2)}\label{assm:CRT.A} The treatment indicators are mutually
independent with $\PP(A_i=1)=\pi\in(0,1)$. For  $(z,m)$ over their valid support,
$
A_i\indep
\left\{C_i,N_i,Z_i(a),M_i(a,z),Y_{ij}(a,z,m):
 a\in\{0,1\},\ j=1,\ldots,N_i\right\}.
$
\myitem{(A3)}\label{assm:ignorability} For every $j$, $a$, $z$
and $m$ over their valid support,
$
M_i(a,z)\indep Y_{ij}(a,z,m)\mid Z_i(a)=z,C_i,N_i.
$
\myitem{(A4)}\label{assm:positivity} The observed clustered data
$O_1,\ldots,O_K$ are mutually independent, with $N$ having a
finite support in $\NN^+$. Conditional on $N$, the observed-data distribution admits 
$
P_{Y|M,Z,A,C,N}\times P_{M|Z,A,C,N}\times P_{Z|A,C,N}
\times P_A\times P_{C|N},
$ 
with $\EE(Y_{\bullet j}^2)<\infty$ across $j\leq N$. For every $a,a'\in\{0,1\}$, suppose
\[
\begin{aligned}
& f_{M|A,Z,C,N}(m|a,z,c,n)>0
\quad\text{if } f_{Z|A,C,N}(z|a,c,n)f_{M|A,C,N}(m|a',c,n)>0,\\
&\text{and } f_{M|A,Z,C,N}(m|1,z,c,n)>0\\
&\qquad\text{if } f_{Z|A,C,N}(z|1,c,n)
 f_{M_{\bullet j}|A,C,N}(m_{\bullet j}|1,c,n)
 f_{M_{\bullet(-j)}|A,C,N}(m_{\bullet(-j)}|0,c,n)>0.
\end{aligned}
\]
In addition, the weight $W$ is nonnegative and bounded. 
\end{enumerate}
\end{assumption}

Assumption~\ref{assm:consistency} links the observed post-treatment variables to their potential values, accommodating arbitrary within-cluster interference while excluding interference across clusters.
Assumption~\ref{assm:CRT.A} specifies independent Bernoulli assignment of clusters. 
Assumption~\ref{assm:ignorability} supplies the additional mediator-outcome exchangeability condition. Conditional on  $(C_i,N_i)$ and the natural confounder values $Z_i(a)=z$, $M(a,z)$ is independent of each $Y_{\bullet j}(a,z,m)$. 
Thus, after accounting for the measured baseline information and the full post-treatment confounder vector, there does not exist remaining confounding of the joint mediator intervention and an individual's outcome. 
The condition allows $Z_i$ to be affected by treatment and to confound the mediator-outcome relationship. %which is similar to the exchangeability conditions for interventional mediation with intermediate confounders in \citet{diaz2021efficient} and \citet{rudolph2024practical}. 
In cluster-randomized trials, this differs from the sequential ignorability condition for identifying natural mediation effects in \citet{cheng2026semiparametric}, which excludes treatment-induced mediator--outcome confounding.  Assumption~\ref{assm:positivity} requires the clustered data are mutually independent and the cluster size is finite. It also requires positivity conditions for the mediator densities.

Before we proceed, we define several key nuisance parameters underlying the observed data distribution. Let $\eta_{\bullet j}(a,z,m,c,n)=\EE[Y_{\bullet j}|A=a,Z=z,M=m,C=c,N=n]$ be the conditional outcome mean.
Let $\xi(m|a,z,c,n)=f_{M|A,Z,C,N}(m|a,z,c,n)$ and $\kappa(m|a,c,n)=f_{M|A,C,N}(m|a,c,n)$ denote the
conditional distributions of the mediator vector given $(A,Z,C,N)$ and $(A,C,N)$, respectively. Also define $\phi(z|a,c,n)=f_{Z|A,C,N}(z|a,c,n)$ as the conditional distribution of the post-treatment confounder vector given $(A,C,N)$. We abbreviate the four nuisance parameters as $h=(\eta,\phi,\kappa,\xi)$. Estimating $h$ can be nontrivial because the nuisance parameters involve multi-dimensional conditional distributions. Section~\ref{sec:nuisance-estimation} of the Appendix describes nuisance estimation details, including an extension of the copula approach of \citet{cheng2026semiparametric} to estimate $(\xi,\kappa,\phi)$. Further let $\kappa_{\bullet j}(m_{\bullet j}|a,c,n)$ and
$\kappa_{\bullet(-j)}(m_{\bullet(-j)}|a,c,n)$ denote the corresponding conditional
 densities of $M_{\bullet j}$ and $M_{\bullet(-j)}$, which can be extracted from $\kappa$ based on 
$\kappa_{\bullet j}(m_{\bullet j}|a,c,n)=\int \kappa(m|a,c,n)dm_{\bullet(-j)}$ and
$\kappa_{\bullet(-j)}(m_{\bullet(-j)}|a,c,n)=\int \kappa(m|a,c,n)dm_{\bullet j}$.
  %The stochastic intervention $G(a)$ uses $\kappa$ and its marginals, while adjustment for mediator--outcome confounding uses $\eta_{\bullet j}$ conditional on $Z$.

\section{Identification and the nonparametric efficiency bounds}
\label{sec:eif}
We focus on identifying $\theta(a,a')$ and $\tau$ as functionals of the observed-data distribution, based on which all interventional mediation effects are identified. The following theorem provides the core identification formulas.

\begin{theorem}\label{thm:identification}
Under Assumptions~\ref{assm:consistency}--\ref{assm:positivity},  $\theta(a,a')$ is identified by
\begin{equation}
\theta(a,a')
=\frac{\EE\left[\frac{W}{N}\displaystyle\sum_{j=1}^N
\iint \eta_{\bullet j}(a,z,m,C,N)\phi(z|a,C,N)
\kappa(m|a',C,N)\,dz\,dm\right]}{\EE[W]},
\label{eq:ident.theta}
\end{equation}
for any $a,a'\in\{0,1\}$. Additionally, $\tau$ is identified by
\begin{equation}
\resizebox{0.95\textwidth}{!}{$
\tau
=\frac{\EE\left[\frac{W}{N}\displaystyle\sum_{j=1}^N
\iint \eta_{\bullet j}(1,z,m,C,N)\phi(z|1,C,N)
\kappa_{\bullet j}(m_{\bullet j}|1,C,N)
\kappa_{\bullet(-j)}(m_{\bullet(-j)}|0,C,N)\,dz\,dm\right]}{\EE[W]}.
$}
\label{eq:ident.tau}
\end{equation}
\end{theorem}
The proof is given in Section~\ref{sec:Identification_proof} of the Appendix. Based on the identification results for our causal targets, their efficient influence functions (EIFs) then characterize the efficiency bounds for estimating these functionals under the nonparametric model.
To express the EIFs, we introduce the auxiliary functions defined in Table~\ref{tab:eif-auxiliary-functions} (Panel A), which depend on the nuisance parameters $(\eta,\xi,\kappa,\phi)$.

\begin{table}[p]
\centering
\begingroup
\renewcommand{\baselinestretch}{1}
\fontsize{9}{11}\selectfont
\caption{Auxiliary functions, nuisance estimation requirements, and properties of the EIF estimators.}
\label{tab:eif-auxiliary-functions}
\label{tab:estimator-summary}
\setlength{\tabcolsep}{4pt}
\renewcommand{\arraystretch}{1.08}

\makebox[\textwidth][l]{\textbf{A. Auxiliary functions}}\par\smallskip
\resizebox{\textwidth}{!}{%
\begin{tabular}{@{}ll@{}}
\toprule
Auxiliary function & Definition \\
\midrule
$\omega^{(1)}(a,a',z,m,c,n)$
& $\displaystyle \frac{\kappa(m\mid a',c,n)}{\xi(m\mid a,z,c,n)}$ \\[6pt]
$\omega_{\bullet j}^{(2)}(a,a',a^*,z,m,c,n)$
& $\displaystyle \frac{\kappa_{\bullet j}(m_{\bullet j}\mid a,c,n)\kappa_{\bullet(-j)}(m_{\bullet(-j)}\mid a^*,c,n)}{\xi(m\mid a',z,c,n)}$ \\[6pt]
$u_{\bullet j}^{(1)}(a,a',z,c,n)$
& $\displaystyle \int \eta_{\bullet j}(a,z,m,c,n)\kappa(m\mid a',c,n)\,dm$ \\[6pt]
$u_{\bullet j}^{(2)}(a,m,c,n)$
& $\displaystyle \int \eta_{\bullet j}(a,z,m,c,n)\phi(z\mid a,c,n)\,dz$ \\[6pt]
$u_{\bullet j}^{(3)}(a,a',c,n)$
& $\displaystyle \iint \eta_{\bullet j}(a,z,m,c,n)\phi(z\mid a,c,n)\kappa(m\mid a',c,n)\,dm\,dz$ \\[6pt]
$r_{\bullet j}^{(1)}(a,a',a^*,z,c,n)$
& $\begin{aligned}
&\int \eta_{\bullet j}(a,z,m,c,n)\kappa_{\bullet j}(m_{\bullet j}\mid a',c,n)\\[-2pt]
&\qquad\times\kappa_{\bullet(-j)}(m_{\bullet(-j)}\mid a^*,c,n)\,dm_{\bullet(-j)}\,dm_{\bullet j}
\end{aligned}$ \\[6pt]
$r_{\bullet j}^{(2)}(a,a',m_{\bullet j},c,n)$
& $\displaystyle \iint \eta_{\bullet j}(a,z,m,c,n)\phi(z\mid a,c,n)\kappa_{\bullet(-j)}(m_{\bullet(-j)}\mid a',c,n)\,dm_{\bullet(-j)}\,dz$ \\[6pt]
$r_{\bullet j}^{(3)}(a,a',m_{\bullet(-j)},c,n)$
& $\displaystyle \iint \eta_{\bullet j}(a,z,m,c,n)\phi(z\mid a,c,n)\kappa_{\bullet j}(m_{\bullet j}\mid a',c,n)\,dm_{\bullet j}\,dz$ \\[6pt]
$r_{\bullet j}^{(4)}(a,a',a^*,c,n)$
& $\begin{aligned}
&\iint \eta_{\bullet j}(a,z,m,c,n)\phi(z\mid a,c,n)\\[-2pt]
&\qquad\times\kappa_{\bullet j}(m_{\bullet j}\mid a',c,n)\kappa_{\bullet(-j)}(m_{\bullet(-j)}\mid a^*,c,n)\,dm\,dz
\end{aligned}$ \\
\bottomrule
\end{tabular}%
}

\vspace{7pt}
\makebox[\textwidth][l]{\textbf{B. Nuisance estimation requirements and properties}}\par\smallskip
\setlength{\tabcolsep}{3pt}
\begin{tabular*}{\textwidth}{@{\extracolsep{\fill}}lcccc@{}}
\toprule
& \multicolumn{2}{c}{Canonical}
& \multicolumn{2}{c}{Low-dimensional} \\
\cmidrule(lr){2-3}\cmidrule(l){4-5}
Nuisance functions
& $\widehat\theta_{\mathrm{cf}}(a,a')$
& $\widehat\tau_{\mathrm{cf}}$
& $\widehat\theta_{\mathrm{ld}}(a,a')$
& $\widehat\tau_{\mathrm{ld}}$ \\
\midrule
$\eta$ & $\checkmark$ & $\checkmark$ & $\checkmark$ & $\checkmark$ \\
$\phi,\kappa,\xi$ & $\checkmark$ & $\checkmark$ & & \\
$d$ & & & $\checkmark$ & $\checkmark$ \\
$s$ & & & $\checkmark$ & \\
$s_{-j},\kappa_{\bullet j},\kappa_{\bullet j}^{\star}$ & & & & $\checkmark$ \\
$u_{\bullet j}^{(1)}$--$u_{\bullet j}^{(3)}$ & Integrated & & Regressed & \\
$r_{\bullet j}^{(1)}$--$r_{\bullet j}^{(4)}$ & & Integrated & & Regressed \\
\midrule
Consistency\textsuperscript{a}
& \begin{tabular}[t]{@{}c@{}}
$(\eta,\phi)$ or\\
$(\eta,\kappa)$ or\\
$(\kappa,\xi)$
\end{tabular}
& \begin{tabular}[t]{@{}c@{}}
$(\eta,\kappa)$ or\\
$(\kappa,\xi)$ or\\
$(\eta,\phi,\kappa_{\bullet j})$ or\\
$(\eta,\phi,\kappa_{\bullet(-j)})$
\end{tabular}
& \begin{tabular}[t]{@{}c@{}}
$\eta$ or\\
$(d,s)$ or\\
$(d,u^{(2)})$
\end{tabular}
& \begin{tabular}[t]{@{}c@{}}
$\eta_{\bullet j}$ or\\
$(e_{\bullet j},s_{-j})$ or\\
$(e_{\bullet j},r_{\bullet j}^{(3)})$
\end{tabular} \\
\addlinespace
Semiparametric efficiency\textsuperscript{b}
& $\checkmark$ & $\checkmark$ & $\checkmark$ & $\checkmark$ \\
\bottomrule
\end{tabular*}

\vspace{3pt}
\begin{minipage}{\textwidth}
\fontsize{8}{10}\selectfont
\textit{Notes.}
The checkmarks in the nuisance rows indicate required estimators.
The auxiliary functions' means are evaluated by integration or conditional regression;
$e_{\bullet j}$ is constructed using \eqref{eq:lowdim_e}.
\textsuperscript{a}Each alternative lists nuisance functions whose correct
limits suffice for consistency under the stated regularity conditions.
Indexed conditions apply to every $j$; the low-dimensional results require
compatible auxiliary limits as defined in Section~\ref{sec:lowdim_reparam}.
\textsuperscript{b}Efficiency requires correct nuisance limits and the
convergence and remainder conditions of
Theorems~\ref{thm:CF-AN-MR} and~\ref{thm:lowdim_practical_AN}
for the canonical and low-dimensional estimators, respectively.
Detailed nuisance estimators appear in Tables~\ref{tab:sim-nuisance-fitting}--\ref{tab:lowdim-scenarios}
of Section~\ref{sec:sim-additional-scenarios} of the Appendix.
\end{minipage}
\endgroup
\end{table}
The ratios $\omega^{(1)}$ and $\omega_{\bullet j}^{(2)}$ compare the mediator distribution specified by an intervention with the observed mediator distribution conditional on treatment and $Z$.
They reweight outcome residuals to the relevant intervention distribution.
The functions $u_{\bullet j}^{(1)}$--$u_{\bullet j}^{(3)}$ and
$r_{\bullet j}^{(1)}$--$r_{\bullet j}^{(4)}$ are integrated outcome regressions
used to express the EIFs of $\theta(a,a')$ and
$\tau$, respectively. 
The next theorem gives the EIFs of $\theta(a,a')$ and $\tau$. 

\begin{theorem}\label{thm:EIF}
Suppose that Assumptions~\ref{assm:consistency}--\ref{assm:positivity} hold and
that the displayed scores below are square-integrable.  The EIF of $\theta(a,a')$ in the nonparametric observed-data model with known treatment probability is
$\mc D_{\theta(a,a')}=\{\psi_\theta(a,a';O)-\theta(a,a')W\}/\EE[W]$, where
\begin{align}
%\begin{split}
\psi_\theta(a,a';O) 
&= \frac{W}{N}\sum_{j=1}^N \Bigg\{
\frac{I(A=a)}{\pi^a (1-\pi)^{1-a}}
\omega^{(1)}(a,a',Z,M,C,N)
\left\{Y_{\bullet j}-\eta_{\bullet j}(a,Z,M,C,N)\right\} \nonumber \\
&\qquad\qquad\qquad + \frac{I(A=a)}{\pi^{a} (1-\pi)^{1-a}}
\left\{u_{\bullet j}^{(1)}(a,a',Z,C,N)-u^{(3)}_{\bullet j}(a,a',C,N)\right\} \nonumber \\
&\qquad\qquad\qquad + \frac{I(A=a')}{\pi^{a'} (1-\pi)^{1-a'}}
\left\{u_{\bullet j}^{(2)}(a,M,C,N)-u^{(3)}_{\bullet j}(a,a',C,N)\right\} \nonumber \\
&\qquad\qquad\qquad +u^{(3)}_{\bullet j}(a,a',C,N)\Bigg\}. \label{eq:EIF_theta_formula}
%\end{split}
\end{align}
The EIF of $\tau$ is
$\mc D_{\tau}=\{\psi_\tau(O)-\tau W\}/\EE[W]$, where
\begin{align}
%\begin{split}
\psi_\tau(O) 
&= \frac{W}{N}\sum_{j=1}^N \Bigg\{
\frac{A}{\pi}\omega_{\bullet j}^{(2)}(1,1,0,Z,M,C,N)
\{Y_{\bullet j}-\eta_{\bullet j}(1,Z,M,C,N)\} \nonumber \\
&\qquad\qquad\qquad +\frac{A}{\pi}
\left\{r_{\bullet j}^{(1)}(1,1,0,Z,C,N)-r^{(4)}_{\bullet j}(1,1,0,C,N)\right\} \nonumber \\
&\qquad\qquad\qquad +\frac{A}{\pi}
\left\{r_{\bullet j}^{(2)}(1,0,M_{\bullet j},C,N)-r^{(4)}_{\bullet j}(1,1,0,C,N)\right\} \nonumber \\
&\qquad\qquad\qquad +\frac{1-A}{1-\pi}
\left\{r_{\bullet j}^{(3)}(1,1,M_{\bullet(-j)},C,N)-r^{(4)}_{\bullet j}(1,1,0,C,N)\right\}\\
&\qquad\qquad\qquad +r^{(4)}_{\bullet j}(1,1,0,C,N)\Bigg\}. \label{eq:EIR_tau_formula}
%\end{split}
\end{align}
\end{theorem}
The proof is provided in Section~\ref{sec:EIF_Proof} of the Appendix. Importantly, the EIFs only depend on the observed data $O$ and nuisance parameters $(\eta, \xi,\kappa, \phi)$. 
%The centering by $\theta(a,a')W$ and $\tau W$ accounts for the denominator $\EE[W]$ in the target parameters. Although the weight function is known, its population mean generally is not. For $W=N$, this correction incorporates estimation of the mean cluster size, as in the individual-average influence functions of \citet{cheng2026semiparametric}; for $W=1$, it reduces to ordinary centering by the target mean. The influence functions combine all individual contributions within a cluster, so their variances include the covariance induced by within-cluster dependence.
Under the conditions of Theorem~\ref{thm:EIF}, the variances of $\mc D_{\theta(a,a')}$ and $\mc D_{\tau}$ give the nonparametric efficiency bounds per independent cluster. 
%The EIFs for the mediation effects follow by taking the same contrasts of these component EIFs as in their definitions. 
In the following Section~\ref{sec:canonical_estimators},
we establish the corresponding robustness and asymptotic results for the canonical EIF estimators directly motivated by Theorem \ref{thm:EIF}.

\section{The canonical EIF estimators and their asymptotic behaviors}
\label{sec:canonical_estimators}

We first consider efficient estimators of $\theta(a,a')$ and $\tau$ based on their canonical EIFs. Here, since we will use machine learning models to estimate the nuisance parameters, we leverage the cross-fitting algorithm to control the over-fitting bias associated with the nuisance estimation. Specifically, let $R\ge2$ be fixed and partition the $K$ clusters into validation folds $\mathcal I_1,\ldots,\mathcal I_R$, with $K_r=|\mathcal I_r|$.
For each fold $r\in\{1,\dots,R\}$, we estimate the nuisance parameters $h=(\eta,\phi,\kappa,\xi)$ using the clusters outside $\mathcal I_r$, yielding $\widehat h^{(-r)}$.
All members of a cluster remain in the same fold, so nuisance estimation excludes every observation in a validation cluster \citep{cheng2026semiparametric}.
Then, the canonical EIF estimators are given by 
\begin{equation*}
\widehat\theta_{\mathrm{cf}}(a,a')
=\frac{K^{-1}\sum_{r=1}^R\sum_{i\in\mathcal I_r}
\psi_\theta\{a,a';O_i;\widehat h^{(-r)}\}}
{\mathbb P_K(W)},\quad \widehat\tau_{\mathrm{cf}}
=\frac{K^{-1}\sum_{r=1}^R\sum_{i\in\mathcal I_r}
\psi_\tau\{O_i;\widehat h^{(-r)}\}}
{\mathbb P_K(W)}, 
\end{equation*}
where $\mathbb P_K(f)=K^{-1}\sum_{i=1}^K f(O_i)$. 

With a slight abuse of notations, we use the subscript $0$ to denote true nuisance parameters so that $h_0 = (\eta_0,\phi_0,\kappa_0,\xi_0)$ are true values of $h$ aligning with the true observed data distribution. Further, let $f_{\bullet j}$ be a generic nuisance parameter or auxiliary function with a possibly individual-specific component $j$, and $f_{\bullet j}^{\star}$ be its convergent value of the cross-fitted estimator $\widehat f_{\bullet j}^{(-r)}$, which may differ from the true value $f_{\bullet j,0}$. Theorems~\ref{thm:MR_theta} and~\ref{thm:MR_tau} in
Section~\ref{sec:canonical-bias-details} of the Appendix establish the multiple robustness conditions for the canonical EIF estimators. To describe the multiple robustness property, we use $f_{\bullet j} = f_{\bullet j,0}$ to denote the correct specification of $\widehat f_{\bullet j}^{(-r)}$ converges to $f_{\bullet j,0}$ in $L_2(P)$-norm for every $j$. 
Let $\nu_0:=\EE(W)>0$, and $N_{\rm max}$ denote the largest possible cluster size.
For $\theta(a,a')$,
$\EE\{\psi_\theta(a,a';O;h)\}=\nu_0\theta_0(a,a')$
if either (i) $(\eta,\phi)=(\eta_0,\phi_0)$,
(ii) $(\eta,\kappa)=(\eta_0,\kappa_0)$, or
(iii) $(\kappa,\xi)=(\kappa_0,\xi_0)$.
For $\tau$, $\EE\{\psi_\tau(O;h)\}=\nu_0\tau_0$
if either (i) $(\kappa,\eta)=(\kappa_0,\eta_0)$,
(ii) $(\kappa,\xi)=(\kappa_0,\xi_0)$,
(iii) $(\eta,\phi,\kappa_{\bullet j})
=(\eta_0,\phi_0,\kappa_{0\bullet j})$, or
(iv) $(\eta,\phi,\kappa_{\bullet(-j)})
=(\eta_0,\phi_0,\kappa_{0\bullet(-j)})$.
Under $L_2(P)$ convergence of
$\psi_\theta\{a,a';\bullet;\widehat h^{(-r)}\}$ and
$\psi_\tau\{\bullet;\widehat h^{(-r)}\}$ to their respective limits
uniformly over the fixed number of folds, $\widehat\theta_{\mathrm{cf}}(a,a')$ and
$\widehat\tau_{\mathrm{cf}}$ are consistent whenever their
corresponding conditions above hold at the limiting nuisance functions, even if the remaining nuisance functions are misspecified.

We need two regularity conditions for the asymptotic linearity and asymptotic normality of the canonical EIF estimators and we describe them below. 

\begin{assumption}
\label{assm:crossfit_common}
The number of folds $R$ is fixed and $K_r/K\to p_r\in(0,1)$ for every $r$.
\end{assumption}

Assumption~\ref{assm:crossfit_common} requires each fold to contain a nonvanishing proportion of clusters. The next assumption specifies the nuisance limits and the additional convergence conditions. To do so, we define $\lambda_{f,r}=\max_{1\le j\le N_{\max}}
\|\widehat f_{\bullet j}^{(-r)}-f_{\bullet j}^{\star}\|_{2,P}$ as its estimation error compared to its convergent value under the $L_{2}(P)$-norm, where $P$ denotes the observed-data distribution. We further define
\begin{align*}
\delta_{\eta,r}&=\|\sqrt W\{\widehat\eta^{(-r)}-\eta_0\}\|_{2,P},
&\delta_{\xi,r}&=\left\|\sqrt W\left\{\frac{\xi_0}{\widehat\xi^{(-r)}}-1\right\}\right\|_{2,P},\\*
\delta_{\eta\phi,r}&=\left\|\sqrt W\left\{\eta_0-\widehat\eta^{(-r)}\frac{\widehat\phi^{(-r)}}{\phi_0}\right\}\right\|_{2,P},
&\delta_{g,r}&=\left\|\sqrt W\left\{\frac{\widehat g^{(-r)}}{g_0}-1\right\}\right\|_{2,P},
\end{align*}
as the estimation errors compared to their true values, where $g\in\{\kappa,\kappa_{\bullet j},\kappa_{\bullet(-j)}\}$, and $\delta_{\kappa_{\bullet j},r}$ and $\delta_{\kappa_{\bullet(-j)},r}$ denote the errors for individual $j$'s mediator distribution and the joint mediator distribution of the other cluster members, respectively.

\begin{assumption}
\label{assm:canonical_rates}
There exists a fixed measurable limit
$h^\star=(\eta^\star,\phi^\star,\kappa^\star,\xi^\star)$ such that, uniformly
over the fixed number of folds, $\max_{1\le r\le R}
\{\lambda_{\eta,r}+\lambda_{\phi,r}+\lambda_{\kappa,r}+\lambda_{\xi,r}\}=o_p(1)$.
Assume $\EE\{\psi_\theta(a,a';O;h^\star)^2\}<\infty$, and $\EE\{\psi_\tau(O;h^\star)^2\}<\infty$.
For each target under consideration, impose the corresponding conditions below.

\begin{enumerate}[label=(\roman*),leftmargin=2.2em]
\item For $\theta(a,a')$, assume
$\max_{1\le r\le R}\left\{
\lambda_{\eta,r}+\lambda_{\omega^{(1)},r}
+\sum_{\ell=1}^{3}\lambda_{u^{(\ell)},r}\right\}=o_p(1)$.
The limit $h^\star$ satisfies one of the three conditions in
Theorem~\ref{thm:MR_theta}, and
\begin{equation}
\label{eq:canonical_rate_theta}
\max_{1\le r\le R}
\{\delta_{\kappa,r}\delta_{\eta\phi,r}
+\delta_{\eta,r}\delta_{\xi,r}\}=o_p(K^{-1/2}).
\end{equation}

\item For $\tau$, assume
$\max_{1\le r\le R}\left\{
\lambda_{\eta,r}+\lambda_{\omega^{(2)},r}
+\sum_{\ell=1}^{4}\lambda_{r^{(\ell)},r}\right\}=o_p(1)$.
The limit $h^\star$ satisfies one of the four conditions in
Theorem~\ref{thm:MR_tau}, and
\begin{equation}
\label{eq:canonical_rate_tau}
\begin{aligned}
\max_{1\le r\le R}\max_{1\le j\le N_{\max}}\Big\{
(\delta_{\kappa_{\bullet j},r}+\delta_{\kappa_{\bullet(-j)},r})\delta_{\eta\phi,r}
+\delta_{\kappa_{\bullet j},r}\delta_{\kappa_{\bullet(-j)},r}+\delta_{\eta,r}\delta_{\xi,r}\Big\}=o_p(K^{-1/2}).
\end{aligned}
\end{equation}
\end{enumerate}
\end{assumption}

Assumption~\ref{assm:canonical_rates} allows nuisance functions to converge to possibly misspecified limits, provided the relevant robustness condition holds. The conditions in \eqref{eq:canonical_rate_theta} and \eqref{eq:canonical_rate_tau}, which are relevant to asymptotic linearity, concern products of errors relative to the truth and are stronger than the corresponding zero-bias conditions in Theorems~\ref{thm:MR_theta} and~\ref{thm:MR_tau}.
If one error has a nonzero limit, the paired error must converge faster for its product to be $o_p(K^{-1/2})$.

Define $\psi_{\theta,\mathrm{can}}^\star(O)
:=\psi_\theta(a,a';O;h^\star)$, and $\psi_{\tau,\mathrm{can}}^\star(O)
=\psi_\tau(O;h^\star)$,  where
\begin{equation*}
\begin{aligned}
D_{\theta,\mathrm{can}}^\star(O)
=\frac{\psi_{\theta,\mathrm{can}}^\star(O)-\theta_0(a,a')W}{\nu_0},\qquad 
D_{\tau,\mathrm{can}}^\star(O)
=\frac{\psi_{\tau,\mathrm{can}}^\star(O)-\tau_0W}{\nu_0}.
\end{aligned}
\end{equation*}
We now state the asymptotic expansions for the two cross-fitted estimators. We assume the support and boundedness conditions stated at the beginning of Section~\ref{sec:proof_AN} of the Appendix.

\begin{theorem}
\label{thm:CF-AN-MR}
Under Assumptions~\ref{assm:crossfit_common} and
\ref{assm:canonical_rates}(i),
\begin{equation}
\sqrt K\{\widehat\theta_{\mathrm{cf}}(a,a')-\theta_0(a,a')\}
=\frac1{\sqrt K}\sum_{i=1}^K D_{\theta,\mathrm{can}}^\star(O_i)+o_p(1).
\label{eq:canonical_AL_theta}
\end{equation}
Under Assumptions~\ref{assm:crossfit_common} and
\ref{assm:canonical_rates}(ii),
\begin{equation}
\sqrt K(\widehat\tau_{\mathrm{cf}}-\tau_0)
=\frac1{\sqrt K}\sum_{i=1}^K D_{\tau,\mathrm{can}}^\star(O_i)+o_p(1).
\label{eq:canonical_AL_tau}
\end{equation}
Consequently, the estimators converge to centered Gaussian limits with variances
$\Sigma_{\theta,\mathrm{can}}^\star=\EE\{(D_{\theta,\mathrm{can}}^\star)^2\}$ and
$\Sigma_{\tau,\mathrm{can}}^\star=\EE\{(D_{\tau,\mathrm{can}}^\star)^2\}$,
respectively.  
If
$h^\star=h_0$, the limiting EIFs are the EIFs in
Theorem~\ref{thm:EIF}, and the corresponding variances attain 
nonparametric efficiency bounds.
\end{theorem}

The proof is given in Section~\ref{sec:proof_AN} of the Appendix. Define $\widehat D_{\theta,\mathrm{can},i}
=\frac{\psi_\theta\{a,a';O_i;\widehat h^{(-r)}\}
-\widehat\theta_{\mathrm{cf}}(a,a')W_i}{\mathbb P_K(W)}$ and $\widehat D_{\tau,\mathrm{can},i}=\frac{\psi_\tau\{O_i;\widehat h^{(-r)}\}
-\widehat\tau_{\mathrm{cf}}W_i}{\mathbb P_K(W)}$, for  $i\in\mathcal I_r$. For the purpose of inference, the asymptotic variances $\Sigma_{\theta,\mathrm{can}}^\star$ and $\Sigma_{\tau,\mathrm{can}}^\star$ can be estimated by the cross-fitted sample variances $\mathbb P_K(\widehat D_{\theta,\mathrm{can}}^2)$ and $\mathbb P_K(\widehat D_{\tau,\mathrm{can}}^2)$, respectively, and Wald-type confidence intervals can be constructed. Once we have $\widehat \theta_{\mathrm{cf}}(a,a')$ and $\widehat \tau_{\mathrm{cf}}$, the interventional mediation effects can be obtained immediately.
% For a finite collection of mediation-effect components, the expansions apply jointly when their respective conditions hold.

\section{Low-dimensional reparameterization and alternative efficient estimators}
\label{sec:lowdim_reparam}

Recall that the canonical nuisance parameters $(\eta,\phi,\kappa,\xi)$
determine the auxiliary functions $\omega^{(1)}$,
$\omega_{\bullet j}^{(2)}$, $u_{\bullet j}^{(1)}$--$u_{\bullet j}^{(3)}$,
and $r_{\bullet j}^{(1)}$--$r_{\bullet j}^{(4)}$, which in turn determine
the EIFs in Theorem~\ref{thm:EIF}.
A direct implementation requires estimating conditional densities
of cluster-level vectors whose dimensions increase with cluster
size. Forming ratios of separately estimated densities can amplify
estimation errors, particularly where the denominator density is
small. Evaluating the auxiliary regressions also requires
multivariate integration, which can be computationally demanding
and introduces numerical approximation error when analytic
expressions are unavailable.
To avoid separately estimating multivariate densities and
evaluating their integrals, we re-express these auxiliary functions
using density ratios, treatment probabilities, univariate mediator
densities, and conditional-mean regressions, similar to 
\citet{cheng2026semiparametric}, \citet{diaz2021efficient},
and \citet{rudolph2024practical}.
We denote $\pi_a=\pi^a(1-\pi)^{1-a}$.  Define $\kappa_{\bullet j}^{\star}
(m_{\bullet j}\mid m_{\bullet(-j)},a,c,n)
=f_{M_{\bullet j}\mid M_{\bullet(-j)},A,C,N}
(m_{\bullet j}\mid m_{\bullet(-j)},a,c,n)$, 
\begin{equation}
 d(a,z,m,c,n)=\frac{\kappa(m\mid a,c,n)}{\xi(m\mid a,z,c,n)}.
 \label{eq:lowdim_d}
\end{equation}
\begin{equation}
 s(a\mid m,c,n)=P(A=a\mid M=m,C=c,N=n),
 \label{eq:lowdim_s}
\end{equation}
and $ s_{-j}(a\mid m_{\bullet(-j)},c,n)
 =P\{A=a\mid M_{\bullet(-j)}=m_{\bullet(-j)},C=c,N=n\}$.
For the spillover functional, further define
\begin{equation}
 e_{\bullet j}(z,m,c,n)
 =d(1,z,m,c,n)
 \frac{\kappa_{\bullet j}(m_{\bullet j}\mid1,c,n)}
 {\kappa_{\bullet j}^{\star}
 (m_{\bullet j}\mid m_{\bullet(-j)},1,c,n)}.
 \label{eq:lowdim_e}
\end{equation}
\begin{proposition}
\label{prop:lowdim_identities}
At the observed data distribution, on the common support of the relevant conditional distributions, the auxiliary density ratios in Theorem~\ref{thm:EIF} satisfy
\begin{equation}
\omega^{(1)}(a,a',z,m,c,n)
=d(a,z,m,c,n)\frac{\pi_a}{\pi_{a'}}
\frac{s(a'\mid m,c,n)}{s(a\mid m,c,n)}.
\label{eq:lowdim_omega1}
\end{equation}
\begin{equation}
\omega_{\bullet j}^{(2)}(1,1,0,z,m,c,n)
=e_{\bullet j}(z,m,c,n)\frac{\pi}{1-\pi}
\frac{s_{-j}(0\mid m_{\bullet(-j)},c,n)}
{s_{-j}(1\mid m_{\bullet(-j)},c,n)}.
\label{eq:lowdim_omega2}
\end{equation}
For any square-integrable working outcome regression $\eta$ for which the following expectations exist, the associated auxiliary regressions, with the conditional distributions evaluated under the observed-data distribution, are
\begin{equation}
 u_{\bullet j}^{(1)}(a,a',z,c,n)
 =\EE\{\eta_{\bullet j}(a,z,M,c,n)\mid A=a',C=c,N=n\}.
 \label{eq:lowdim_u1}
\end{equation}
\begin{equation}
 u_{\bullet j}^{(2)}(a,m,c,n)
 =\EE\{\eta_{\bullet j}(a,Z,m,c,n)\mid A=a,C=c,N=n\}.
 \label{eq:lowdim_u2}
\end{equation}
\begin{equation}
\begin{split}
 u_{\bullet j}^{(3)}(a,a',c,n)
 &=\EE\{u_{\bullet j}^{(1)}(a,a',Z,C,N)
 \mid A=a,C=c,N=n\}\\
 &=\EE\{u_{\bullet j}^{(2)}(a,M,C,N)
 \mid A=a',C=c,N=n\}.
\end{split}
\label{eq:lowdim_u3}
\end{equation}
The corresponding representations of $r_{\bullet j}^{(1)}$--$r_{\bullet j}^{(4)}$ use conditional regressions and the univariate densities $\kappa_{\bullet j}$ and $\kappa_{\bullet j}^{\star}$, as given in \eqref{eq:lowdim_r1}--\eqref{eq:lowdim_r4} of the Appendix. At $\eta=\eta_0$, these regressions coincide with the true auxiliary functions defined in Section~\ref{sec:eif}.
\end{proposition}
\noindent The proof is given in Section~\ref{sec: proof:lowdim_identities} of the Appendix. This reparameterization reduces the dimension of the density responses and replaces multivariate integration by conditional regression. Corresponding estimation of the nuisance and auxiliary functions is described in Section~\ref{sec:nuisance-estimation} of the Appendix. In particular, the additional ratio $d$ accounts for conditioning on the post-treatment confounder vector $Z$. We extend the augmented method proposed by \citet{rudolph2024practical} to estimate $d$ from two classifiers distinguishing observed cluster records from reference copies, without separately estimating $\kappa$ and $\xi$; see Section~\ref{sec:augmented_classification} of the Appendix for further details.

In our setting, a nuisance tuple is called \emph{compatible} when its auxiliary
regressions are obtained from the same working outcome regression
$\eta$ through Proposition~\ref{prop:lowdim_identities}, using the
true conditional distributions of $Z$ and $M$ under the
observed-data distribution.
We construct $\widehat\theta_{\mathrm{ld}}(a,a')$ and
$\widehat\tau_{\mathrm{ld}}$ using cluster-level cross-fitting,
as in Section~\ref{sec:canonical_estimators}.
Theorem~\ref{thm:lowdim_drift} establishes multiple robustness
for compatible tuples.
For compatible fitted tuples, Assumptions~\ref{assm:crossfit_common}
and~\ref{assm:lowdim_rates}, together with the stated regularity
conditions, yield root-$K$ asymptotic normality by
Theorem~\ref{thm:lowdim_AN}, with the nonparametric efficiency
bounds attained at the true nuisance limits.
Both theorems and Assumption~\ref{assm:lowdim_rates} are
presented in Section~\ref{sec:lowdim added details} of the Appendix. For separately estimated auxiliary regressions converging to
compatible limits, Theorem~\ref{thm:lowdim_practical_drift}
identifies their additional contributions to
$\EE\{\widehat\psi_{\theta,r}(O)\}-\nu_0\theta_0(a,a')$
and $\EE\{\widehat\psi_{\tau,r}(O)\}-\nu_0\tau_0$.
Theorem~\ref{thm:lowdim_practical_AN} gives sufficient conditions
for consistency and asymptotic normality.
Both theorems, their assumptions, and the corresponding constructions are presented in Section~\ref{sec:lowdim_practical} of the Appendix, along with a discussion of the related compatibility considerations in \citet{rudolph2024practical}.

Panel B in Table~\ref{tab:estimator-summary} summarizes the nuisance estimation
requirements and robustness and efficiency properties of the proposed
EIF estimators.

\section{Simulation studies}
\label{sec:simulation}

We conducted two simulation studies with shared data-generating process to examine the finite-sample performance of canonical and low-dimensional EIF estimators.  
The complete code for implementing the estimators and reproducing the simulation studies is available at \url{https://github.com/daeyoungspam-stat/CRT-interventional-mediation}.
At each
\(K\in\{50,75,100,150\}\), we generated 2,000 independent
cluster-randomized trials. The data-generating model for $O_i$ and true values of interventional mediation effects are given in Section~\ref{sec:sim-data-generating-model} of the Appendix. The nuisance parameters were estimated and auxiliary functions were evaluated using five-fold cluster-level cross-fitting. Both studies used the nuisance fits specified in Table~\ref{tab:sim-nuisance-fitting} of Section~\ref{sec:sim-additional-scenarios} of the Appendix. 
% For numerical stability, we truncated the estimated treatment-assignment probabilities ($s$, $s_{-j}$), the estimated class-membership
% probabilities in the augmented data ($\ell_a$, $\ell_a^M$),
% and the estimated ratios multiplying the outcome residuals
% ($\omega^{(1)}$, $\omega_{\bullet j}^{(2)}$), including the
% intermediate density and odds ratios used to construct them.
We also applied the stabilization motivated by
\citet{cheng2026semiparametric}; see Appendix Section~\ref{sec:sim-stabilization} for implementation considerations including these numerical stabilization procedures
and the construction of EIF-based confidence intervals.
We expect the asymptotic analysis of the outcome regression stabilization described in Appendix Section~\ref{sec:sim-stabilization} (excluding the truncation discussed below) to follow arguments similar to those used to prove Propositions S1 and S2 of \citet{cheng2026semiparametric}, and therefore omit the details here.
%%%%%%%%%%%%%%%%%%%%%%%%%%%%%%%%%%%%%%%
%%%%%%%%%%%%%%%%%%%%%%%%%%%%%%%%%%%%%%%
% Chao: I deleted the following sentence, which I think is a  minor point
% but this can introduce some unnecessary concerns from the reviewer.
% {\color{blue}We additionally apply truncation to improve numerical stability in the presence of the extra density-ratio terms arising in our setting; this practical modification is not covered by our asymptotic theory.
We compared our EIF-based estimators to the g-computation-based plug-in method, which is calculated based on the identification formulas in Theorem \ref{thm:identification} directly.

\subsection{Simulation study for the canonical EIF estimators}
\label{sec:sim-canonical}

The canonical study evaluated all 16 correct/incorrect combinations of
\((\eta,\phi,\kappa,\xi)\), denoted C-\(\eta\phi\kappa\xi\). 
 The correctly specified models used the covariates and interactions listed for each nuisance in Table~\ref{tab:sim-nuisance-fitting} in Section~\ref{sec:sim-additional-scenarios} of the Appendix. The misspecification scenarios omitted baseline covariates and their interactions while considering treatment, cluster size, and the relevant mediator or post-treatment-confounder arguments.
As in Appendix B of \citet{cheng2026semiparametric}, the residual scale and exchangeable correlation
were estimated in each training fold.
We compared stabilized canonical EIF estimators using canonical parametric regression and flexible nuisance estimation.  The latter supplied the same
prespecified correct or misspecified feature sets to Super Learner with
\texttt{SL.library} set to \texttt{c("SL.glm","SL.ranger")}. 
The parametric g-computation used the corresponding
\(\eta,\phi,\kappa\) fits, and exploited a
200-resample cluster bootstrap for inference.

Table~\ref{tab:canonical-scenarios} in
Section~\ref{sec:sim-additional-scenarios} summarizes the guarantees
for point consistency, asymptotic linearity for each estimator, in accordance with
Theorems~\ref{thm:MR_theta} and~\ref{thm:MR_tau} (see Section~\ref{sec:canonical-bias-details} of the Appendix) for point consistency, and Theorem~\ref{thm:CF-AN-MR} for asymptotic linearity. 
Among the scenarios considered here, the asymptotic linearity guarantees in Table~\ref{tab:canonical-scenarios} apply to the theoretical canonical EIF estimators under C-1111 and to parametric g-computation under both C-1111 and C-1110.

\begin{figure}[p]
\centering
\includegraphics[width=0.96\linewidth]{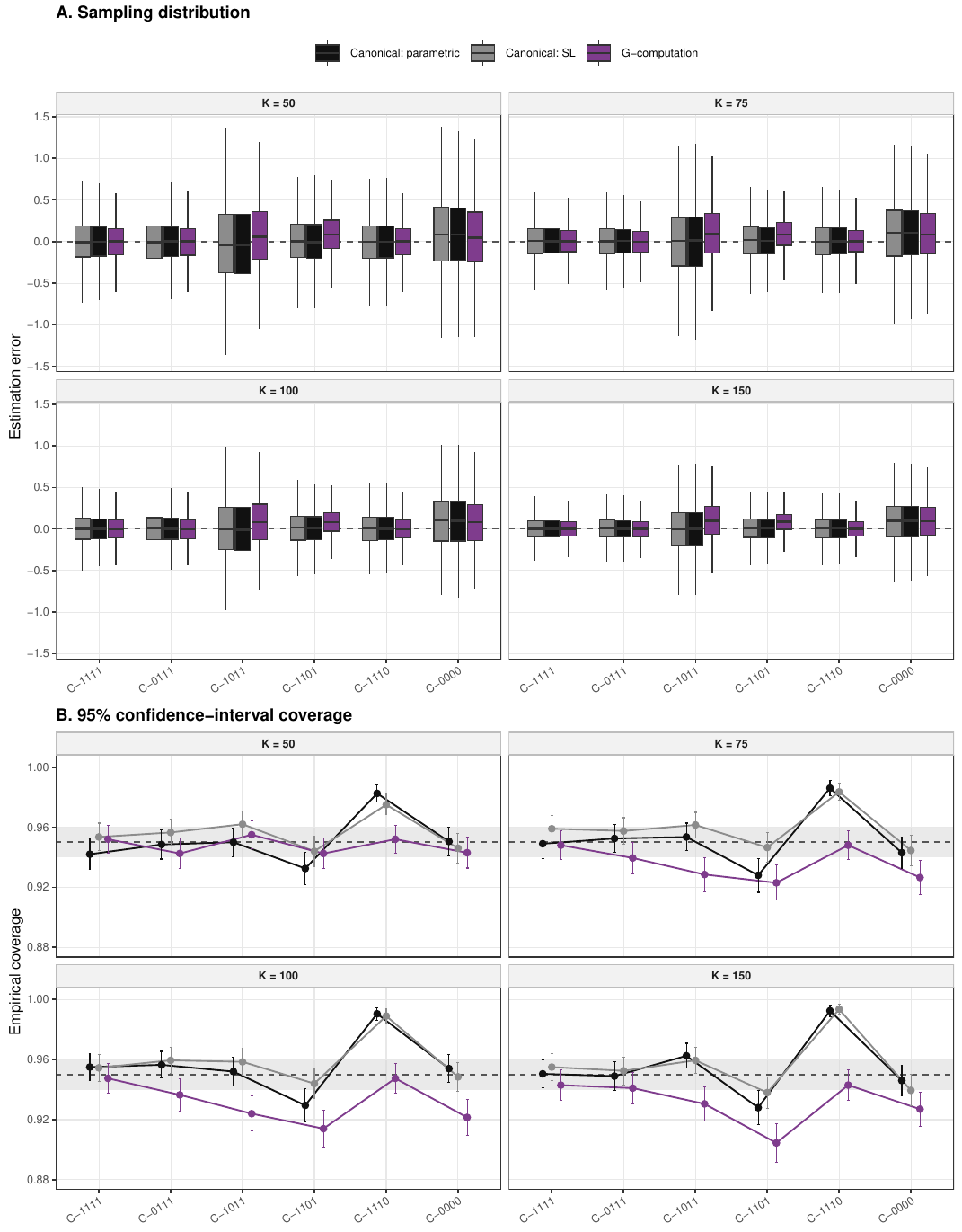}
\caption{Canonical-study performance comparison for the cluster-average IDE in C-1111,
C-0000, and the four scenarios with exactly one wrong nuisance.  Panel A
shows estimation error over 2,000 replications.  Panel B shows empirical coverage with Monte Carlo error bars.}
\label{fig:sim-main-can-IDE-C}
\end{figure}

\begin{figure}[p]
\centering
\includegraphics[width=0.96\linewidth]{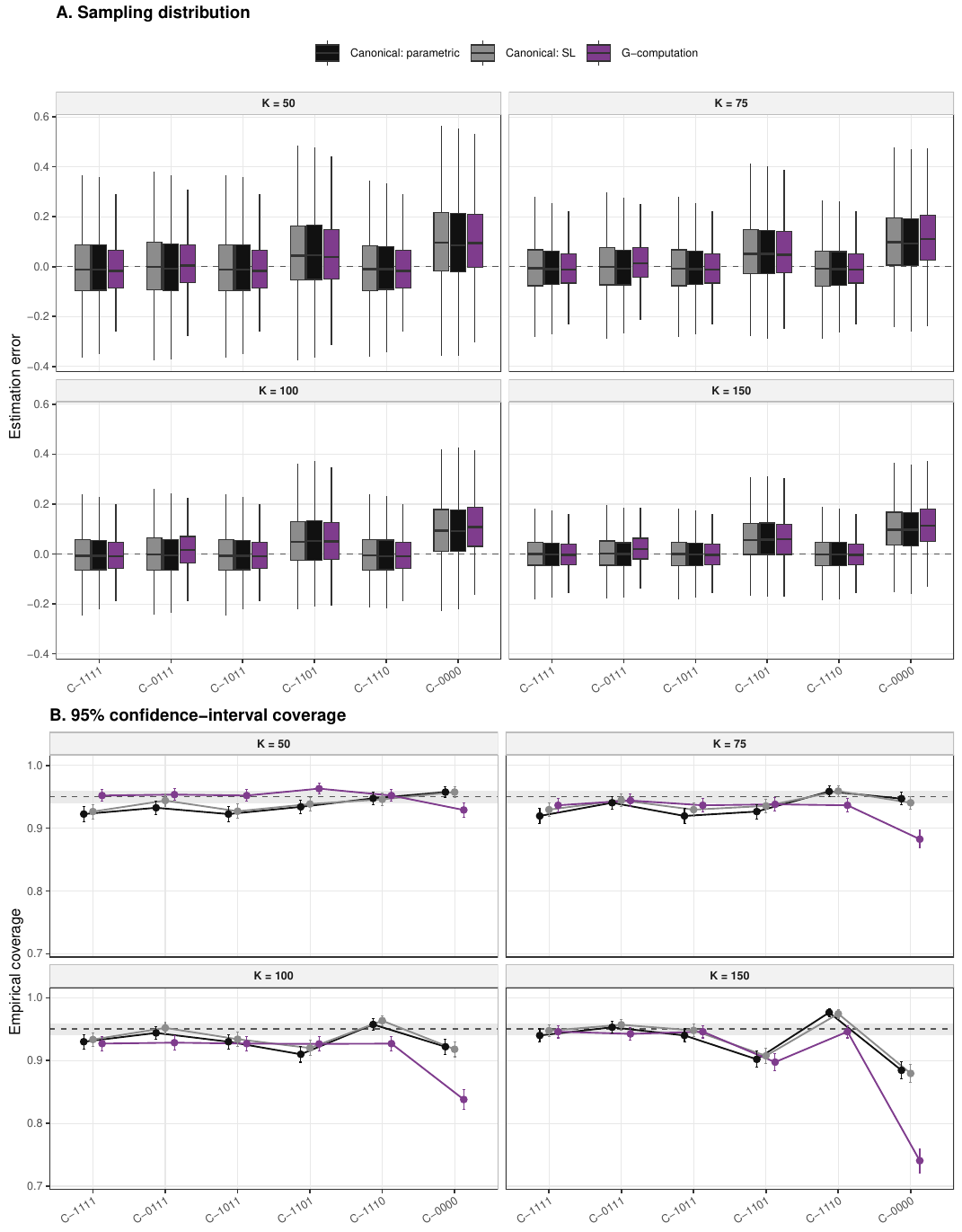}
\caption{Canonical-study performance comparison for the cluster-average ISME in the six scenarios as in
Figure~\ref{fig:sim-main-can-IDE-C}.}
\label{fig:sim-main-can-ISME-C}
\end{figure}

Figures~\ref{fig:sim-main-can-IDE-C} and~\ref{fig:sim-main-can-ISME-C} display the cluster-average IDE and ISME results, respectively. Under C-1111, both canonical EIF methods remained approximately unbiased, and their performance improved as $K$ increased.  The IDE coverage remained close to nominal, whereas the mild ISME undercoverage at smaller $K$ diminished as $K$
increased. The g-computation was also approximately unbiased with coverage close to nominal, reflecting the correct specification of its parametric models. When only $\eta$ was misspecified (C-0111), both EIF estimators remained approximately unbiased for IDE and ISME, as expected from the point-consistency condition requiring correct $(\kappa,\xi)$. Their coverage was generally close to nominal at larger $K$. The g-computation exhibited little IDE bias, but its ISME bias persisted as $K$ increased.  When only $\phi$ was misspecified (C-1011), both EIF estimators again remained approximately unbiased, now through correct $(\eta,\kappa)$.
%The loss of precision relative to C-1111 was much more pronounced for IDE than for ISME.  The IDE estimation remained less precise than under C-1111. The G-computation still showed IDE bias, with coverage tending to deteriorate as $K$ increased. 

The distinction between point consistency and inference was
particularly evident when $\kappa$ or $\xi$ was misspecified. Under C-1101, correct $(\eta,\phi)$ sufficed for IDE point consistency. Accordingly, both EIF estimators remained close to the IDE target and became more concentrated with $K$, but their intervals undercovered, more noticeably for the
parametric implementation. For ISME, the same nuisance
combination did not satisfy a point-consistency condition:
bias persisted, and coverage deteriorated as $K$ increased.
Under C-1110, both IDE and ISME estimates remained unbiased, but coverage became conservative, especially for IDE.

Under C-0000, the sampling distributions narrowed with $K$
but remained displaced from the targets. For ISME, coverage
deteriorated for all three estimators, with a substantially larger under-coverage for g-computation. The IDE coverage for the EIF estimators remained comparatively close to nominal despite persistent bias and considerably greater variation than under C-1111. We note that Figures~\ref{fig:sim-app-can-IDE-C}--\ref{fig:sim-app-can-ISME-C} in Section~\ref{sec:sim-additional-cluster-results} extend these comparisons to all 16 nuisance scenarios and include IIME. The individual-average results in Figures~\ref{fig:sim-app-can-IDE-I}--\ref{fig:sim-app-can-ISME-I} of
Section~\ref{sec:sim-individual-results} show similar patterns.

\subsection{Simulation study for the low-dimensional EIF estimators}
\label{sec:sim-lowdim}

To estimate the density ratio $d$, we used the
augmented-classification approach of \citet{rudolph2024practical},
adapted to clustered data.
For each observed training cluster, we created random copies
by replacing its mediator vector with a draw from an auxiliary
distribution $\widetilde\lambda$, while keeping its treatment,
post-treatment confounders, baseline covariates, and cluster size
unchanged.
The distribution $\widetilde\lambda$ could depend on treatment,
baseline covariates, and cluster size, but not on the
post-treatment confounders.
Two binary classifiers were then trained to distinguish the generated copies from the observed data, and their fitted probabilities were combined to estimate $d$.
We considered two constructions based on fitted arm-specific
mediator means: Gaussian residual draws (G) and independent
empirical residual-bootstrap draws (RB).
Each mechanism was combined with glmnet, random forest (RF), or Super Learner (SL), yielding six different versions.
The definition of the reference distribution and the
classification-based representation of $d$ are provided in
Section~\ref{sec:augmented_classification}.
The nuisance estimation stage used the same cluster-level cross-fitting scheme as the canonical study.
The working models and implementation details are provided in
Section~\ref{sec:sim-additional-scenarios}.

The scenario label LD-AC denotes all-correct specification.
The labels LD-T2--LD-T6 vary the specification of the nuisances for
$\theta$ while holding correct specification for $\tau$;
LD-U2--LD-U6 reverse these roles.
Table~\ref{tab:lowdim-scenarios} in Section~\ref{sec:sim-additional-scenarios}
summarizes the scenario-specific guarantees for point
consistency and asymptotic linearity, following Theorems~\ref{thm:lowdim_drift} and~\ref{thm:lowdim_AN}, together with Theorem~\ref{thm:lowdim_practical_AN} for estimated auxiliary functions.
Among these simulation scenarios, asymptotic linearity with
the corresponding EIFs, and hence the corresponding interval
validity, is guaranteed only under LD-AC (as in the canonical study), subject to the
stated rate conditions.

\begin{figure}[p]
\centering
\includegraphics[width=0.96\linewidth]{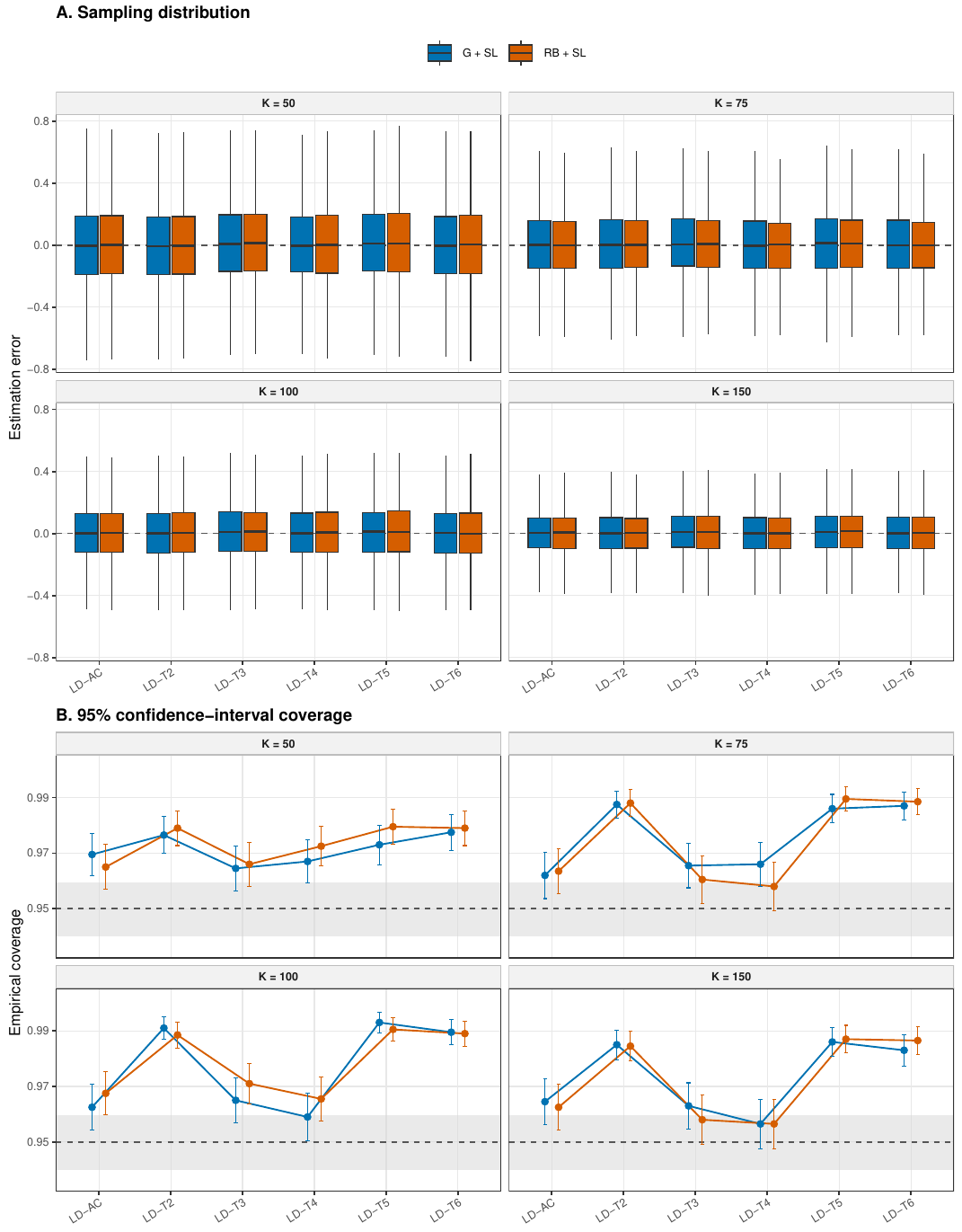}
\caption{Low-dimensional-study performance comparison for the cluster-average IDE,
comparing G+SL and RB+SL in LD-AC and LD-T2--LD-T6.  Plot elements are as in
Figure~\ref{fig:sim-main-can-IDE-C}.}
\label{fig:sim-main-ld-IDE-C}
\end{figure}

\begin{figure}[p]
\centering
\includegraphics[width=0.96\linewidth]{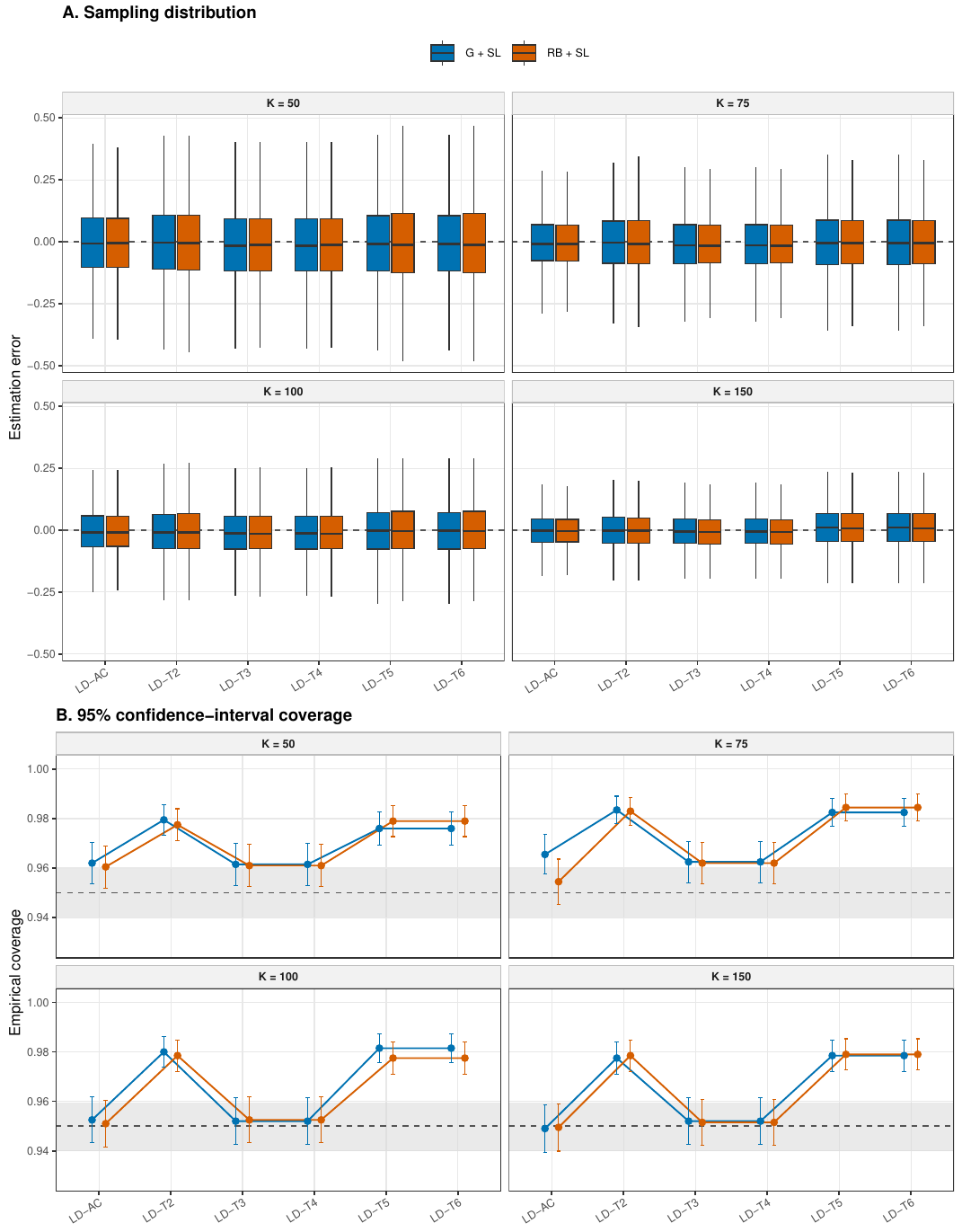}
\caption{Low-dimensional-study performance comparison for the cluster-average ISME for
the same estimators and scenarios as
Figure~\ref{fig:sim-main-ld-IDE-C}.}
\label{fig:sim-main-ld-ISME-C}
\end{figure}

Figures~\ref{fig:sim-main-ld-IDE-C} and~\ref{fig:sim-main-ld-ISME-C}
display the cluster-average IDE and ISME
results for G+SL and RB+SL under LD-AC and LD-T2--LD-T6.
Under LD-AC, both estimators remained approximately centered,
and their sampling distributions became more concentrated
as $K$ increased.
The ISME coverage approached nominal at larger $K$, whereas
IDE intervals remained mildly conservative.
The two reference distributions yielded very similar
centering, dispersion, and coverage throughout the displayed
scenarios, indicating little sensitivity to this choice
in the present Gaussian design.

Under LD-T2 and LD-T3, both effects satisfied the
point-consistency conditions.
The estimates remained close to their targets.
Their coverage patterns nevertheless differed:
LD-T2 intervals remained more conservative than those
under LD-AC, whereas LD-T3 coverage was closer to nominal
at larger $K$, particularly for ISME.
Thus, the two scenarios exhibited similar point-estimation
performance but different behavior of the confidence intervals.

LD-T4 distinguished the nuisance requirements for the two
effects.
ISME has a point-consistency guarantee and remained
close to its target, with coverage approaching nominal
as $K$ increased.
IDE also exhibited little bias and increasingly concentrated
estimates, although its nuisance specification did not
satisfy a point-consistency condition.
Under LD-T5 and LD-T6, neither displayed effect had a
point-consistency guarantee.
The observed biases were modest over the examined values
of $K$, but the intervals remained more conservative than
under LD-AC, especially for IDE.

Additional scenarios and IIME results are reported in
Figures~\ref{fig:sim-app-ld-IDE-C}--\ref{fig:sim-app-ld-ISME-C}
of Section~\ref{sec:sim-additional-cluster-results}.
These results further illustrate the effect-specific
point-consistency guarantees.
The individual-average results in
Figures~\ref{fig:sim-app-ld-IDE-I}--\ref{fig:sim-app-ld-ISME-I}
of Section~\ref{sec:sim-individual-results} showed similar patterns.

Across the six estimators under LD-AC, glmnet and RF had
similar RMSE, and the SL implementations achieved comparable
point-estimation performance.
The SL implementations generally produced wider intervals
than glmnet and RF, particularly for IDE and ISME at smaller
$K$.
The comprehensive simulation reports can be found in Section~\ref{sec:sim-additional-results}. 

\section{Data application: Analysis of the PPACT cluster-randomized trial}
\label{sec:ppact-application}

We reanalyzed the Pain Program for Active Coping and Training (PPACT) trial \citep{debar2022primary}, a pragmatic cluster-randomized trial to evaluate a primary care-based cognitive behavioral therapy intervention among adults with chronic pain receiving long-term opioid therapy. The PPACT trial comprises 850 participants within 106 primary care provider clusters, with cluster sizes ranging from 3 to 13 participants. In our mediation analysis, the outcome is pain impact at 12 months, measured using the 0--10 PEGS scale, for which higher values indicate greater pain impact. We consider the Roland--Morris Disability Questionnaire (RMDQ) score at 6 months (range 0--1, with higher values indicating greater pain-related disability) as the mediator of interest. Satisfaction with primary care services at 6 months (range 1--5) was treated as the post-treatment confounder of the mediator--outcome relationship. We adjusted for cluster size and the following individual-level baseline covariates: sex, age, diagnosis of at least two chronic conditions, count of pain types, benzodiazepine receipt during the 6 months before randomization, baseline PEGS score, baseline RMDQ score, and average daily opioid dose during the 6 months before randomization. % Missing values were imputed once by predictive mean matching using the \texttt{mice} package, so the analysis pertains to this single completed data set.

We applied the five estimators described in the simulation study (Section~\ref{sec:simulation}) to investigate the cluster-average interventional mediation effects, including parametric g-computation, the stabilized canonical EIF estimator with parametric nuisance estimation, the stabilized canonical EIF estimator with flexible nuisance estimation, and the low-dimensional EIF estimators based on either a Gaussian reference distribution or a residual-bootstrap reference distribution. The g-computation and parametric canonical EIF estimators used generalized linear models only. For the flexible canonical estimator, the Super Learner library contained generalized linear models, elastic-net regression, and random forests; the low-dimensional estimators used the corresponding prespecified Super Learner libraries. %Confidence intervals for g-computation were obtained from 1,000 cluster bootstrap samples; the confidence intervals used cluster-level influence-function variances and a $t_{K-3}$ critical value.

\begin{figure}[t]
  \centering
  \includegraphics[width=\textwidth]{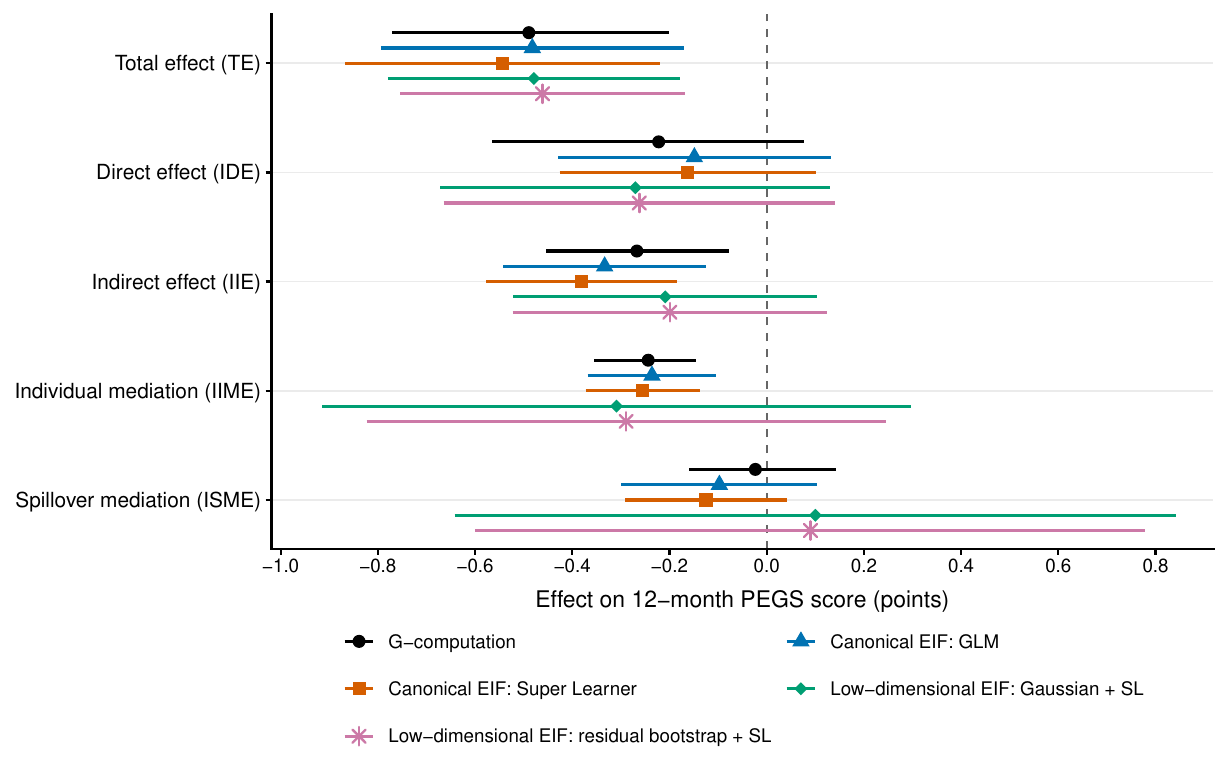}
  \caption{Estimated effects of the PPACT cognitive behavioral therapy intervention on the 12-month PEGS score under cluster-average weighting ($W_i=1$). Points are estimates and horizontal bars are 95\% confidence intervals. TE denotes the total effect, IDE the interventional direct effect, IIE the interventional indirect effect, IIME the interventional individual mediation effect, and ISME the interventional spillover mediation effect. Negative values favor the intervention because higher PEGS scores indicate worse pain impact.}
  \label{fig:ppact-effects}
\end{figure}

The mediation analysis results are shown in Figure~\ref{fig:ppact-effects}. First, all five estimators indicated that the intervention reduced pain impact at 12 months. The TEs ranged from $-0.54$ to $-0.46$ PEGS points, and every 95\% confidence interval excluded zero. Second, IDEs ranged from $-0.27$ to $-0.15$, which are all statistically insignificant. The indirect effects were consistently negative. For the IIEs, the g-computation and both canonical EIF intervals excluded zero, whereas the intervals from the two low-dimensional estimators were wider and included zero. Third, the decomposition effects from IIE (i.e., the IIME and ISME) suggested that the mediated pathway operated primarily through the participant's own RMDQ score rather than through same-cluster spillover. The IIMEs ranged from $-0.31$ to $-0.24$; their confidence intervals excluded zero for g-computation and the two canonical EIF estimators but not for the low-dimensional EIF estimators. By comparison, the ISMEs were smaller and less stable, ranging from $-0.12$ to $0.10$, and all intervals included zero. Taken together, the results provide evidence of a protective total effect of the intervention, which also suggests some evidence for existence of indirect effect. However, the results provide little evidence of spillover mediation through interactions with other participants in the same provider cluster. The wider intervals from the low-dimensional EIF estimators, particularly for the IIME and ISME, warrant cautious interpretation of the detailed decomposition. We remark that the low-dimensional EIF estimators also yielded wider intervals than their canonical EIF counterparts in the simulation studies, particularly for ISME (see Section~\ref{sec:sim-additional-cluster-results} in the Appendix).

Based on the canonical EIF estimator with machine learning nuisance estimation, we compared interventional mediation effects, which account for satisfaction with primary care services as a post-treatment confounder, with the corresponding natural mediation effects, which do not adjust for this confounder. As shown in Supplementary Table~\ref{tab:ppact-comparison} in Appendix~\ref{sec:ppact-comparison}, the estimated indirect effect, individual mediation effect, and spillover mediation effect were similar across the two analyses. However, the direct effect differed substantially. The interventional direct effect was $-0.163$ (95\% CI: $-0.426$, $0.101$), whereas the natural direct effect was $-0.277$ (95\% CI: $-0.502$, $-0.052$). This suggests that accounting for post-treatment confounding has little influence on the substantive conclusion regarding mediation through RMDQ but materially changes the magnitude and significance of the direct  effect. Because the two analyses target different estimands, however, this difference should be interpreted descriptively and not a formal estimate of confounding bias.

\section{Conclusion}
We develop, to our knowledge, the first semiparametric framework for mediation analysis in cluster-randomized trials that jointly accommodates treatment-induced mediator–outcome confounding and within-cluster spillover.
The central barrier is that treatment can alter common causes of the mediator and outcome, thereby hindering identification of natural mediation effects. By defining stochastic interventions on the joint within-cluster mediator distribution, we identify interventional direct and indirect effects without requiring cross-world independence assumptions and decompose the indirect effect into individual and spillover components. 
The proposed estimands quantify the consequences of hypothetical interventions on the joint mediator distribution and can differ from treatment effects carried through individual causal pathways, thereby providing an intervention-oriented perspective on the role of mediators that complements natural effects.
For these estimands, we derive efficient influence functions and construct cluster-level cross-fitted estimators with multiple robustness properties, establishing asymptotic normality and semiparametric efficiency under suitable nuisance estimation conditions.
A lower-dimensional reparameterization facilitates flexible nuisance estimation while reducing the burden of multivariate density estimation and integration. 
Extensions to multiple mediators and sensitivity analyses for residual mediator–outcome confounding merit further investigation.

\section*{Acknowledgement}
Research in this article was supported by the Patient-Centered Outcomes Research Institute\textsuperscript{\textregistered} (PCORI\textsuperscript{\textregistered} Award ME-2023C1-31350). The statements presented in this article are solely the responsibility of the authors and do not necessarily represent the views of PCORI\textsuperscript{\textregistered}, its Board of Governors or Methodology Committee.

\bibliography{allref}

% PRESERVED MANUSCRIPT END

\clearpage
\phantomsection
\pdfbookmark[0]{Supplementary Material}{supplement}
\begin{center}
{\Large\bfseries Supplementary Material\par}
\medskip
{\large Semiparametric causal mediation analysis in cluster-randomized trials with post-treatment confounding\par}
\medskip
Daeyoung Ham, Chao Cheng, and Fan Li
\end{center}
\setcounter{figure}{0}
\setcounter{table}{0}
\renewcommand{\thefigure}{S\arabic{figure}}
\renewcommand{\thetable}{S\arabic{table}}
\renewcommand{\theHfigure}{supp.\arabic{figure}}
\renewcommand{\theHtable}{supp.\arabic{table}}
% PRESERVED APPENDIX START
\appendix

\section{Additional technical details for the low-dimensional estimator}\label{sec:lowdim added details}

This section gives the identities and explicit score and estimator formulas underlying the low-dimensional reparameterization, its exact bias decompositions and multiple-robustness conditions, and the assumptions and asymptotic results for cross-fitted inference.

We give the spillover-regression representations stated in Proposition~\ref{prop:lowdim_identities} of Section~\ref{sec:lowdim_reparam}. For any square-integrable working regression $\eta$, the compatible partial means for $\tau$ can be written using conditional regressions and densities with the scalar response $M_{\bullet j}$, with the conditional distributions evaluated under the observed-data distribution:
\begin{equation}
\begin{split}
r_{\bullet j}^{(1)}(1,1,0,z,c,n)=\EE\left[
 \eta_{\bullet j}(1,z,M,c,n)
 \frac{\kappa_{\bullet j}(M_{\bullet j}\mid1,c,n)}
 {\kappa_{\bullet j}^{\star}
 (M_{\bullet j}\mid M_{\bullet(-j)},0,c,n)}
 \ \middle|\ A=0,C=c,N=n\right].
\end{split}
\label{eq:lowdim_r1}
\end{equation}
\begin{equation}
\begin{split}
&r_{\bullet j}^{(2)}(1,0,m_{\bullet j},c,n)\\
&\quad=\EE\left[
 u_{\bullet j}^{(2)}(1,M,C,N)
 \frac{\kappa_{\bullet j}(M_{\bullet j}\mid0,C,N)}
 {\kappa_{\bullet j}^{\star}
 (M_{\bullet j}\mid M_{\bullet(-j)},0,C,N)}
 \ \middle|\ A=0,M_{\bullet j}=m_{\bullet j},C=c,N=n\right].
\end{split}
\label{eq:lowdim_r2}
\end{equation}
\begin{equation}
\begin{split}
&r_{\bullet j}^{(3)}(1,1,m_{\bullet(-j)},c,n)\\
&\quad=\EE\left[
 u_{\bullet j}^{(2)}(1,M,C,N)
 \frac{\kappa_{\bullet j}(M_{\bullet j}\mid1,C,N)}
 {\kappa_{\bullet j}^{\star}
 (M_{\bullet j}\mid M_{\bullet(-j)},1,C,N)}
 \ \middle|\ A=1,M_{\bullet(-j)}=m_{\bullet(-j)},C=c,N=n\right].
\end{split}
\label{eq:lowdim_r3}
\raisetag{-12pt}
\end{equation}
\begin{equation}
\begin{split}
 r_{\bullet j}^{(4)}(1,1,0,c,n)
 &=\int r_{\bullet j}^{(2)}(1,0,m_{\bullet j},c,n)
 \kappa_{\bullet j}(m_{\bullet j}\mid1,c,n)\,dm_{\bullet j}\\
 &=\EE\{r_{\bullet j}^{(3)}(1,1,M_{\bullet(-j)},C,N)
 \mid A=0,C=c,N=n\}.
\end{split}
\label{eq:lowdim_r4}
\end{equation}
When $\eta=\eta_0$, these functions coincide with the auxiliary functions in
Section~\ref{sec:eif}.

\noindent The proof is given in Section~\ref{sec: proof:lowdim_identities} of the Appendix.

Substituting the auxiliary functions introduced in Section~\ref{sec:lowdim_reparam} into the canonical scores, we obtain
\begin{equation}
\label{eq:lowdim_score_theta}
\resizebox{\linewidth}{!}{$
\begin{aligned}
&\psi_{\theta,\mathrm{ld}}(a,a';O)\\
&=\frac{W}{N}\sum_{j=1}^{N}\Bigg[
 \frac{I(A=a)}{\pi_a}
 d(a,Z,M,C,N)\frac{\pi_a}{\pi_{a'}}
 \frac{s(a'\mid M,C,N)}{s(a\mid M,C,N)}
 \{Y_{\bullet j}-\eta_{\bullet j}(a,Z,M,C,N)\}\\
&\qquad\qquad\qquad+\frac{I(A=a)}{\pi_a}
 \{u_{\bullet j}^{(1)}(a,a',Z,C,N)
 -u_{\bullet j}^{(3)}(a,a',C,N)\}\\
&\qquad\qquad\qquad+\frac{I(A=a')}{\pi_{a'}}
 \{u_{\bullet j}^{(2)}(a,M,C,N)
 -u_{\bullet j}^{(3)}(a,a',C,N)\}
 +u_{\bullet j}^{(3)}(a,a',C,N)
\Bigg],
\end{aligned}
$}
\end{equation}
and
\begin{equation}
\label{eq:lowdim_score_tau}
\resizebox{\linewidth}{!}{$
\begin{aligned}
&\psi_{\tau,\mathrm{ld}}(O)\\
&=\frac{W}{N}\sum_{j=1}^{N}\Bigg[
 \frac{A}{\pi}
 e_{\bullet j}(Z,M,C,N)\frac{\pi}{1-\pi}
 \frac{s_{-j}(0\mid M_{\bullet(-j)},C,N)}
 {s_{-j}(1\mid M_{\bullet(-j)},C,N)}
 \{Y_{\bullet j}-\eta_{\bullet j}(1,Z,M,C,N)\}\\
&\qquad\qquad\qquad+\frac{A}{\pi}
 \{r_{\bullet j}^{(1)}(1,1,0,Z,C,N)
 -r_{\bullet j}^{(4)}(1,1,0,C,N)\}\\
&\qquad\qquad\qquad+\frac{A}{\pi}
 \{r_{\bullet j}^{(2)}(1,0,M_{\bullet j},C,N)
 -r_{\bullet j}^{(4)}(1,1,0,C,N)\}\\
&\qquad\qquad\qquad+\frac{1-A}{1-\pi}
 \{r_{\bullet j}^{(3)}(1,1,M_{\bullet(-j)},C,N)
 -r_{\bullet j}^{(4)}(1,1,0,C,N)\}
 +r_{\bullet j}^{(4)}(1,1,0,C,N)
\Bigg].
\end{aligned}
$}
\end{equation}

The following lemma shows that the reparameterized scores recover the canonical scores at the true nuisance functions.

\begin{lemma}
\label{lem:lowdim_canonical}
At the true nuisance functions,
$\psi_{\theta,\mathrm{ld}}(a,a';O)$ equals
$\psi_\theta(a,a';O)$ in \eqref{eq:EIF_theta_formula}, and
$\psi_{\tau,\mathrm{ld}}(O)$ equals
$\psi_\tau(O)$ in \eqref{eq:EIR_tau_formula}.
\end{lemma}

\noindent The proof is provided in Section~\ref{sec: lem:lowdim_canonical proof} of the Appendix.
Thus, the reparameterization preserves the canonical gradients at the true nuisance functions.

\subsection{Bias decomposition and multiple robustness}\label{sec: lowdim MR appendix}

The next theorem gives exact bias decompositions for the reparameterized scores within the compatible class.

\begin{theorem}
\label{thm:lowdim_drift}
For every compatible candidate nuisance tuple,
\begin{equation}
\label{eq:lowdim_drift_theta}
\resizebox{\linewidth}{!}{$
\begin{aligned}
&\EE\{\psi_{\theta,\mathrm{ld}}(a,a';O)\}
-\EE(W)\theta_0(a,a')\\
&=\EE\Bigg[\frac{W}{N}\sum_{j=1}^{N}
\EE\Bigg[
\frac{\pi_a}{\pi_{a'}}
\frac{s(a'\mid M,C,N)}{s(a\mid M,C,N)}
(d-d_0)(a,Z,M,C,N)
(\eta_{0\bullet j}-\eta_{\bullet j})(a,Z,M,C,N)\\
&\qquad \qquad \qquad \qquad 
+\frac{\pi_a}{\pi_{a'}}
\left\{
\frac{s(a'\mid M,C,N)}{s(a\mid M,C,N)}
-\frac{s_0(a'\mid M,C,N)}{s_0(a\mid M,C,N)}
\right\}
\{u_{0\bullet j}^{(2)}-u_{\bullet j}^{(2)}\}(a,M,C,N)
\ \Big|\ A=a,C,N
\Bigg]\Bigg].
\end{aligned}
$}
\end{equation}
Moreover,
\begin{equation}
\label{eq:lowdim_drift_tau}
\resizebox{\linewidth}{!}{$
\begin{aligned}
&\EE\{\psi_{\tau,\mathrm{ld}}(O)\}-\EE(W)\tau_0\\
&=\EE\Bigg[\frac{W}{N}\sum_{j=1}^{N}
\EE\Bigg[
\frac{\pi}{1-\pi}
\frac{s_{-j}(0\mid M_{\bullet(-j)},C,N)}
{s_{-j}(1\mid M_{\bullet(-j)},C,N)}
(e_{\bullet j}-e_{0\bullet j})(Z,M,C,N)
(\eta_{0\bullet j}-\eta_{\bullet j})(1,Z,M,C,N)\\
&\qquad \qquad \qquad \qquad 
+\frac{\pi}{1-\pi}
\left\{
\frac{s_{-j}(0\mid M_{\bullet(-j)},C,N)}
{s_{-j}(1\mid M_{\bullet(-j)},C,N)}
-\frac{s_{0,-j}(0\mid M_{\bullet(-j)},C,N)}
{s_{0,-j}(1\mid M_{\bullet(-j)},C,N)}
\right\}
\{r_{0\bullet j}^{(3)}-r_{\bullet j}^{(3)}\}
(1,1,M_{\bullet(-j)},C,N)
\ \Big|\ A=1,C,N
\Bigg]\Bigg].
\end{aligned}
$}
\end{equation}
\end{theorem}

\noindent The proof is given in Section~\ref{sec: proof thm:lowdim_drift} of the Appendix.
Each decomposition consists of two products of nuisance errors.
Theorem~\ref{thm:lowdim_drift} gives the following sufficient conditions for zero population bias for $\theta(a,a')$:
\begin{equation}
\label{eq:lowdim_MR_theta}
\resizebox{\linewidth}{!}{$
\displaystyle
\left\{d=d_0\ \text{or}\ \eta=\eta_0\right\}
\quad\text{and}\quad
\left\{
\frac{s(a'\mid M,C,N)}{s(a\mid M,C,N)}
=\frac{s_0(a'\mid M,C,N)}{s_0(a\mid M,C,N)}
\ \text{or}\ u^{(2)}=u_0^{(2)}
\right\}.
$}
\end{equation}
For $\tau$, the corresponding conditions are, for every $j$,
\begin{equation}
\label{eq:lowdim_MR_tau}
\resizebox{\linewidth}{!}{$
\displaystyle
\left\{e_{\bullet j}=e_{0\bullet j}\ \text{or}\
\eta_{\bullet j}=\eta_{0\bullet j}\right\}
\quad\text{and}\quad
\left\{
\frac{s_{-j}(0\mid M_{\bullet(-j)},C,N)}
{s_{-j}(1\mid M_{\bullet(-j)},C,N)}
=\frac{s_{0,-j}(0\mid M_{\bullet(-j)},C,N)}
{s_{0,-j}(1\mid M_{\bullet(-j)},C,N)}
\ \text{or}\ r_{\bullet j}^{(3)}=r_{0\bullet j}^{(3)}
\right\}.
$}
\end{equation}
Under compatibility, correct $\eta$ also gives the correct $u^{(2)}$ and $r_{\bullet j}^{(3)}$, making both population bias terms zero.
Otherwise, correct $d$ or $e_{\bullet j}$ removes the first term, while the second term vanishes if either the corresponding classifier odds or integrated regression is correct.
For $a=a'$, the classifier-odds ratio is identically one, so the second term for $\theta(a,a)$ is zero.
When $e_{\bullet j}$ is constructed through \eqref{eq:lowdim_e}, correctness of $d(1,\bullet)$ and $\kappa_{\bullet j}/\kappa_{\bullet j}^{\star}$ is sufficient for its correctness.

\subsection{Asymptotic properties}\label{sec: lowdim AN appendix}

Use the same cluster-level folds as in Section~\ref{sec:canonical_estimators}.
For each fold $\mathcal I_r$, estimate the low-dimensional nuisance functions and their associated auxiliary regressions outside the fold and evaluate the scores on its validation clusters.

The results through Theorem~\ref{thm:lowdim_AN} concern fitted tuples
in the compatible class. Section~\ref{sec:lowdim_practical} gives
conditions for \eqref{eq:lowdim_theta_est} and
\eqref{eq:lowdim_tau_est} with estimated auxiliary regressions.
Define
\begin{equation}
\widehat\theta_{\mathrm{ld}}(a,a')
=\frac{K^{-1}\sum_{r=1}^{R}\sum_{i\in\mathcal I_r}
\widehat\psi_{\theta,r}(O_i)}{\mathbb P_K(W)}.
\label{eq:lowdim_theta_est}
\end{equation}
\begin{equation}
\widehat\tau_{\mathrm{ld}}
=\frac{K^{-1}\sum_{r=1}^{R}\sum_{i\in\mathcal I_r}
\widehat\psi_{\tau,r}(O_i)}{\mathbb P_K(W)}.
\label{eq:lowdim_tau_est}
\end{equation}
Let
$h_\theta=(\eta,d,s,u^{(1)},u^{(2)},u^{(3)})$ and
$h_\tau=(\eta,e,s_{-j},r^{(1)},r^{(2)},r^{(3)},r^{(4)})$.
These tuples include both the nuisance functions $(\eta,d,s)$ or $(\eta,e,s_{-j})$ and the associated auxiliary regressions $u^{(1)},u^{(2)},u^{(3)}$ or $r^{(1)},r^{(2)},r^{(3)},r^{(4)}$, so that the convergence requirements below cover both.
Again, for vector-indexed nuisance or auxiliary functions, every norm below is understood as the
maximum over $j\le N_{\max}$.

The following assumption specifies the nuisance limits and error products for the reparameterized scores.
Denote $\nu_0=\EE(W)\in(0,\infty)$.
\begin{assumption}
\label{assm:lowdim_rates}
The weight $W$ is nonnegative and bounded.
For each target, there exists a fixed compatible nuisance limit and the
following conditions hold uniformly over the fixed number of folds.
\begin{enumerate}[label=(\roman*),leftmargin=2.2em]
\item For $\theta(a,a')$,
\begin{equation}
\label{eq:lowdim_limit_theta}
\begin{aligned}
&\|\widehat\eta^{(-r)}-\eta^\star\|_{2,P}
+\|\widehat d^{(-r)}-d^\star\|_{2,P}
+\left\|
\frac{\widehat s^{(-r)}(a'\mid M,C,N)}
{\widehat s^{(-r)}(a\mid M,C,N)}
-\frac{s^\star(a'\mid M,C,N)}{s^\star(a\mid M,C,N)}
\right\|_{2,P}\\
&\qquad
+\sum_{\ell=1}^{3}
\|\widehat u^{(\ell),(-r)}-u^{(\ell),\star}\|_{2,P}
=o_p(1).
\end{aligned}
\end{equation}
The limit satisfies \eqref{eq:lowdim_MR_theta}, and
\begin{equation}
\label{eq:lowdim_rate_theta}
\resizebox{\linewidth}{!}{$
\begin{aligned}
\max_{1\le r\le R}\Bigg[&
\|\sqrt W\{\widehat d^{(-r)}-d_0\}\|_{2,P}
\|\sqrt W\{\widehat\eta^{(-r)}-\eta_0\}\|_{2,P}\\
&+\left\|\sqrt W\left\{
\frac{\widehat s^{(-r)}(a'\mid M,C,N)}
{\widehat s^{(-r)}(a\mid M,C,N)}
-\frac{s_0(a'\mid M,C,N)}{s_0(a\mid M,C,N)}
\right\}\right\|_{2,P}
\|\sqrt W\{\widehat u^{(2),(-r)}-u_0^{(2)}\}\|_{2,P}
\Bigg]=o_p(K^{-1/2}).
\end{aligned}
$}
\end{equation}

\item For $\tau$,
\begin{equation}
\label{eq:lowdim_limit_tau}
\resizebox{\linewidth}{!}{$
\begin{aligned}
&\|\widehat\eta^{(-r)}-\eta^\star\|_{2,P}
+\|\widehat e^{(-r)}-e^\star\|_{2,P}
+\max_{1\le j\le N_{\max}}
\left\|
\frac{\widehat s_{-j}^{(-r)}(0\mid M_{\bullet(-j)},C,N)}
{\widehat s_{-j}^{(-r)}(1\mid M_{\bullet(-j)},C,N)}
-\frac{s_{-j}^\star(0\mid M_{\bullet(-j)},C,N)}
{s_{-j}^\star(1\mid M_{\bullet(-j)},C,N)}
\right\|_{2,P}\\
&\qquad
+\sum_{\ell=1}^{4}
\|\widehat r^{(\ell),(-r)}-r^{(\ell),\star}\|_{2,P}
=o_p(1).
\end{aligned}
$}
\end{equation}
The limit satisfies \eqref{eq:lowdim_MR_tau} for every $j$, and
\begin{equation}
\label{eq:lowdim_rate_tau}
\resizebox{\linewidth}{!}{$
\begin{aligned}
\max_{1\le r\le R}\max_{1\le j\le N_{\max}}\Bigg[&
\|\sqrt W\{\widehat e_{\bullet j}^{(-r)}-e_{0\bullet j}\}\|_{2,P}
\|\sqrt W\{\widehat\eta_{\bullet j}^{(-r)}-\eta_{0\bullet j}\}\|_{2,P}\\
&+\left\|\sqrt W\left\{
\frac{\widehat s_{-j}^{(-r)}(0\mid M_{\bullet(-j)},C,N)}
{\widehat s_{-j}^{(-r)}(1\mid M_{\bullet(-j)},C,N)}
-\frac{s_{0,-j}(0\mid M_{\bullet(-j)},C,N)}
{s_{0,-j}(1\mid M_{\bullet(-j)},C,N)}
\right\}\right\|_{2,P}
\|\sqrt W\{\widehat r_{\bullet j}^{(3),(-r)}-r_{0\bullet j}^{(3)}\}\|_{2,P}
\Bigg]=o_p(K^{-1/2}).
\end{aligned}
$}
\end{equation}
\end{enumerate}
\end{assumption}
\noindent Assumption~\ref{assm:lowdim_rates} includes convergence of all auxiliary regressions entering the scores.
The conditions in \eqref{eq:lowdim_rate_theta} and \eqref{eq:lowdim_rate_tau}, which are required for asymptotic linearity, require each paired error product to be $o_p(K^{-1/2})$.
As in Section~\ref{sec:canonical_estimators}, zero population bias at the nuisance limit does not by itself establish these rates.

Define
\begin{equation*}
\begin{aligned}
\psi_{\theta,\mathrm{ld}}^\star(O)
&=\psi_{\theta,\mathrm{ld}}(a,a';O;h_\theta^\star),
&D_{\theta,\mathrm{ld}}^\star(O)
&=\frac{\psi_{\theta,\mathrm{ld}}^\star(O)-\theta_0(a,a')W}{\nu_0},\\
\psi_{\tau,\mathrm{ld}}^\star(O)
&=\psi_{\tau,\mathrm{ld}}(O;h_\tau^\star),
&D_{\tau,\mathrm{ld}}^\star(O)
&=\frac{\psi_{\tau,\mathrm{ld}}^\star(O)-\tau_0W}{\nu_0}.
\end{aligned}
\end{equation*}
Assume the boundedness conditions stated at the beginning of Section~\ref{sec: thm:lowdim_AN proof}.
The following theorem states the asymptotic expansions for the low-dimensional estimators.

\begin{theorem}
\label{thm:lowdim_AN}
Under Assumptions~\ref{assm:crossfit_common} and
\ref{assm:lowdim_rates}(i),
\begin{equation}
\sqrt K\{\widehat\theta_{\mathrm{ld}}(a,a')-\theta_0(a,a')\}
=\frac1{\sqrt K}\sum_{i=1}^{K}D_{\theta,\mathrm{ld}}^\star(O_i)+o_p(1).
\label{eq:lowdim_AL_theta}
\end{equation}
Under Assumptions~\ref{assm:crossfit_common} and
\ref{assm:lowdim_rates}(ii),
\begin{equation}
\sqrt K(\widehat\tau_{\mathrm{ld}}-\tau_0)
=\frac1{\sqrt K}\sum_{i=1}^{K}D_{\tau,\mathrm{ld}}^\star(O_i)+o_p(1).
\label{eq:lowdim_AL_tau}
\end{equation}
Consequently, the estimators converge to centered normal limits with variances
$\Sigma_{\theta,\mathrm{ld}}^\star
=\EE\{(D_{\theta,\mathrm{ld}}^\star)^2\}$ and
$\Sigma_{\tau,\mathrm{ld}}^\star
=\EE\{(D_{\tau,\mathrm{ld}}^\star)^2\}$, respectively. 
If all
nuisance limits are correct, the
corresponding variances attain the nonparametric efficiency bounds.
\end{theorem}

\noindent The proof is deferred to Section~\ref{sec: thm:lowdim_AN proof} of the Appendix.
At the true nuisance limits, Lemma~\ref{lem:lowdim_canonical} identifies the limiting scores with the canonical scores, giving the same efficiency bounds as in Section~\ref{sec:canonical_estimators}.

For $i\in\mathcal I_r$, define $\widehat D_{\theta,\mathrm{ld},i}=\frac{\widehat\psi_{\theta,r}(O_i)
-\widehat\theta_{\mathrm{ld}}(a,a')W_i}{\mathbb P_K(W)}$ and $\widehat D_{\tau,\mathrm{ld},i}
=\frac{\widehat\psi_{\tau,r}(O_i)
-\widehat\tau_{\mathrm{ld}}W_i}{\mathbb P_K(W)}$.
% \begin{equation*}
% \begin{aligned}
% \widehat D_{\theta,\mathrm{ld},i}
% &=\frac{\widehat\psi_{\theta,r}(O_i)
% -\widehat\theta_{\mathrm{ld}}(a,a')W_i}{\mathbb P_K(W)},\\
% \widehat D_{\tau,\mathrm{ld},i}
% &=\frac{\widehat\psi_{\tau,r}(O_i)
% -\widehat\tau_{\mathrm{ld}}W_i}{\mathbb P_K(W)}.
% \end{aligned}
% \end{equation*}
The variance estimators associated with Theorem~\ref{thm:lowdim_AN} are $\mathbb P_K(\widehat D_{\theta,\mathrm{ld}}^2)$ and $\mathbb P_K(\widehat D_{\tau,\mathrm{ld}}^2)$.

\subsection{Consistency and asymptotic normality with estimated auxiliary regressions}
\label{sec:lowdim_practical}

We study \eqref{eq:lowdim_theta_est} and \eqref{eq:lowdim_tau_est}
with every fitted function in \eqref{eq:lowdim_score_theta} and
\eqref{eq:lowdim_score_tau} estimated outside $\mathcal I_r$.
Expectations involving these functions integrate a new cluster under $P$ conditional on the training observations.
The sampling and moment conditions of Assumption~\ref{assm:positivity},
the data-independent folds of Assumption~\ref{assm:crossfit_common}, and the boundedness
requirement in Section~\ref{sec: thm:lowdim_AN proof} apply to the
asymptotic results below.
We maintain $h_\theta^\star$, $h_\tau^\star$,
$\psi_{\theta,\mathrm{ld}}^\star$, $\psi_{\tau,\mathrm{ld}}^\star$,
$D_{\theta,\mathrm{ld}}^\star$, and $D_{\tau,\mathrm{ld}}^\star$
as specified in Assumption~\ref{assm:lowdim_rates} and
Theorem~\ref{thm:lowdim_AN}.
The functions $\widehat u^{(1),(-r)},\widehat u^{(2),(-r)},
\widehat u^{(3),(-r)}$ and $\widehat r^{(1),(-r)},\ldots,
\widehat r^{(4),(-r)}$ may be fitted separately.

\begin{theorem}
\label{thm:lowdim_practical_drift}
Suppose the support conditions in Assumption~\ref{assm:positivity} hold,
the fitted ratios below are finite on the support of $P$, and all
displayed expectations and integrals are absolutely integrable.
For every fold $r$, \eqref{eq:lowdim_score_theta} gives the following bias decomposition:
\begin{equation}
\label{eq:practical_drift_theta}
\resizebox{\ifdim\width>\linewidth\linewidth\else\width\fi}{!}{$
\begin{aligned}
&\EE\{\widehat\psi_{\theta,r}(O)\}-\nu_0\theta_0(a,a')\\
&=\EE\Bigg[\frac{W}{N}\sum_{j=1}^N
\EE\Bigg[
\frac{\pi_a}{\pi_{a'}}\frac{\widehat s^{(-r)}(a'\mid M,C,N)}{\widehat s^{(-r)}(a\mid M,C,N)}\{\widehat d^{(-r)}-d_0\}(a,Z,M,C,N)\{\eta_{0\bullet j}-\widehat\eta_{\bullet j}^{(-r)}\}(a,Z,M,C,N)\\
&\qquad\quad+
\Bigg\{\frac{\pi_a}{\pi_{a'}}\frac{\widehat s^{(-r)}(a'\mid M,C,N)}{\widehat s^{(-r)}(a\mid M,C,N)}-\frac{\pi_a}{\pi_{a'}}\frac{s_0(a'\mid M,C,N)}{s_0(a\mid M,C,N)}\Bigg\}\{u_{0\bullet j}^{(2)}-\widehat u_{\bullet j}^{(2),(-r)}\}(a,M,C,N)
\,\Bigm|\, A=a,C,N\Bigg]\Bigg]\\
&\quad+\EE\Bigg[\frac{W}{N}\sum_{j=1}^N\Bigg\{
\EE\{\widehat u_{\bullet j}^{(1),(-r)}(a,a',Z,C,N)\mid A=a,C,N\}
-\widehat u_{\bullet j}^{(3),(-r)}(a,a',C,N)\\
&\qquad+\int\kappa_0(m\mid a,C,N)
\frac{\pi_a}{\pi_{a'}}\frac{\widehat s^{(-r)}(a'\mid m,C,N)}{\widehat s^{(-r)}(a\mid m,C,N)}\Bigg[
\widehat u_{\bullet j}^{(2),(-r)}(a,m,C,N)-\int\widehat\eta_{\bullet j}^{(-r)}(a,z,m,C,N)\phi_0(z\mid a,C,N)\,dz
\Bigg]dm\Bigg\}\Bigg].
\end{aligned}
$}
\end{equation}
Likewise, \eqref{eq:lowdim_score_tau} gives
\begin{equation}
\label{eq:practical_drift_tau}
\resizebox{\ifdim\width>\linewidth\linewidth\else\width\fi}{!}{$
\begin{aligned}
&\EE\{\widehat\psi_{\tau,r}(O)\}-\nu_0\tau_0\\
&=\EE\Bigg[\frac{W}{N}\sum_{j=1}^N
\EE\Bigg[
\frac{\pi}{1-\pi}\frac{\widehat s_{-j}^{(-r)}(0\mid M_{\bullet(-j)},C,N)}{\widehat s_{-j}^{(-r)}(1\mid M_{\bullet(-j)},C,N)}\{\widehat e_{\bullet j}^{(-r)}-e_{0\bullet j}\}(Z,M,C,N)\{\eta_{0\bullet j}-\widehat\eta_{\bullet j}^{(-r)}\}(1,Z,M,C,N)\\
&\qquad\quad+
\Bigg\{\frac{\pi}{1-\pi}\frac{\widehat s_{-j}^{(-r)}(0\mid M_{\bullet(-j)},C,N)}{\widehat s_{-j}^{(-r)}(1\mid M_{\bullet(-j)},C,N)}-\frac{\pi}{1-\pi}\frac{s_{0,-j}(0\mid M_{\bullet(-j)},C,N)}{s_{0,-j}(1\mid M_{\bullet(-j)},C,N)}\Bigg\}\{r_{0\bullet j}^{(3)}-\widehat r_{\bullet j}^{(3),(-r)}\}(1,1,M_{\bullet(-j)},C,N)
\,\Bigm|\, A=1,C,N\Bigg]\Bigg]\\
&\quad+\EE\Bigg[\frac{W}{N}\sum_{j=1}^N\Bigg\{
\EE\{\widehat r_{\bullet j}^{(1),(-r)}(1,1,0,Z,C,N)\mid A=1,C,N\}\\
&\qquad+\EE\{\widehat r_{\bullet j}^{(2),(-r)}(1,0,M_{\bullet j},C,N)\mid A=1,C,N\}
-2\widehat r_{\bullet j}^{(4),(-r)}(1,1,0,C,N)\\
&\qquad+\int\kappa_{0\bullet(-j)}(m_{\bullet(-j)}\mid1,C,N)
\frac{\pi}{1-\pi}\frac{\widehat s_{-j}^{(-r)}(0\mid m_{\bullet(-j)},C,N)}{\widehat s_{-j}^{(-r)}(1\mid m_{\bullet(-j)},C,N)}\\
&\qquad\qquad\qquad\times\Bigg[
\widehat r_{\bullet j}^{(3),(-r)}(1,1,m_{\bullet(-j)},C,N)-\iint\widehat\eta_{\bullet j}^{(-r)}(1,z,m,C,N)
\phi_0(z\mid1,C,N)\kappa_{0\bullet j}(m_{\bullet j}\mid1,C,N)
\,dz\,dm_{\bullet j}\Bigg]dm_{\bullet(-j)}\Bigg\}\Bigg].
\end{aligned}
$}
\end{equation}
\end{theorem}
The proof is given in Section~\ref{sec:proof_lowdim_practical_drift}.
Under the integral equalities defining the compatible tuples in
Section~\ref{sec:lowdim_reparam} and
\eqref{eq:lowdim_r1}--\eqref{eq:lowdim_r4},
\eqref{eq:practical_drift_theta} and \eqref{eq:practical_drift_tau}
reduce to \eqref{eq:lowdim_drift_theta} and \eqref{eq:lowdim_drift_tau}.
The next result allows the fitted functions to approach those equalities
through \eqref{eq:lowdim_limit_theta} and \eqref{eq:lowdim_limit_tau}.

\begin{theorem}
\label{thm:lowdim_practical_AN}
Under the sampling, moment, fold, and boundedness requirements stated
in Section~\ref{sec:lowdim_practical}, the following assertions hold.
For $h_\theta^\star$ satisfying \eqref{eq:lowdim_MR_theta},
\eqref{eq:lowdim_limit_theta} implies
$\widehat\theta_{\mathrm{ld}}(a,a')\to_p\theta_0(a,a')$ and
\begin{equation}
\label{eq:practical_aux_theta}
\resizebox{\ifdim\width>\linewidth\linewidth\else\width\fi}{!}{$
\begin{aligned}
\max_{1\le r\le R}\Bigg|&\EE\Bigg[\frac{W}{N}\sum_{j=1}^N\Bigg\{
\EE\{\widehat u_{\bullet j}^{(1),(-r)}(a,a',Z,C,N)\mid A=a,C,N\}
-\widehat u_{\bullet j}^{(3),(-r)}(a,a',C,N)\\
&\qquad+\int\kappa_0(m\mid a,C,N)
\frac{\pi_a}{\pi_{a'}}\frac{\widehat s^{(-r)}(a'\mid m,C,N)}{\widehat s^{(-r)}(a\mid m,C,N)}\Bigg[
\widehat u_{\bullet j}^{(2),(-r)}(a,m,C,N)-\int\widehat\eta_{\bullet j}^{(-r)}(a,z,m,C,N)\phi_0(z\mid a,C,N)\,dz
\Bigg]dm\Bigg\}\Bigg]\Bigg|=o_p(1).
\end{aligned}
$}
\end{equation}
If, in addition, \eqref{eq:lowdim_rate_theta} holds and the left-hand
side of \eqref{eq:practical_aux_theta} is $o_p(K^{-1/2})$, then
\eqref{eq:lowdim_AL_theta} holds and
\begin{equation}\label{eq:practical_normal_theta}
\sqrt K\{\widehat\theta_{\mathrm{ld}}(a,a')-\theta_0(a,a')\}
\rightsquigarrow N(0,\Sigma_{\theta,\mathrm{ld}}^\star),
\qquad
\mathbb P_K(\widehat D_{\theta,\mathrm{ld}}^2)
\to_p\Sigma_{\theta,\mathrm{ld}}^\star.
\end{equation}

For $h_\tau^\star$ satisfying \eqref{eq:lowdim_MR_tau} for every $j$,
\eqref{eq:lowdim_limit_tau} implies
$\widehat\tau_{\mathrm{ld}}\to_p\tau_0$ and
\begin{equation}
\label{eq:practical_aux_tau}
\resizebox{\ifdim\width>\linewidth\linewidth\else\width\fi}{!}{$
\begin{aligned}
\max_{1\le r\le R}\Bigg|&\EE\Bigg[\frac{W}{N}\sum_{j=1}^N\Bigg\{
\EE\{\widehat r_{\bullet j}^{(1),(-r)}(1,1,0,Z,C,N)\mid A=1,C,N\}+\EE\{\widehat r_{\bullet j}^{(2),(-r)}(1,0,M_{\bullet j},C,N)\mid A=1,C,N\}
-2\widehat r_{\bullet j}^{(4),(-r)}(1,1,0,C,N)\\
&\qquad\qquad\qquad+\int\kappa_{0\bullet(-j)}(m_{\bullet(-j)}\mid1,C,N)
\frac{\pi}{1-\pi}\frac{\widehat s_{-j}^{(-r)}(0\mid m_{\bullet(-j)},C,N)}{\widehat s_{-j}^{(-r)}(1\mid m_{\bullet(-j)},C,N)}\\
&\qquad\qquad\qquad\qquad\times\Bigg[
\widehat r_{\bullet j}^{(3),(-r)}(1,1,m_{\bullet(-j)},C,N)-\iint\widehat\eta_{\bullet j}^{(-r)}(1,z,m,C,N)
\phi_0(z\mid1,C,N)\kappa_{0\bullet j}(m_{\bullet j}\mid1,C,N)
\,dz\,dm_{\bullet j}\Bigg]dm_{\bullet(-j)}\Bigg\}\Bigg]\Bigg|=o_p(1).
\end{aligned}
$}
\end{equation}
If, in addition, \eqref{eq:lowdim_rate_tau} holds and the left-hand
side of \eqref{eq:practical_aux_tau} is $o_p(K^{-1/2})$, then
\eqref{eq:lowdim_AL_tau} holds and
\begin{equation}\label{eq:practical_normal_tau}
\sqrt K(\widehat\tau_{\mathrm{ld}}-\tau_0)
\rightsquigarrow N(0,\Sigma_{\tau,\mathrm{ld}}^\star),
\qquad
\mathbb P_K(\widehat D_{\tau,\mathrm{ld}}^2)
\to_p\Sigma_{\tau,\mathrm{ld}}^\star.
\end{equation}
For $\Sigma_{\theta,\mathrm{ld}}^\star>0$ and
$\Sigma_{\tau,\mathrm{ld}}^\star>0$, respectively,
\begin{equation}
\label{eq:practical_studentized}
\frac{\sqrt K\{\widehat\theta_{\mathrm{ld}}(a,a')-\theta_0(a,a')\}}
{\{\mathbb P_K(\widehat D_{\theta,\mathrm{ld}}^2)\}^{1/2}}
\rightsquigarrow N(0,1),
\qquad
\frac{\sqrt K(\widehat\tau_{\mathrm{ld}}-\tau_0)}
{\{\mathbb P_K(\widehat D_{\tau,\mathrm{ld}}^2)\}^{1/2}}
\rightsquigarrow N(0,1).
\end{equation}
At the true nuisance limits, the variances in
\eqref{eq:practical_normal_theta} and \eqref{eq:practical_normal_tau}
equal the corresponding bounds in Theorem~\ref{thm:EIF}.
\end{theorem}

The proof is given in Section~\ref{sec:proof_lowdim_practical_AN}.
The requirements in \eqref{eq:lowdim_limit_theta} and
\eqref{eq:lowdim_limit_tau} concern all components of
$h_\theta^\star$ and $h_\tau^\star$, including their auxiliary
regressions. Together with \eqref{eq:lowdim_MR_theta} or
\eqref{eq:lowdim_MR_tau}, they give both consistency and
\eqref{eq:practical_aux_theta} or \eqref{eq:practical_aux_tau}.
The additional requirement for \eqref{eq:lowdim_AL_theta} or
\eqref{eq:lowdim_AL_tau} is the $o_p(K^{-1/2})$ rate in
\eqref{eq:practical_aux_theta} or \eqref{eq:practical_aux_tau},
together with the existing \eqref{eq:lowdim_rate_theta} or
\eqref{eq:lowdim_rate_tau}.

The following construction gives some sufficient conditions for
\eqref{eq:practical_aux_theta} and \eqref{eq:practical_aux_tau}.
Choose fitted conditional densities
$\widehat\phi^{(-r)}$ and $\widehat\kappa_{\bullet j}^{(-r)}$
with unit integrals, satisfying
$\widehat\phi^{(-r)}(z\mid a,c,n)=0$ wherever
$\phi_0(z\mid a,c,n)=0$ and
$\widehat\kappa_{\bullet j}^{(-r)}(m_{\bullet j}\mid1,c,n)=0$
wherever $\kappa_{0\bullet j}(m_{\bullet j}\mid1,c,n)=0$,
for $a\in\{0,1\}$ and almost every $(c,n)$.
Take the ratios below to be one where both numerator and denominator
are zero.
For $\widehat\theta_{\mathrm{ld}}(a,a')$, choose
$\widehat u^{(2),(-r)}$ and $\widehat u^{(3),(-r)}$ to satisfy
\begin{equation}
\label{eq:practical_centered_theta}
\begin{aligned}
\widehat u_{\bullet j}^{(2),(-r)}(a,m,c,n)
&=\int\widehat\eta_{\bullet j}^{(-r)}(a,z,m,c,n)
       \widehat\phi^{(-r)}(z\mid a,c,n)\,dz,\\
\widehat u_{\bullet j}^{(3),(-r)}(a,a',c,n)
&=\int\widehat u_{\bullet j}^{(1),(-r)}(a,a',z,c,n)
       \widehat\phi^{(-r)}(z\mid a,c,n)\,dz.
\end{aligned}
\end{equation}
The regression $\widehat u^{(1),(-r)}$ may be fitted separately.
Under \eqref{eq:practical_centered_theta}, a sufficient condition
for \eqref{eq:practical_aux_theta} is
\begin{equation}
\label{eq:practical_centered_rate_theta}
\resizebox{\ifdim\width>\linewidth\linewidth\else\width\fi}{!}{$
\begin{aligned}
\max_{1\le r\le R}\max_{1\le j\le N_{\max}}&
\left\|\sqrt W\left\{
\frac{\widehat\phi^{(-r)}(Z\mid a,C,N)}{\phi_0(Z\mid a,C,N)}-1
\right\}\right\|_{2,P}\\
&\times\Bigg[
\left\|\sqrt W\{\widehat u_{\bullet j}^{(1),(-r)}-u_{0\bullet j}^{(1)}\}
          (a,a',Z,C,N)\right\|_{2,P}
+\left\|\sqrt W\{\widehat\eta_{\bullet j}^{(-r)}-\eta_{0\bullet j}\}
          (a,Z,M,C,N)\right\|_{2,P}\\
&\qquad\quad+
\left\|\sqrt W\left\{
\frac{\widehat s^{(-r)}(a'\mid M,C,N)}{\widehat s^{(-r)}(a\mid M,C,N)}
-\frac{s_0(a'\mid M,C,N)}{s_0(a\mid M,C,N)}
\right\}\right\|_{2,P}\Bigg]=o_p(1).
\end{aligned}
$}
\end{equation}
Replacing $o_p(1)$ in \eqref{eq:practical_centered_rate_theta} by
$o_p(K^{-1/2})$ gives the $o_p(K^{-1/2})$ rate in
\eqref{eq:practical_aux_theta}.
For example, under the boundedness requirement in
Section~\ref{sec: thm:lowdim_AN proof},
\begin{equation}
\label{eq:practical_phi_consistent}
\max_{1\le r\le R}
\left\|\sqrt W\left\{
\frac{\widehat\phi^{(-r)}(Z\mid a,C,N)}{\phi_0(Z\mid a,C,N)}-1
\right\}\right\|_{2,P}=o_p(1)
\end{equation}
implies \eqref{eq:practical_centered_rate_theta}.
If \eqref{eq:practical_phi_consistent} instead has rate
$O_p(K^{-1/2})$, and \eqref{eq:lowdim_limit_theta} holds with
$(\eta^\star,s^\star)=(\eta_0,s_0)$, then
\eqref{eq:practical_centered_rate_theta} has rate $o_p(K^{-1/2})$.
Alternatively, $o_p(K^{-1/4})$ for every norm in
\eqref{eq:practical_centered_rate_theta} gives that rate.

For $\widehat\tau_{\mathrm{ld}}$, choose
$\widehat r^{(3),(-r)}$ and $\widehat r^{(4),(-r)}$ to satisfy
\begin{equation}
\label{eq:practical_centered_tau}
\resizebox{\ifdim\width>\linewidth\linewidth\else\width\fi}{!}{$
\begin{aligned}
\widehat r_{\bullet j}^{(3),(-r)}(1,1,m_{\bullet(-j)},c,n)
&=\iint\widehat\eta_{\bullet j}^{(-r)}(1,z,m,c,n)
       \widehat\phi^{(-r)}(z\mid1,c,n)
       \widehat\kappa_{\bullet j}^{(-r)}(m_{\bullet j}\mid1,c,n)
       \,dz\,dm_{\bullet j},\\
2\widehat r_{\bullet j}^{(4),(-r)}(1,1,0,c,n)
&=\int\widehat r_{\bullet j}^{(1),(-r)}(1,1,0,z,c,n)
          \widehat\phi^{(-r)}(z\mid1,c,n)\,dz\\
&\quad+\int\widehat r_{\bullet j}^{(2),(-r)}(1,0,m_{\bullet j},c,n)
          \widehat\kappa_{\bullet j}^{(-r)}(m_{\bullet j}\mid1,c,n)
          \,dm_{\bullet j}.
\end{aligned}
$}
\end{equation}
The regressions $\widehat r^{(1),(-r)}$ and
$\widehat r^{(2),(-r)}$ may be fitted separately.
Under \eqref{eq:practical_centered_tau}, a sufficient condition
for \eqref{eq:practical_aux_tau} is
\begin{equation}
\label{eq:practical_centered_rate_tau}
\resizebox{\ifdim\width>\linewidth\linewidth\else\width\fi}{!}{$
\begin{aligned}
\max_{1\le r\le R}\max_{1\le j\le N_{\max}}\Bigg[&
\left\|\sqrt W\left\{
\frac{\widehat\phi^{(-r)}(Z\mid1,C,N)}{\phi_0(Z\mid1,C,N)}-1
\right\}\right\|_{2,P}
\left\|\sqrt W\{\widehat r_{\bullet j}^{(1),(-r)}-r_{0\bullet j}^{(1)}\}
          (1,1,0,Z,C,N)\right\|_{2,P}\\
&+\left\|\sqrt W\left\{
\frac{\widehat\kappa_{\bullet j}^{(-r)}(M_{\bullet j}\mid1,C,N)}
     {\kappa_{0\bullet j}(M_{\bullet j}\mid1,C,N)}-1
\right\}\right\|_{2,P}
\left\|\sqrt W\{\widehat r_{\bullet j}^{(2),(-r)}-r_{0\bullet j}^{(2)}\}
          (1,0,M_{\bullet j},C,N)\right\|_{2,P}\\
&+\Bigg(
\left\|\sqrt W\left\{
\frac{\widehat\phi^{(-r)}(Z\mid1,C,N)}{\phi_0(Z\mid1,C,N)}-1
\right\}\right\|_{2,P}
+\left\|\sqrt W\left\{
\frac{\widehat\kappa_{\bullet j}^{(-r)}(M_{\bullet j}\mid1,C,N)}
     {\kappa_{0\bullet j}(M_{\bullet j}\mid1,C,N)}-1
\right\}\right\|_{2,P}\Bigg)\\
&\qquad\times\Bigg(
\left\|\sqrt W\{\widehat\eta_{\bullet j}^{(-r)}-\eta_{0\bullet j}\}
          (1,Z,M,C,N)\right\|_{2,P}
+\left\|\sqrt W\left\{
\frac{\widehat s_{-j}^{(-r)}(0\mid M_{\bullet(-j)},C,N)}
     {\widehat s_{-j}^{(-r)}(1\mid M_{\bullet(-j)},C,N)}
-\frac{s_{0,-j}(0\mid M_{\bullet(-j)},C,N)}
      {s_{0,-j}(1\mid M_{\bullet(-j)},C,N)}
\right\}\right\|_{2,P}\Bigg)\\
&+\left\|\sqrt W\left\{
\frac{\widehat\phi^{(-r)}(Z\mid1,C,N)}{\phi_0(Z\mid1,C,N)}-1
\right\}\right\|_{2,P}
\left\|\sqrt W\left\{
\frac{\widehat\kappa_{\bullet j}^{(-r)}(M_{\bullet j}\mid1,C,N)}
     {\kappa_{0\bullet j}(M_{\bullet j}\mid1,C,N)}-1
\right\}\right\|_{2,P}\Bigg]=o_p(1).
\end{aligned}
$}
\end{equation}
Replacing $o_p(1)$ in \eqref{eq:practical_centered_rate_tau} by
$o_p(K^{-1/2})$ gives the $o_p(K^{-1/2})$ rate in
\eqref{eq:practical_aux_tau}.
In particular, \eqref{eq:practical_phi_consistent} with $a=1$ and
\begin{equation}
\label{eq:practical_kappaj_consistent}
\max_{1\le r\le R}\max_{1\le j\le N_{\max}}
\left\|\sqrt W\left\{
\frac{\widehat\kappa_{\bullet j}^{(-r)}(M_{\bullet j}\mid1,C,N)}
     {\kappa_{0\bullet j}(M_{\bullet j}\mid1,C,N)}-1
\right\}\right\|_{2,P}=o_p(1)
\end{equation}
imply \eqref{eq:practical_centered_rate_tau} under the same
boundedness requirement.
If \eqref{eq:practical_phi_consistent} with $a=1$ and
\eqref{eq:practical_kappaj_consistent} instead have rates
$O_p(K^{-1/2})$, and \eqref{eq:lowdim_limit_tau} holds with
$(\eta^\star,s_{-j}^\star)=(\eta_0,s_{0,-j})$ for every $j$, then
\eqref{eq:practical_centered_rate_tau} has rate $o_p(K^{-1/2})$.
Alternatively, $o_p(K^{-1/4})$ for every norm in
\eqref{eq:practical_centered_rate_tau} gives that rate.
Section~\ref{sec:proof_lowdim_practical_AN} derives
\eqref{eq:practical_centered_rate_theta} and
\eqref{eq:practical_centered_rate_tau} from exact integral identities.
Combining their $o_p(K^{-1/2})$ versions with
\eqref{eq:lowdim_rate_theta} or \eqref{eq:lowdim_rate_tau}
and the respective component convergence requirements gives
\eqref{eq:lowdim_AL_theta} and \eqref{eq:practical_normal_theta},
or \eqref{eq:lowdim_AL_tau} and \eqref{eq:practical_normal_tau}.

The augmented-classification estimator of $d$ in
Section~\ref{sec:augmented_classification} follows
\citet{rudolph2024practical}.
The condition in \citet[Assumption~1]{rudolph2024practical}
controls agreement between fitted integral representations at the
rates required by \citet[Theorems~4.2 and~4.3]{rudolph2024practical}.
Here, \eqref{eq:practical_centered_theta} and
\eqref{eq:practical_centered_tau} give explicit relationships among
the estimated auxiliary regressions, and
\eqref{eq:practical_centered_rate_theta} and
\eqref{eq:practical_centered_rate_tau} give sufficient nuisance rates
for \eqref{eq:practical_aux_theta} and \eqref{eq:practical_aux_tau}.

\section{Proof of Theorem~\ref{thm:identification}}
\label{sec:Identification_proof}

\begin{proof}
Fix $a,a'\in\{0,1\}$.  Conditional on $(C,N)$, $G(a')$ is an independent
random draw from the conditional distribution of $M(a')$.  Therefore, by iterated
expectation, the numerator of \eqref{def:theta} is
\begin{equation*}
\begin{aligned}
\EE\left[\frac{W}{N}\sum_{j=1}^N
Y_{\bullet j}\{a,G(a')\}\right]=\EE\left[\frac{W}{N}\sum_{j=1}^N
\int\EE\{Y_{\bullet j}(a,m)\mid C,N\}
\kappa(m\mid a',C,N)\,dm\right],
\end{aligned}
\end{equation*}
where randomization and consistency identify the conditional distribution of $M(a')$
given $(C,N)$ with $\kappa(\bullet\mid a',C,N)$.  By the composition,
$Y_{\bullet j}(a,m)=Y_{\bullet j}\{a,Z(a),m\}$, and hence
\begin{equation*}
\EE\{Y_{\bullet j}(a,m)\mid C,N\}
=\int\EE\{Y_{\bullet j}(a,z,m)\mid Z(a)=z,C,N\}
\phi(z\mid a,C,N)\,dz.
\end{equation*}
For each $(z,m)$ in the relevant support, randomization gives
\begin{equation*}
\begin{split}
\EE\{Y_{\bullet j}(a,z,m)\mid Z(a)=z,C,N\}=\EE\{Y_{\bullet j}(a,z,m)
\mid A=a,Z(a)=z,C,N\}.
\end{split}
\end{equation*}
Assumption~\ref{assm:ignorability} then permits conditioning on
$M(a,z)=m$, and Assumption~\ref{assm:consistency} maps the resulting
counterfactual variables to their observed counterparts:
\begin{equation*}
\begin{aligned}
&\EE\{Y_{\bullet j}(a,z,m)
\mid A=a,Z(a)=z,C,N\}\\
&=\EE\{Y_{\bullet j}(a,z,m)
\mid A=a,Z(a)=z,M(a,z)=m,C,N\}\\
&=\EE(Y_{\bullet j}\mid A=a,Z=z,M=m,C,N)
=\eta_{\bullet j}(a,z,m,C,N).
\end{aligned}
\end{equation*}
Substitution and division by $\EE(W)$ prove \eqref{eq:ident.theta}.

For \eqref{def:tau}, $G(1)$ and $G(0)$ are conditionally independent draws.
Consequently, $G_{\bullet j}(1)$ and $G_{\bullet(-j)}(0)$ have the product distribution
with densities $\kappa_{\bullet j}(\bullet\mid1,C,N)$ and
$\kappa_{\bullet(-j)}(\bullet\mid0,C,N)$, and
\begin{equation*}
\begin{aligned}
&\EE\left[\frac{W}{N}\sum_{j=1}^N
Y_{\bullet j}\{1,G_{\bullet j}(1),G_{\bullet(-j)}(0)\}\right]\\
&\quad=\EE\Bigg[\frac{W}{N}\sum_{j=1}^N
\int\EE\{Y_{\bullet j}(1,m)\mid C,N\}
\kappa_{\bullet j}(m_{\bullet j}\mid1,C,N)
\kappa_{\bullet(-j)}(m_{\bullet(-j)}\mid0,C,N)\,dm\Bigg].
\end{aligned}
\end{equation*}
In addition, we have
\begin{equation*}
\EE\{Y_{\bullet j}(1,m)\mid C,N\}
=\int\eta_{\bullet j}(1,z,m,C,N)\phi(z\mid1,C,N)\,dz.
\end{equation*}
Substitution and division by $\EE(W)$ prove \eqref{eq:ident.tau}.
\end{proof}
\section{Proof of Theorem~\ref{thm:EIF}}
\label{sec:EIF_Proof}

\begin{proof}
Because the treatment mechanism is known by design, an arbitrary regular
one-dimensional parametric submodel through the observed-data distribution can be written as
\begin{equation*}
\begin{split}
f_t(O)=f_t(Y\mid A,Z,M,C,N)f_t(M\mid A,Z,C,N) f_t(Z\mid A,C,N)f(A)f_t(C,N),
\end{split}
\end{equation*}
with score $S(O)$ at $t=0$. 
We first differentiate the numerator of \eqref{eq:ident.theta}.  Its value along
the submodel is
\begin{equation*}
\int\frac{w}{n}\sum_{j=1}^n
\iint\eta_{\bullet j,t}(a,z,m,c,n)
\phi_t(z\mid a,c,n)\kappa_t(m\mid a',c,n)
\,dz\,dm\,f_t(c,n)\,dc\,dn,
\end{equation*}
where
\begin{equation*}
\kappa_t(m\mid a',c,n)
=\int\xi_t(m\mid a',z,c,n)\phi_t(z\mid a',c,n)\,dz
\end{equation*}
is the marginal conditional distribution of $M$ in arm $a'$.

For the outcome regression, the standard conditional-mean derivative calculation yields
\begin{equation*}
\begin{split}
&\left.\frac{\partial}{\partial t}\eta_{\bullet j,t}(a,z,m,c,n)
\right|_{t=0}\\
&\qquad=\EE\left[\{Y_{\bullet j}-\eta_{\bullet j}(a,z,m,c,n)\}S(O)
\mid A=a,Z=z,M=m,C=c,N=n\right].
\end{split}
\end{equation*}
Changing measure from
$\phi(z\mid a,c,n)\kappa(m\mid a',c,n)$ to the observed conditional distribution
$\phi(z\mid a,c,n)\xi(m\mid a,z,c,n)$ therefore yields
\begin{equation*}
\EE\Bigg[\frac{W}{N}\sum_{j=1}^N
\frac{I(A=a)}{\pi^a(1-\pi)^{1-a}}
\omega^{(1)}(a,a',Z,M,C,N)
\{Y_{\bullet j}-\eta_{\bullet j}(a,Z,M,C,N)\}S(O)\Bigg].
\end{equation*}

The conditional score of $Z$ in arm $a$ is represented by
$\EE\{S(O)\mid A=a,Z,C,N\}-\EE\{S(O)\mid A=a,C,N\}$.  Differentiating the
explicit factor $\phi_t(z\mid a,c,n)$ thus contributes
\begin{equation*}
\EE\Bigg[\frac{W}{N}\sum_{j=1}^N
\frac{I(A=a)}{\pi^a(1-\pi)^{1-a}}
\{u_{\bullet j}^{(1)}(a,a',Z,C,N)-u_{\bullet j}^{(3)}(a,a',C,N)\}S(O)\Bigg].
\end{equation*}
Likewise, because $\kappa_t(\bullet\mid a',c,n)$ is the marginal distribution of $M$
in arm $a'$, its conditional score is represented by
$\EE\{S(O)\mid A=a',M,C,N\}-\EE\{S(O)\mid A=a',C,N\}$.  Its derivative
therefore contributes
\begin{equation*}
\EE\Bigg[\frac{W}{N}\sum_{j=1}^N
\frac{I(A=a')}{\pi^{a'}(1-\pi)^{1-a'}}
\{u_{\bullet j}^{(2)}(a,M,C,N)-u_{\bullet j}^{(3)}(a,a',C,N)\}S(O)\Bigg].
\end{equation*}
This representation remains valid when $a=a'$; in that case the two
factor-specific derivatives are both present and add in the same treatment
arm.  Finally, differentiation of the baseline distribution $f_t(c,n)$ gives
\begin{equation*}
\EE\Bigg[\left\{\frac{W}{N}\sum_{j=1}^N
u_{\bullet j}^{(3)}(a,a',C,N)-\EE(W)\theta(a,a')\right\}S(O)\Bigg],
\end{equation*}
Combining the four derivatives shows that the
influence function of the numerator is
$\psi_\theta(a,a';O)-\EE(W)\theta(a,a')$.

For the numerator of \eqref{eq:ident.tau}, the same argument
applies.  Differentiation of the outcome regression and change of measure to
the product intervention give
\begin{equation*}
\EE\Bigg[\frac{W}{N}\sum_{j=1}^N\frac{A}{\pi}
\omega_{\bullet j}^{(2)}(1,1,0,Z,M,C,N)
\{Y_{\bullet j}-\eta_{\bullet j}(1,Z,M,C,N)\}S(O)\Bigg].
\end{equation*}
Differentiation of the explicit treated-arm $Z$ distribution contributes
\begin{equation*}
\EE\Bigg[\frac{W}{N}\sum_{j=1}^N\frac{A}{\pi}
\{r_{\bullet j}^{(1)}(1,1,0,Z,C,N)-r_{\bullet j}^{(4)}(1,1,0,C,N)\}S(O)\Bigg].
\end{equation*}
The score of the treated-arm marginal distribution of $M_{\bullet j}$ is represented by
conditioning $S(O)$ on $(A=1,M_{\bullet j},C,N)$ and then centering given
$(A=1,C,N)$; its derivative contributes
\begin{equation*}
\EE\Bigg[\frac{W}{N}\sum_{j=1}^N\frac{A}{\pi}
\{r_{\bullet j}^{(2)}(1,0,M_{\bullet j},C,N)-r_{\bullet j}^{(4)}(1,1,0,C,N)\}S(O)\Bigg].
\end{equation*}
Similarly, differentiating the control-arm marginal distribution of
$M_{\bullet(-j)}$ contributes
\begin{equation*}
\EE\Bigg[\frac{W}{N}\sum_{j=1}^N\frac{1-A}{1-\pi}
\{r_{\bullet j}^{(3)}(1,1,M_{\bullet(-j)},C,N)-r_{\bullet j}^{(4)}(1,1,0,C,N)\}S(O)\Bigg].
\end{equation*}
Differentiation of $f_t(c,n)$ gives
\begin{equation*}
\EE\Bigg[\left\{\frac{W}{N}\sum_{j=1}^N
r_{\bullet j}^{(4)}(1,1,0,C,N)-\EE(W)\tau\right\}S(O)\Bigg].
\end{equation*}
Thus the influence function of this numerator is
$\psi_\tau(O)-\EE(W)\tau$.

The influence function of $\EE(W)$ is $W-\EE(W)$.
We get
\begin{equation*}
\frac{\psi_\theta-\EE(W)\theta-\theta\{W-\EE(W)\}}{\EE(W)}
=\frac{\psi_\theta-\theta W}{\EE(W)}
\end{equation*}
and the analogous expression for $\tau$. 
\end{proof}

\section{Proofs of Theorems~\ref{thm:MR_theta} and \ref{thm:MR_tau}}
\label{sec:proof_MR}

\subsection{Bias decompositions}
\label{sec:canonical-bias-details}

We use the subscript $0$ to denote true parameter and nuisance values.
For a candidate tuple $(\eta,\phi,\kappa,\xi)$, the scores use the density ratios and integrated outcome regressions obtained from that tuple as defined in Section~\ref{sec:eif}.
The following theorem gives the exact bias decomposition for $\theta(a,a')$.

\begin{theorem}
\label{thm:MR_theta}
Fix arbitrary candidate nuisance functions $(\eta,\phi,\kappa,\xi)$ satisfying
the support conditions in Assumption~\ref{assm:positivity}, then
\begin{equation}
\resizebox{\linewidth}{!}{$
\begin{aligned}
&\EE\{\psi_\theta(a,a';O;\eta,\phi,\kappa,\xi)\}
-\EE(W)\theta_0(a,a')\\
&=\EE\Bigg[\frac{W}{N}\sum_{j=1}^N\Bigg\{
\iint\{\kappa-\kappa_0\}(m\mid a',C,N)
\left[\eta_{0\bullet j}(a,z,m,C,N)\phi_0(z\mid a,C,N)
-\eta_{\bullet j}(a,z,m,C,N)\phi(z\mid a,C,N)\right]dz\,dm\\
&\qquad\qquad\qquad\qquad+\iint\{\eta_{0\bullet j}-\eta_{\bullet j}\}(a,z,m,C,N)
\kappa(m\mid a',C,N)\phi_0(z\mid a,C,N)
\left\{\frac{\xi_0(m\mid a,z,C,N)}{\xi(m\mid a,z,C,N)}-1\right\}
\,dz\,dm\Bigg\}\Bigg].
\end{aligned}
$}
\label{eq:bias_decomp_theta}
\end{equation}
Consequently, if at least one of the following conditions holds for every $j$:
(i) $(\eta_{\bullet j},\phi)=(\eta_{0\bullet j},\phi_0)$;\quad
(ii) $(\eta_{\bullet j},\kappa)=(\eta_{0\bullet j},\kappa_0)$ for every $j$;\quad
(iii) $(\kappa,\xi)=(\kappa_0,\xi_0)$.
then
\[
\EE\{\psi_\theta(a,a';O;\eta,\phi,\kappa,\xi)\}
=\EE(W)\,\theta_0(a,a').
\]
\end{theorem}
For $\tau$, we establish the following bias decomposition.
\begin{theorem}
\label{thm:MR_tau}
Fix arbitrary candidate nuisance functions $(\eta,\phi,\kappa,\xi)$ satisfying
the support conditions in Assumption~\ref{assm:positivity}.  Then
\begin{equation}
\label{eq:MR-tau-bias}
\resizebox{\linewidth}{!}{$
\begin{aligned}
&\EE\{\psi_\tau(O;\eta,\phi,\kappa,\xi)\}-\EE(W)\tau_0\\
&=\EE\Bigg[\frac{W}{N}\sum_{j=1}^N\iint\Bigg\{
\{\kappa_{\bullet j}-\kappa_{0\bullet j}\}(m_{\bullet j}\mid1,C,N)
\kappa_{\bullet(-j)}(m_{\bullet(-j)}\mid0,C,N)
\left\{\eta_{0\bullet j}(1,z,m,C,N)\phi_0(z\mid1,C,N)
-\eta_{\bullet j}(1,z,m,C,N)\phi(z\mid1,C,N)\right\}\\
&\qquad\qquad\qquad\qquad+\{\kappa_{\bullet(-j)}-\kappa_{0\bullet(-j)}\}
(m_{\bullet(-j)}\mid0,C,N)
\kappa_{\bullet j}(m_{\bullet j}\mid1,C,N)
\left\{\eta_{0\bullet j}(1,z,m,C,N)\phi_0(z\mid1,C,N)
-\eta_{\bullet j}(1,z,m,C,N)\phi(z\mid1,C,N)\right\}\\
&\qquad\qquad\qquad\qquad-\eta_{0\bullet j}(1,z,m,C,N)\phi_0(z\mid1,C,N)
\{\kappa_{\bullet j}-\kappa_{0\bullet j}\}(m_{\bullet j}\mid1,C,N)
\{\kappa_{\bullet(-j)}-\kappa_{0\bullet(-j)}\}
(m_{\bullet(-j)}\mid0,C,N)\\
&\qquad\qquad\qquad\qquad+\{\eta_{0\bullet j}-\eta_{\bullet j}\}(1,z,m,C,N)
\kappa_{\bullet j}(m_{\bullet j}\mid1,C,N)
\kappa_{\bullet(-j)}(m_{\bullet(-j)}\mid0,C,N)\phi_0(z\mid1,C,N)
\left\{\frac{\xi_0(m\mid1,z,C,N)}{\xi(m\mid1,z,C,N)}-1\right\}
\Bigg\}\,dm\,dz\Bigg].
\end{aligned}
$}
\end{equation}
Consequently, if at least one of the following conditions holds for every $j$:
(i) $(\kappa,\eta_{\bullet j})=(\kappa_0,\eta_{0\bullet j})$;\quad
(ii) $(\kappa,\xi)=(\kappa_0,\xi_0)$;\quad
(iii) $(\eta_{\bullet j},\phi,\kappa_{\bullet j})
=(\eta_{0\bullet j},\phi_0,\kappa_{0\bullet j})$;\quad
(iv) $(\eta_{\bullet j},\phi,\kappa_{\bullet(-j)})
=(\eta_{0\bullet j},\phi_0,\kappa_{0\bullet(-j)})$, then $\EE\{\psi_\tau(O;\eta,\phi,\kappa,\xi)\}=\EE(W)\,\tau_0$.
\end{theorem}
\noindent The proofs of Theorems~\ref{thm:MR_theta} and \ref{thm:MR_tau} are given in Sections~\ref{sec:proof_MR_theta} and \ref{sec:proof_MR_tau} of the Appendix.

Each listed combination of correct nuisance functions makes the corresponding population bias zero.
For $\theta(a,a')$, correct $\eta$ and $\phi$ suffice even when both mediator density models are misspecified.
For $\tau$, the product of the two mediator marginal errors remains, so correctness of $\eta$ and $\phi$ alone does not generally suffice; the last two robustness conditions additionally require one of the relevant mediator marginals to be correct.

\subsection{Multiple robustness of \texorpdfstring{$\theta(a,a')$}{theta(a,a')}}
\label{sec:proof_MR_theta}

\begin{proof}
By iterated expectation, we have
\begin{equation*}
\resizebox{\linewidth}{!}{$
\begin{aligned}
&\EE\{\psi_\theta(a,a';O;\eta,\phi,\kappa,\xi)\}\\
&=\EE\Bigg[\frac{W}{N}\sum_{j=1}^N\Bigg\{
\EE\big[
\omega^{(1)}(a,a',Z,M,C,N)
\{\eta_{0\bullet j}-\eta_{\bullet j}\}(a,Z,M,C,N)
\mid A=a,C,N
\big]\\
&\qquad\qquad\qquad\qquad+\iint
\eta_{\bullet j}(a,z,m,C,N)
\phi_0(z\mid a,C,N)
\kappa(m\mid a',C,N)
\,dz\,dm\\
&\qquad\qquad\qquad\qquad+\iint
\eta_{\bullet j}(a,z,m,C,N)
\phi(z\mid a,C,N)
\kappa_0(m\mid a',C,N)
\,dz\,dm\\
&\qquad\qquad\qquad\qquad-\iint
\eta_{\bullet j}(a,z,m,C,N)
\phi(z\mid a,C,N)
\kappa(m\mid a',C,N)
\,dz\,dm
\Bigg\}\Bigg].
\end{aligned}
$}
\end{equation*}
Subtracting $\EE(W)\theta_0(a,a')$, adding and subtracting the true ratio
$\omega_0^{(1)}$, and using
\begin{equation*}
\EE\big[\omega_0^{(1)}(\eta_{0\bullet j}-\eta_{\bullet j})
\mid A=a,C,N\big]
=\iint(\eta_{0\bullet j}-\eta_{\bullet j})
\phi_0\kappa_0\,dz\,dm
\end{equation*}
yield
\begin{equation*}
\resizebox{\linewidth}{!}{$
\begin{aligned}
&\EE\{\psi_\theta(a,a';O;\eta,\phi,\kappa,\xi)\}
-\EE(W)\theta_0(a,a')\\
&=\EE\Bigg[\frac{W}{N}\sum_{j=1}^N\Bigg\{
\EE\big[
(\omega^{(1)}-\omega_0^{(1)})
(\eta_{0\bullet j}-\eta_{\bullet j})(a,Z,M,C,N)
\mid A=a,C,N
\big]\\
&\qquad\qquad
+\iint
\eta_{\bullet j}(a,z,m,C,N)
\{\phi_0-\phi\}(z\mid a,C,N)
\{\kappa-\kappa_0\}(m\mid a',C,N)
\,dz\,dm
\Bigg\}\Bigg].
\end{aligned}
$}
\end{equation*}
Now
\begin{equation*}
(\omega^{(1)}-\omega_0^{(1)})\xi_0
=(\kappa-\kappa_0)+\kappa\left(\frac{\xi_0}{\xi}-1\right).
\end{equation*}
Substitution and rearrangement gives the exact identity
\begin{equation*}
\resizebox{\linewidth}{!}{$
\begin{aligned}
&\EE\{\psi_\theta(a,a';O;\eta,\phi,\kappa,\xi)\}
-\EE(W)\theta_0(a,a')\\
&=\EE\Bigg[\frac{W}{N}\sum_{j=1}^N\Bigg\{
\iint\{\kappa-\kappa_0\}(m\mid a',C,N)
\left[
\eta_{0\bullet j}(a,z,m,C,N)\phi_0(z\mid a,C,N)
-\eta_{\bullet j}(a,z,m,C,N)\phi(z\mid a,C,N)
\right]
\,dz\,dm\\
&\qquad\qquad\qquad\qquad+\iint
\{\eta_{0\bullet j}-\eta_{\bullet j}\}(a,z,m,C,N)
\kappa(m\mid a',C,N)\phi_0(z\mid a,C,N)
\left\{
\frac{\xi_0(m\mid a,z,C,N)}
{\xi(m\mid a,z,C,N)}
-1
\right\}
\,dz\,dm
\Bigg\}\Bigg].
\end{aligned}
$}
\end{equation*}
Each of the three configurations in Theorem~\ref{thm:MR_theta} makes both
terms in \eqref{eq:bias_decomp_theta} equal to zero, proving the theorem.
\end{proof}

\subsection{Multiple robustness of \texorpdfstring{$\tau$}{tau}}
\label{sec:proof_MR_tau}

\begin{proof}
Again by iterated expectation, the conditional mean of the outcome-residual
term can be expressed as
\begin{equation*}
\begin{aligned}&\iint\{\eta_{0\bullet j}-\eta_{\bullet j}\}(1,z,m,C,N)
\kappa_{\bullet j}(m_{\bullet j}\mid1,C,N)
\kappa_{\bullet(-j)}(m_{\bullet(-j)}\mid0,C,N)\\
&\hspace{31mm}\times\phi_0(z\mid1,C,N)
\frac{\xi_0(m\mid1,z,C,N)}{\xi(m\mid1,z,C,N)}\,dm\,dz.
\end{aligned}
\end{equation*}
The conditional means of the four remaining score components are, respectively,
\begin{equation*}
\begin{aligned}&\iint\eta_{\bullet j}\phi_0\kappa_{\bullet j}\kappa_{\bullet(-j)}\,dm\,dz,
\qquad
\iint\eta_{\bullet j}\phi\kappa_{0\bullet j}\kappa_{\bullet(-j)}\,dm\,dz,\\
&\iint\eta_{\bullet j}\phi\kappa_{\bullet j}\kappa_{0\bullet(-j)}\,dm\,dz,
\qquad
-2\iint\eta_{\bullet j}\phi\kappa_{\bullet j}\kappa_{\bullet(-j)}\,dm\,dz,
\end{aligned}
\end{equation*}
where every function is evaluated at the arguments shown in
\eqref{eq:EIR_tau_formula}: $\phi$ is under arm $1$,
$\kappa_{\bullet j}$ under arm $1$, and $\kappa_{\bullet(-j)}$ under arm $0$.
Write $\xi_0/\xi=1+(\xi_0/\xi-1)$, subtract the true numerator, and collect
terms.  This yields
\begin{equation*}
\resizebox{\linewidth}{!}{$
\begin{aligned}
&\EE\{\psi_\tau(O;\eta,\phi,\kappa,\xi)\}-\EE(W)\tau_0\\
&=\EE\Bigg[\frac{W}{N}\sum_{j=1}^N\iint\Bigg\{
\{\kappa_{\bullet j}-\kappa_{0\bullet j}\}(m_{\bullet j}\mid1,C,N)
\kappa_{\bullet(-j)}(m_{\bullet(-j)}\mid0,C,N)
\left\{
\eta_{0\bullet j}(1,z,m,C,N)\phi_0(z\mid1,C,N)
-\eta_{\bullet j}(1,z,m,C,N)\phi(z\mid1,C,N)
\right\}\\
&\qquad\qquad\qquad\qquad\quad+\{\kappa_{\bullet(-j)}-\kappa_{0\bullet(-j)}\}
(m_{\bullet(-j)}\mid0,C,N)
\kappa_{\bullet j}(m_{\bullet j}\mid1,C,N)
\left\{
\eta_{0\bullet j}(1,z,m,C,N)\phi_0(z\mid1,C,N)
-\eta_{\bullet j}(1,z,m,C,N)\phi(z\mid1,C,N)
\right\}\\
&\qquad\qquad\qquad\qquad\quad-\eta_{0\bullet j}(1,z,m,C,N)\phi_0(z\mid1,C,N)
\{\kappa_{\bullet j}-\kappa_{0\bullet j}\}(m_{\bullet j}\mid1,C,N)
\{\kappa_{\bullet(-j)}-\kappa_{0\bullet(-j)}\}
(m_{\bullet(-j)}\mid0,C,N)\\
&\qquad\qquad\qquad\qquad\quad+\{\eta_{0\bullet j}-\eta_{\bullet j}\}(1,z,m,C,N)
\kappa_{\bullet j}(m_{\bullet j}\mid1,C,N)
\kappa_{\bullet(-j)}(m_{\bullet(-j)}\mid0,C,N)
\phi_0(z\mid1,C,N)
\left\{
\frac{\xi_0(m\mid1,z,C,N)}
{\xi(m\mid1,z,C,N)}
-1
\right\}
\Bigg\}\,dm\,dz\Bigg].
\end{aligned}
$}
\end{equation*}
Each configuration in Theorem~\ref{thm:MR_tau} makes every term in
\eqref{eq:MR-tau-bias} vanish, proving the theorem.
\end{proof}
\section{Generic cross-fitting machinery}
\label{sec:generic_crossfit}

The following lemma isolates the argument shared by the canonical and
low-dimensional estimators.  It is stated at the cluster level, so the
independent sample size is $K$.

\begin{lemma}
\label{lem:generic_crossfit}
Let $\zeta_0$ be a generic scalar estimand, let $\nu_0=\EE(W)\in(0,\infty)$, and set
$\mu_0=\nu_0\zeta_0$.  For $r=1,\ldots,R$, let $\widehat\psi_r$ be a random
function measurable with respect to the training clusters outside
$\mathcal I_r$, and define
\begin{equation*}
\widehat\mu
=\sum_{r=1}^R\frac{K_r}{K}\mathbb P_{K,r}(\widehat\psi_r),
\qquad
\widehat\zeta=\frac{\widehat\mu}{\mathbb P_K(W)},
\end{equation*}
where $\mathbb P_{K,r}$ is the empirical mean over $\mathcal I_r$.
Suppose Assumption~\ref{assm:crossfit_common} holds and there is a fixed
square-integrable function $\psi^\star$ such that, uniformly over $r$,
\begin{equation}
\|\widehat\psi_r-\psi^\star\|_{2,P}=o_p(1).
\label{eq:generic_score_conv}
\end{equation}
\begin{equation}
\EE\psi^\star=\mu_0.
\label{eq:generic_centering}
\end{equation}
\begin{equation}
|\EE(\widehat\psi_r-\psi^\star)|=o_p(K^{-1/2}).
\label{eq:generic_drift}
\end{equation}
Then
\begin{equation}
\sqrt K(\widehat\zeta-\zeta_0)
=\frac1{\sqrt K}\sum_{i=1}^K
\frac{\psi^\star(O_i)-\zeta_0W_i}{\nu_0}+o_p(1).
\label{eq:generic_AL}
\end{equation}
Consequently, if
$D^\star(O)=\{\psi^\star(O)-\zeta_0W\}/\nu_0$, then
$\sqrt K(\widehat\zeta-\zeta_0)\rightsquigarrow
N\{0,\EE(D^{\star 2})\}$.

If, for $i\in\mathcal I_r$,
\begin{equation*}
\widehat D_i
=\frac{\widehat\psi_r(O_i)-\widehat\zeta W_i}{\mathbb P_K(W)},
\end{equation*}
then $\mathbb P_K(\widehat D^2)\to_p\EE(D^{\star 2})$.
\end{lemma}

\begin{proof}
Let $p_r=K_r/K$.  Adding and subtracting $\psi^\star$ gives the exact
numerator decomposition
\begin{equation}
\widehat\mu-\mu_0
=(\mathbb P_K-\EE)\psi^\star+R_{1K}+R_{2K},
\label{eq:generic_num_decomp}
\end{equation}
where
\begin{equation*}
R_{1K}=\sum_{r=1}^R p_r(\mathbb P_{K,r}-\EE)
(\widehat\psi_r-\psi^\star),
\qquad
R_{2K}=\sum_{r=1}^R p_r\EE(\widehat\psi_r-\psi^\star).
\end{equation*}
For a fixed fold $r$, condition on the sigma-field generated by the training
clusters.  The function $g_r=\widehat\psi_r-\psi^\star$ is then fixed and
independent of the validation clusters, and
\begin{equation*}
K\,\Var\{p_r(\mathbb P_{K,r}-\EE)g_r\mid\mathcal F_{-r}\}
\le p_r\EE(g_r^2).
\end{equation*}
By \eqref{eq:generic_score_conv} and Chebyshev's inequality, we obtain
$\sqrt Kp_r(\mathbb P_{K,r}-\EE)g_r=o_p(1)$.  Since $R$ is fixed, we have
\begin{equation}
\sqrt KR_{1K}=o_p(1).
\label{eq:generic_R1}
\end{equation}
Condition~\eqref{eq:generic_drift} similarly gives
\begin{equation}
\sqrt KR_{2K}=o_p(1).
\label{eq:generic_R2}
\end{equation}
Combining \eqref{eq:generic_num_decomp}--\eqref{eq:generic_R2}, we get
\begin{equation}
\sqrt K(\widehat\mu-\mu_0)
=\sqrt K(\mathbb P_K-\EE)\psi^\star+o_p(1).
\label{eq:generic_num_AL}
\end{equation}
Thus $\widehat\mu-\mu_0=O_p(K^{-1/2})$.  Also,
$\widehat\nu=\mathbb P_K(W)$ satisfies
$\widehat\nu-\nu_0=O_p(K^{-1/2})$ and $\widehat\nu\to_p\nu_0$.
The identity
\begin{equation*}
\widehat\zeta-\zeta_0
=\frac{(\widehat\mu-\mu_0)-\zeta_0(\widehat\nu-\nu_0)}{\nu_0}
+\left(\frac1{\widehat\nu}-\frac1{\nu_0}\right)
\{(\widehat\mu-\mu_0)-\zeta_0(\widehat\nu-\nu_0)\}
\end{equation*}
shows that the second term is $O_p(K^{-1})$.
Substitution of \eqref{eq:generic_num_AL} proves \eqref{eq:generic_AL}; the
cluster-level central limit theorem concludes the Gaussian limit claim.

For variance estimation, score convergence and the same conditional Markov
argument imply
\begin{equation*}
\frac1K\sum_{r=1}^R\sum_{i\in\mathcal I_r}
\{\widehat\psi_r(O_i)-\psi^\star(O_i)\}^2=o_p(1).
\end{equation*}
Together with $\widehat\zeta\to_p\zeta_0$,
$\widehat\nu\to_p\nu_0$, and $\mathbb P_K(W^2)=O_p(1)$, this yields
$\mathbb P_K\{(\widehat D-D^\star)^2\}=o_p(1)$.  Hence, by
the Cauchy--Schwarz inequality, we obtain
\begin{equation*}
\left|\mathbb P_K(\widehat D^2)-\mathbb P_K(D^{\star 2})\right|=o_p(1),
\end{equation*}
and the distribution of large numbers yields
$\mathbb P_K(D^{\star 2})\to_p\EE(D^{\star 2})$.
\end{proof}

\section{Proof of Theorem~\ref{thm:CF-AN-MR}}
\label{sec:proof_AN}

We assume that the outcome-residual density ratios in
\eqref{eq:EIF_theta_formula}--\eqref{eq:EIR_tau_formula}, the auxiliary
regressions entering the applicable score, and the fitted-to-true ratios that
appear in the product-rate conditions below are well-defined on their
corresponding supports.  
Moreover, we assume that there is a finite constant $B$ such that, with
probability tending to one, the true, limiting, and fold-specific values of the
outcome regressions, the outcome-residual density ratios, the auxiliary
regressions entering the applicable score, and the fitted-to-true ratios in
\eqref{eq:canonical_rate_theta}--\eqref{eq:canonical_rate_tau} are bounded in
absolute value by $B$.

\begin{proof}
We verify the hypotheses of Lemma~\ref{lem:generic_crossfit} separately for
$\theta(a,a')$ and $\tau$.  Throughout, condition on the training clusters
outside a fixed validation fold $\mathcal I_r$.  The fold-specific nuisance
functions are then fixed, while the validation clusters remain independent
cluster-level observations with distribution $P$.

We first compute the elementary weighted truncation step used to derive score
convergence from the componentwise conditions in Assumption~\ref{assm:canonical_rates}.
If $G$ is nonnegative integrable, $|f_r|\leq B$, and $\EE(f_r^2)=o_p(1)$, then for every
$L>0$,
\begin{equation}
\EE(Gf_r^2)\leq L \EE(f_r^2)+B^2\EE\{G I(G>L)\}.
\label{eq:canonical_truncation}
\end{equation}
Letting first $K\to\infty$ and then $L\to\infty$ gives
$\EE(Gf_r^2)=o_p(1)$.  We apply this with
\begin{equation*}
G=W^2\left\{1+\frac1N\sum_{j=1}^{N}Y_{\bullet j}^2\right\},
\end{equation*}
which is integrable by Assumptions~\ref{assm:positivity} and \ref{assm:canonical_rates}. 

For $\theta(a,a')$, set
\begin{equation*}
\widehat\psi_r(O)=\psi_\theta\{a,a';O;\widehat h^{(-r)}\},
\qquad
\psi^\star(O)=\psi_{\theta,\mathrm{can}}^\star(O).
\end{equation*}
Since $h^\star$ satisfies one of the configurations in
Theorem~\ref{thm:MR_theta}, the exact identity
\eqref{eq:bias_decomp_theta} gives
$\EE\psi^\star=\nu_0\theta_0(a,a')$.

We next derive score convergence.  Expanding the
finite cluster-level sum in \eqref{eq:EIF_theta_formula}, the difference
$\widehat\psi_r-\psi^\star$ is a finite sum, over at most $N_{\max}$ indices,
of terms of the following forms:
\begin{equation*}
W(\widehat\omega^{(1),(-r)}-\omega^{(1),\star})Y_{\bullet j},
\quad
W\widehat\omega^{(1),(-r)}(\widehat\eta_{\bullet j}^{(-r)}-\eta_{\bullet j}^\star),
\quad
W\{\widehat u_{\bullet j}^{(\ell),(-r)}-u_{\bullet j}^{(\ell),\star}\},
\end{equation*}
up to fixed treatment-probability constants and signs, with $\ell=1,2,3$.
The second and third forms are controlled directly by boundedness and Assumption C(i).  The first form contains
$Y_{\bullet j}$ and is controlled by \eqref{eq:canonical_truncation}.  Since the
number of terms is finite by Assumption~\ref{assm:positivity}, we get
\begin{equation}
\|\widehat\psi_r-\psi_{\theta,\mathrm{can}}^\star\|_{2,P}=o_p(1)
\label{eq:canonical_derived_score_theta}
\end{equation}
uniformly over folds.  
Thus the score-convergence condition in Lemma~\ref{lem:generic_crossfit} is verified for the canonical $\theta$ score.

It remains to verify the conditional population drift.  Insert
$\widehat h^{(-r)}$ into the exact remainder
\eqref{eq:bias_decomp_theta}.  Fix $(C,N)$ and $j$.  For the first product in
that identity, insert the true product-intervention density and apply the
Cauchy--Schwarz inequality under the $(z,m)$ distribution:
\begin{equation*}
\resizebox{\linewidth}{!}{$
\begin{aligned}
&\left|\iint
\{\widehat\kappa^{(-r)}-\kappa_0\}(m\mid a',C,N)
\left\{
\eta_{0\bullet j}\phi_0-
\widehat\eta_{\bullet j}^{(-r)}\widehat\phi^{(-r)}
\right\}
\,dz\,dm\right|\\
&\qquad=
\left|\iint
\phi_0(z\mid a,C,N)\kappa_0(m\mid a',C,N)
\left\{
\frac{\widehat\kappa^{(-r)}(m\mid a',C,N)}
{\kappa_0(m\mid a',C,N)}-1
\right\}
\left\{
\eta_{0\bullet j}-
\widehat\eta_{\bullet j}^{(-r)}
\frac{\widehat\phi^{(-r)}(z\mid a,C,N)}
{\phi_0(z\mid a,C,N)}
\right\}
\,dz\,dm\right|\\
&\qquad\le
\left[
\iint
\phi_0\kappa_0
\left\{
\frac{\widehat\kappa^{(-r)}}{\kappa_0}-1
\right\}^2
\,dz\,dm
\right]^{1/2}
\left[
\iint
\phi_0\kappa_0
\left\{
\eta_{0\bullet j}
-
\widehat\eta_{\bullet j}^{(-r)}
\frac{\widehat\phi^{(-r)}}{\phi_0}
\right\}^2
\,dz\,dm
\right]^{1/2}.
\end{aligned}
$}
\end{equation*}
Multiplying by $W$, averaging over $(C,N)$ and over $j$, and applying another Cauchy--Schwarz inequality in the outer cluster-level distribution yields
\begin{equation}
\begin{split}
&\left|\EE\Bigg[\frac{W}{N}\sum_{j=1}^N
\iint
\{\widehat\kappa^{(-r)}-\kappa_0\}
\{\eta_{0\bullet j}\phi_0-
\widehat\eta_{\bullet j}^{(-r)}\widehat\phi^{(-r)}\}
\,dz\,dm\Bigg]\right|\\
&\qquad\leq C
\left\|\sqrt W\left\{
\frac{\widehat\kappa^{(-r)}}{\kappa_0}-1\right\}\right\|_{2,P}
\left\|\sqrt W\left\{\eta_0-
\widehat\eta^{(-r)}\frac{\widehat\phi^{(-r)}}{\phi_0}
\right\}\right\|_{2,P}.
\label{eq:canonical_theta_first_bound}
\end{split}
\end{equation}
The constant $C$ depends only on the treatment probabilities, $N_{\max}$, and the
bounded nuisance ratios.

For the second product in \eqref{eq:bias_decomp_theta}, rewrite
\begin{equation*}
\begin{split}
&\iint
\{\eta_{0\bullet j}-\widehat\eta_{\bullet j}^{(-r)}\}
\widehat\kappa^{(-r)}\phi_0
\left\{\frac{\xi_0}{\widehat\xi^{(-r)}}-1\right\}
\,dz\,dm\\
&\qquad=
\iint\phi_0\kappa_0
\{\eta_{0\bullet j}-\widehat\eta_{\bullet j}^{(-r)}\}
\frac{\widehat\kappa^{(-r)}}{\kappa_0}
\left\{\frac{\xi_0}{\widehat\xi^{(-r)}}-1\right\}
\,dz\,dm.
\end{split}
\end{equation*}
By the boundedness of $\widehat\kappa^{(-r)}/\kappa_0$, a similar Cauchy-Schwarz inequality argument provides
\begin{equation}
\begin{split}
&\left|\EE\Bigg[\frac{W}{N}\sum_{j=1}^N
\iint
\{\eta_{0\bullet j}-\widehat\eta_{\bullet j}^{(-r)}\}
\widehat\kappa^{(-r)}\phi_0
\left\{\frac{\xi_0}{\widehat\xi^{(-r)}}-1\right\}
\,dz\,dm\Bigg]\right|\\
&\qquad\leq C
\|\sqrt W\{\widehat\eta^{(-r)}-\eta_0\}\|_{2,P}
\left\|\sqrt W\left\{
\frac{\xi_0}{\widehat\xi^{(-r)}}-1\right\}\right\|_{2,P}.
\label{eq:canonical_theta_second_bound}
\end{split}
\end{equation}
Adding \eqref{eq:canonical_theta_first_bound} and
\eqref{eq:canonical_theta_second_bound}, and then using
\eqref{eq:canonical_rate_theta}, gives
\begin{equation*}
\left|\EE\widehat\psi_r-\EE(W)\theta_0(a,a')\right|=o_p(K^{-1/2})
\end{equation*}
uniformly over folds.  Since $\EE\psi^\star=\EE(W)\theta_0(a,a')$, we have
\begin{equation}
|\EE(\widehat\psi_r-\psi^\star)|=o_p(K^{-1/2}),
\label{eq:canonical_derived_drift_theta}
\end{equation}
which verifies the condition for Lemma~\ref{lem:generic_crossfit} regarding
$\theta(a,a')$.  Lemma~\ref{lem:generic_crossfit} proves
\eqref{eq:canonical_AL_theta} and variance consistency.

For $\tau$, set
\begin{equation*}
\widehat\psi_r(O)=\psi_\tau\{O;\widehat h^{(-r)}\},
\qquad
\psi^\star(O)=\psi_{\tau,\mathrm{can}}^\star(O).
\end{equation*}
Because $h^\star$ satisfies one of the configurations in
Theorem~\ref{thm:MR_tau}, the exact identity \eqref{eq:MR-tau-bias} gives
$\EE\psi^\star=\nu_0\tau_0$.
Expanding
\eqref{eq:EIR_tau_formula}, the difference
$\widehat\psi_r-\psi_{\tau,\mathrm{can}}^\star$ is a finite sum of terms of the
forms
\begin{equation*}
W(\widehat\omega_{\bullet j}^{(2),(-r)}-
\omega_{\bullet j}^{(2),\star})Y_{\bullet j},
\quad
W\widehat\omega_{\bullet j}^{(2),(-r)}
(\widehat\eta_{\bullet j}^{(-r)}-\eta_{\bullet j}^\star),
\quad
W\{\widehat r_{\bullet j}^{(\ell),(-r)}-r_{\bullet j}^{(\ell),\star}\},
\end{equation*}
with $\ell=1,2,3,4$, up to fixed treatment-probability constants and signs.
Boundedness, Assumption C(ii), the finite support of
$N$, and \eqref{eq:canonical_truncation} therefore imply
\begin{equation}
\|\widehat\psi_r-\psi_{\tau,\mathrm{can}}^\star\|_{2,P}=o_p(1)
\label{eq:canonical_derived_score_tau}
\end{equation}
uniformly over folds.

It remains to verify the drift.  Insert $\widehat h^{(-r)}$ into
\eqref{eq:MR-tau-bias}.  For the first product, write the integrand relative to
$\phi_0(z\mid1,C,N)\kappa_{0\bullet j}(m_{\bullet j}\mid1,C,N)
\kappa_{0\bullet(-j)}(m_{\bullet(-j)}\mid0,C,N)$:
\begin{equation*}
\begin{split}
&\{\widehat\kappa_{\bullet j}^{(-r)}-\kappa_{0\bullet j}\}
\widehat\kappa_{\bullet(-j)}^{(-r)}
\{\eta_{0\bullet j}\phi_0-
\widehat\eta_{\bullet j}^{(-r)}\widehat\phi^{(-r)}\}\\
&\qquad=
\phi_0\kappa_{0\bullet j}\kappa_{0\bullet(-j)}
\left\{\frac{\widehat\kappa_{\bullet j}^{(-r)}}{\kappa_{0\bullet j}}-1\right\}
\frac{\widehat\kappa_{\bullet(-j)}^{(-r)}}{\kappa_{0\bullet(-j)}}
\left\{\eta_{0\bullet j}-
\widehat\eta_{\bullet j}^{(-r)}\frac{\widehat\phi^{(-r)}}{\phi_0}\right\}.
\end{split}
\end{equation*}
By boundedness of
$\widehat\kappa_{\bullet(-j)}^{(-r)}/\kappa_{0\bullet(-j)}$ and the Cauchy--Schwarz inequality, we get
\begin{equation}
\begin{split}
&\left|\EE\left[\frac{W}{N}\sum_{j=1}^N\iint
\{\widehat\kappa_{\bullet j}^{(-r)}-\kappa_{0\bullet j}\}
\widehat\kappa_{\bullet(-j)}^{(-r)}
\{\eta_{0\bullet j}\phi_0-
\widehat\eta_{\bullet j}^{(-r)}\widehat\phi^{(-r)}\}
\,dm\,dz\right]\right|\\
&\qquad\leq C\max_{1\leq j\leq N_{\max}}
\left\|\sqrt W\left\{
\frac{\widehat\kappa_{\bullet j}^{(-r)}}{\kappa_{0\bullet j}}-1
\right\}\right\|_{2,P}
\left\|\sqrt W\left\{\eta_0-
\widehat\eta^{(-r)}\frac{\widehat\phi^{(-r)}}{\phi_0}
\right\}\right\|_{2,P}.
\label{eq:canonical_tau_first_bound}
\end{split}
\end{equation}
The second product is symmetric and implies
\begin{equation}
\begin{split}
&\left|\EE\left[\frac{W}{N}\sum_{j=1}^N\iint
\{\widehat\kappa_{\bullet(-j)}^{(-r)}-\kappa_{0\bullet(-j)}\}
\widehat\kappa_{\bullet j}^{(-r)}
\{\eta_{0\bullet j}\phi_0-
\widehat\eta_{\bullet j}^{(-r)}\widehat\phi^{(-r)}\}
\,dm\,dz\right]\right|\\
&\qquad\leq C\max_{1\leq j\leq N_{\max}}
\left\|\sqrt W\left\{
\frac{\widehat\kappa_{\bullet(-j)}^{(-r)}}{\kappa_{0\bullet(-j)}}-1
\right\}\right\|_{2,P}
\left\|\sqrt W\left\{\eta_0-
\widehat\eta^{(-r)}\frac{\widehat\phi^{(-r)}}{\phi_0}
\right\}\right\|_{2,P}.
\label{eq:canonical_tau_second_bound}
\end{split}
\end{equation}
For the third product, insert the same true product density:
\begin{equation*}
\resizebox{\linewidth}{!}{$
\displaystyle
\eta_{0\bullet j}\phi_0
\{\widehat\kappa_{\bullet j}^{(-r)}-\kappa_{0\bullet j}\}
\{\widehat\kappa_{\bullet(-j)}^{(-r)}-\kappa_{0\bullet(-j)}\}
=
\eta_{0\bullet j}\phi_0
\kappa_{0\bullet j}\kappa_{0\bullet(-j)}
\left\{
\frac{\widehat\kappa_{\bullet j}^{(-r)}}
{\kappa_{0\bullet j}}-1
\right\}
\left\{
\frac{\widehat\kappa_{\bullet(-j)}^{(-r)}}
{\kappa_{0\bullet(-j)}}-1
\right\}.
$}
\end{equation*}
The boundedness of $\eta_0$ and the  Cauchy--Schwarz inequality yield
\begin{equation}
\begin{split}
&\left|\EE\left[\frac{W}{N}\sum_{j=1}^N\iint
\eta_{0\bullet j}\phi_0
\{\widehat\kappa_{\bullet j}^{(-r)}-\kappa_{0\bullet j}\}
\{\widehat\kappa_{\bullet(-j)}^{(-r)}-\kappa_{0\bullet(-j)}\}
\,dm\,dz\right]\right|\\
&\qquad\leq C\max_{1\leq j\leq N_{\max}}
\left\|\sqrt W\left\{
\frac{\widehat\kappa_{\bullet j}^{(-r)}}{\kappa_{0\bullet j}}-1
\right\}\right\|_{2,P}
\left\|\sqrt W\left\{
\frac{\widehat\kappa_{\bullet(-j)}^{(-r)}}{\kappa_{0\bullet(-j)}}-1
\right\}\right\|_{2,P}.
\label{eq:canonical_tau_third_bound}
\end{split}
\end{equation}
For the fourth product, rewrite
\begin{equation*}
\begin{split}
&\{\eta_{0\bullet j}-\widehat\eta_{\bullet j}^{(-r)}\}
\widehat\kappa_{\bullet j}^{(-r)}
\widehat\kappa_{\bullet(-j)}^{(-r)}\phi_0
\left\{\frac{\xi_0}{\widehat\xi^{(-r)}}-1\right\}\\
&\qquad=
\phi_0\kappa_{0\bullet j}\kappa_{0\bullet(-j)}
\{\eta_{0\bullet j}-\widehat\eta_{\bullet j}^{(-r)}\}
\frac{\widehat\kappa_{\bullet j}^{(-r)}}{\kappa_{0\bullet j}}
\frac{\widehat\kappa_{\bullet(-j)}^{(-r)}}{\kappa_{0\bullet(-j)}}
\left\{\frac{\xi_0}{\widehat\xi^{(-r)}}-1\right\}.
\end{split}
\end{equation*}
Absorbing the two bounded fitted-to-true mediator ratios into $C$ gives
\begin{equation}
\begin{split}
&\left|\EE\left[\frac{W}{N}\sum_{j=1}^N\iint
\{\eta_{0\bullet j}-\widehat\eta_{\bullet j}^{(-r)}\}
\widehat\kappa_{\bullet j}^{(-r)}
\widehat\kappa_{\bullet(-j)}^{(-r)}\phi_0
\left\{\frac{\xi_0}{\widehat\xi^{(-r)}}-1\right\}
\,dm\,dz\right]\right|\\
&\qquad\leq C\max_{1\leq j\leq N_{\max}}
\|\sqrt W\{\widehat\eta_{\bullet j}^{(-r)}-\eta_{0\bullet j}\}\|_{2,P}
\left\|\sqrt W\left\{
\frac{\xi_0}{\widehat\xi^{(-r)}}-1\right\}\right\|_{2,P}.
\label{eq:canonical_tau_fourth_bound}
\end{split}
\end{equation}
Adding \eqref{eq:canonical_tau_first_bound}--
\eqref{eq:canonical_tau_fourth_bound} and using
\eqref{eq:canonical_rate_tau} yields
\begin{equation*}
\left|\EE\widehat\psi_r-\EE(W)\tau_0\right|=o_p(K^{-1/2})
\end{equation*}
uniformly over folds.  Since $\EE\psi^\star=\EE(W)\tau_0$, we obtain
\begin{equation}
|\EE(\widehat\psi_r-\psi^\star)|=o_p(K^{-1/2}),
\label{eq:canonical_derived_drift_tau}
\end{equation}
which verifies Lemma~\ref{lem:generic_crossfit}'s drift condition for $\tau$.
Together with \eqref{eq:canonical_derived_score_tau}, Lemma~\ref{lem:generic_crossfit}
proves \eqref{eq:canonical_AL_tau} and variance consistency.

If $h^\star=h_0$, Theorem~\ref{thm:EIF} identifies the limiting
influence functions with the canonical gradients, so their variances attain the
nonparametric efficiency bounds.
\end{proof}

\section{Proofs for Section~\ref{sec:lowdim_reparam}}
\label{sec:ratio_implementation}

\subsection{Augmented-classification estimation of \texorpdfstring{\(d\)}{d}}
\label{sec:augmented_classification}

The ratio \(d\) in \eqref{eq:lowdim_d} can also be written as
\[
d(a,z,m,c,n)
=
\frac{\phi(z\mid a,c,n)}
     {f_{Z\mid A,M,C,N}(z\mid a,m,c,n)}
=
\frac{\kappa(m\mid a,c,n)}
     {\xi(m\mid a,z,c,n)} .
\]
Thus \(d\) is the conditional density ratio of the distribution of \(Z\) given
\((A,C,N)\) to the distribution of \(Z\) given \((A,M,C,N)\). We estimate this
ratio using the augmented-classification method proposed by
\citet{rudolph2024practical}, adapted to the cluster-level observed data.

For each treatment arm \(a\), let \(\lambda_a(\bullet\mid c,n)\) be a
user-specified reference distribution on mediator vectors of dimension \(n\), with support
covering the support of \(M\mid A=a,C=c,N=n\). The reference draw is required
to be conditionally independent of \(Z\) given \((A,C,N)\). In a training fold,
construct an augmented sample by creating, for each observed training cluster,
one observed copy with label \(\Gamma=0\) and one reference copy with label
\(\Gamma=1\). The observed copy keeps \((Z,M,C,N,A)\) unchanged. The reference
copy keeps \((Z,C,N,A)\) unchanged but replaces \(M\) by
\[
\widetilde M \sim \lambda_A(\bullet\mid C,N),
\]
drawn independently of \(Z\) conditional on \((A,C,N)\).

Let \(P_\lambda\) denote the distribution of the augmented data. Define the two binary
classification functions
\[
\ell_a(z,m,c,n)
=
P_\lambda(\Gamma=1\mid A=a,Z=z,M=m,C=c,N=n),
\]
and
\[
\ell^M_a(m,c,n)
=
P_\lambda(\Gamma=1\mid A=a,M=m,C=c,N=n).
\]
Then, on the common support, one has
\[
d(a,z,m,c,n)
=
\frac{\ell_a(z,m,c,n)}{1-\ell_a(z,m,c,n)}
\frac{1-\ell^M_a(m,c,n)}{\ell^M_a(m,c,n)} .
\]
To verify this identity, note that under \(P_\lambda\), conditional on
\((A=a,C=c,N=n)\), the observed copy has joint density
\[
\phi(z\mid a,c,n)\xi(m\mid a,z,c,n),
\]
whereas the reference copy has joint density
\[
\phi(z\mid a,c,n)\lambda_a(m\mid c,n).
\]
Therefore,
\[
\frac{\ell_a(z,m,c,n)}{1-\ell_a(z,m,c,n)}
=
\frac{\lambda_a(m\mid c,n)}
     {\xi(m\mid a,z,c,n)} .
\]
Similarly, after marginalizing over \(Z\),
\[
\frac{\ell^M_a(m,c,n)}{1-\ell^M_a(m,c,n)}
=
\frac{\lambda_a(m\mid c,n)}
     {\kappa(m\mid a,c,n)} .
\]
Taking the ratio of these two odds eliminates the user-specified reference distribution
\(\lambda_a\) and yields
\[
\frac{\ell_a(z,m,c,n)}{1-\ell_a(z,m,c,n)}
\frac{1-\ell^M_a(m,c,n)}{\ell^M_a(m,c,n)}
=
\frac{\kappa(m\mid a,c,n)}
     {\xi(m\mid a,z,c,n)}
=
d(a,z,m,c,n).
\]
Multiple reference copies may be generated for each training cluster. Provided that the two classifiers use the same class weights, the common class-prior odds cancel from their odds ratio.

In implementation, for each validation fold \(I_r\), the two classifiers
\(\ell_a\) and \(\ell^M_a\) are trained only on clusters outside \(I_r\). The
fold-specific estimator is
\[
\widehat d^{(-r)}(a,z,m,c,n)
=
\frac{\widehat\ell^{(-r)}_a(z,m,c,n)}
     {1-\widehat\ell^{(-r)}_a(z,m,c,n)}
\frac{1-\widehat\ell^{M,(-r)}_a(m,c,n)}
     {\widehat\ell^{M,(-r)}_a(m,c,n)} .
\]

\subsection{Proof of Proposition~\ref{prop:lowdim_identities}}\label{sec: proof:lowdim_identities}

\begin{proof}
Bayes' rule and randomization give
\begin{equation*}
\frac{\pi_a}{\pi_{a'}}
\frac{s(a'\mid m,c,n)}{s(a\mid m,c,n)}
=\frac{\kappa(m\mid a',c,n)}{\kappa(m\mid a,c,n)}.
\end{equation*}
Multiplication by $d(a,z,m,c,n)$ proves
\eqref{eq:lowdim_omega1}.  Applying Bayes' rule to
$A\mid(M_{\bullet(-j)},C,N)$ gives
\begin{equation*}
\frac{\pi}{1-\pi}
\frac{s_{-j}(0\mid m_{\bullet(-j)},c,n)}
{s_{-j}(1\mid m_{\bullet(-j)},c,n)}
=\frac{\kappa_{\bullet(-j)}(m_{\bullet(-j)}\mid0,c,n)}
{\kappa_{\bullet(-j)}(m_{\bullet(-j)}\mid1,c,n)}.
\end{equation*}
Since
\begin{equation*}
\kappa(m\mid1,c,n)
=\kappa_{\bullet j}^{\star}
(m_{\bullet j}\mid m_{\bullet(-j)},1,c,n)
\kappa_{\bullet(-j)}(m_{\bullet(-j)}\mid1,c,n),
\end{equation*}
substitution into \eqref{eq:lowdim_e} proves
\eqref{eq:lowdim_omega2}.

Displays~\eqref{eq:lowdim_u1}--\eqref{eq:lowdim_u3} follow directly from
iterated expectation, with $z$ or $m$ held fixed when it appears as an
argument of the working regression.  We verify the spillover regressions.
Under $A=0,C=c,N=n$, the joint mediator density factors as
\begin{equation*}
\kappa(m\mid0,c,n)
=\kappa_{\bullet j}^{\star}
(m_{\bullet j}\mid m_{\bullet(-j)},0,c,n)
\kappa_{\bullet(-j)}(m_{\bullet(-j)}\mid0,c,n).
\end{equation*}
Therefore, the right-hand side of \eqref{eq:lowdim_r1} equals
\begin{equation*}
\int \eta_{\bullet j}(1,z,m,c,n)
\kappa_{\bullet j}(m_{\bullet j}\mid1,c,n)
\kappa_{\bullet(-j)}(m_{\bullet(-j)}\mid0,c,n)\,dm,
\end{equation*}
which is $r_{\bullet j}^{(1)}(1,1,0,z,c,n)$ with the working $\eta$.
Conditional on $A=0,M_{\bullet j}=m_{\bullet j},C=c,N=n$, the density of
$M_{\bullet(-j)}$ is
\begin{equation*}
\frac{\kappa_{\bullet j}^{\star}
(m_{\bullet j}\mid m_{\bullet(-j)},0,c,n)
\kappa_{\bullet(-j)}(m_{\bullet(-j)}\mid0,c,n)}
{\kappa_{\bullet j}(m_{\bullet j}\mid0,c,n)}.
\end{equation*}
The scalar density ratio in \eqref{eq:lowdim_r2} cancels the first and last
factors, proving \eqref{eq:lowdim_r2}.  Similarly, conditional on
$A=1,M_{\bullet(-j)}=m_{\bullet(-j)},C=c,N=n$, the density of
$M_{\bullet j}$ is
$\kappa_{\bullet j}^{\star}
(m_{\bullet j}\mid m_{\bullet(-j)},1,c,n)$; hence
\eqref{eq:lowdim_r3} follows by cancellation.  Integrating
\eqref{eq:lowdim_r2} against
$\kappa_{\bullet j}(m_{\bullet j}\mid1,c,n)$, or averaging
\eqref{eq:lowdim_r3} over
$M_{\bullet(-j)}\mid(A=0,C=c,N=n)$, gives the two expressions in
\eqref{eq:lowdim_r4}.
\end{proof}

\subsection{Proof of Lemma~\ref{lem:lowdim_canonical}}\label{sec: lem:lowdim_canonical proof}

\begin{proof}
At $\eta=\eta_0$, the partial means in
Proposition~\ref{prop:lowdim_identities} are exactly the functions
$u^{(1)},u^{(2)},u^{(3)}$ and
$r^{(1)},r^{(2)},r^{(3)},r^{(4)}$ defined in Section~\ref{sec:eif}.  
Substitution of
\eqref{eq:lowdim_omega1} into \eqref{eq:lowdim_score_theta} gives
\eqref{eq:EIF_theta_formula} term by term.  Substitution of
\eqref{eq:lowdim_omega2} into \eqref{eq:lowdim_score_tau} gives
\eqref{eq:EIR_tau_formula} term by term.  Both equalities are pointwise.
\end{proof}

\subsection{Proof of Theorem~\ref{thm:lowdim_drift}}\label{sec: proof thm:lowdim_drift}

\begin{proof}
We first prove \eqref{eq:lowdim_drift_theta}.  In this proof, denote
\begin{equation*}
\rho(M,C,N)=\frac{\pi_a}{\pi_{a'}}
\frac{s(a'\mid M,C,N)}{s(a\mid M,C,N)},
\qquad
\rho_0(M,C,N)=\frac{\pi_a}{\pi_{a'}}
\frac{s_0(a'\mid M,C,N)}{s_0(a\mid M,C,N)}.
\end{equation*}
Fix $(C,N)$ and one $j$.  By iterated expectation, the conditional mean of the
$j$th summand in \eqref{eq:lowdim_score_theta} is
\begin{equation*}
\begin{split}
&\EE\left[d(a,Z,M,C,N)\rho(M,C,N)
\{\eta_{0\bullet j}-\eta_{\bullet j}\}(a,Z,M,C,N)
\mid A=a,C,N\right]\\
&\quad+\EE\{u_{\bullet j}^{(1)}(a,a',Z,C,N)\mid A=a,C,N\}\\
&\quad+\EE\{u_{\bullet j}^{(2)}(a,M,C,N)\mid A=a',C,N\}
-u_{\bullet j}^{(3)}(a,a',C,N).
\end{split}
\end{equation*}
Compatibility and Proposition~\ref{prop:lowdim_identities} imply that the last
three terms equal
\begin{equation*}
\iint\eta_{\bullet j}(a,z,m,C,N)
\phi_0(z\mid a,C,N)\kappa_0(m\mid a',C,N)\,dz\,dm.
\end{equation*}
The Radon--Nikodym identities for $d_0$ and the true classifier odds show that
this display equals
\begin{equation*}
\EE\left[d_0(a,Z,M,C,N)\rho_0(M,C,N)
\eta_{\bullet j}(a,Z,M,C,N)\mid A=a,C,N\right].
\end{equation*}
After subtracting the true conditional target, the conditional drift is
therefore
\begin{equation*}
\EE\left[\{d\rho-d_0\rho_0\}
\{\eta_{0\bullet j}-\eta_{\bullet j}\}
\mid A=a,C,N\right].
\end{equation*}
The algebraic identity
\begin{equation*}
d\rho-d_0\rho_0=\rho(d-d_0)+d_0(\rho-\rho_0)
\end{equation*}
gives the first product in \eqref{eq:lowdim_drift_theta}.
It remains to identify the second product in \eqref{eq:lowdim_drift_theta}. Since
\(\rho(M,C,N)-\rho_0(M,C,N)\) is measurable with respect to
\((M,C,N)\), iterated expectation gives
\[
\resizebox{\linewidth}{!}{$
\begin{aligned}
& E\Big[
d_0(a,Z,M,C,N)\{\rho(M,C,N)-\rho_0(M,C,N)\}
\{\eta_{0\bullet j}-\eta_{\bullet j}\}(a,Z,M,C,N)
\,\Big|\, A=a,C,N
\Big] \\
&\quad =
E\Big[
\{\rho(M,C,N)-\rho_0(M,C,N)\}
E\Big\{
d_0(a,Z,M,C,N)
\{\eta_{0\bullet j}-\eta_{\bullet j}\}(a,Z,M,C,N)
\,\Big|\, A=a,M,C,N
\Big\}
\,\Big|\, A=a,C,N
\Big].
\end{aligned}
$}
\]
For fixed \((m,c,n)\), the inner conditional expectation is
\[
\begin{aligned}
& E\{d_0(a,Z,M,C,N)
\{\eta_{0\bullet j}-\eta_{\bullet j}\}(a,Z,M,C,N)
\,\big|\, A=a,M=m,C=c,N=n\} \\
&\quad =
\int
\frac{\kappa_0(m\mid a,c,n)}
     {\xi_0(m\mid a,z,c,n)}
\{\eta_{0\bullet j}-\eta_{\bullet j}\}(a,z,m,c,n)
\frac{\phi_0(z\mid a,c,n)\xi_0(m\mid a,z,c,n)}
     {\kappa_0(m\mid a,c,n)}
\,dz \\
&\quad =
\int
\{\eta_{0\bullet j}-\eta_{\bullet j}\}(a,z,m,c,n)
\phi_0(z\mid a,c,n)\,dz \\
&\quad =
\{u^{(2)}_{0\bullet j}-u^{(2)}_{\bullet j}\}(a,m,c,n),
\end{aligned}
\]
where the last equality follows from compatibility of \(u^{(2)}_{\bullet j}\) with the
same working regression \(\eta_{\bullet j}\). Therefore, we obtain
\[
\resizebox{\linewidth}{!}{$
\begin{aligned}
& E\Big[
d_0(a,Z,M,C,N)\{\rho(M,C,N)-\rho_0(M,C,N)\}
\{\eta_{0\bullet j}-\eta_{\bullet j}\}(a,Z,M,C,N)
\,\Big|\, A=a,C,N
\Big] \\
&\quad =
E\Big[
\{\rho(M,C,N)-\rho_0(M,C,N)\}
\{u^{(2)}_{0\bullet j}-u^{(2)}_{\bullet j}\}(a,M,C,N)
\,\Big|\, A=a,C,N
\Big].
\end{aligned}
$}
\]
Combining this display with the first product and then averaging over
\((C,N)\), multiplying by \(W/N\), and summing over \(j\) proves \eqref{eq:lowdim_drift_theta}.

We next prove \eqref{eq:lowdim_drift_tau}. Since $\rho_j-\rho_{0j}$ is measurable with respect to
$(M_{\bullet(-j)},C,N)$, the second term in
\[
\EE\left[
\left\{\rho_j(e_{\bullet j}-e_{0\bullet j})
+e_{0\bullet j}(\rho_j-\rho_{0j})\right\}
\{\eta_{0\bullet j}-\eta_{\bullet j}\}
\mid A=1,C,N\right]
\]
can be reduced by conditioning further on $(M_{\bullet(-j)},C,N)$.  We now
identify the resulting conditional mean.  Fix
$(m_{\bullet(-j)},c,n)$ and write $m=(m_{\bullet j},m_{\bullet(-j)})$.  Under the
true distribution, one has
\begin{equation*}
\begin{split}
p_0(z,m_{\bullet j}
\mid A=1,M_{\bullet(-j)}=m_{\bullet(-j)},C=c,N=n)=
\frac{
\phi_0(z\mid1,c,n)\xi_0(m\mid1,z,c,n)}
{\kappa_{0\bullet(-j)}(m_{\bullet(-j)}\mid1,c,n)} .
\end{split}
\end{equation*}
Also, by the definition of $e_{0\bullet j}$, we get
\begin{equation*}
\begin{split}
e_{0\bullet j}(z,m,c,n)
&=
d_0(1,z,m,c,n)
\frac{\kappa_{0\bullet j}(m_{\bullet j}\mid1,c,n)}
{\kappa_{0\bullet j}^{\star}
(m_{\bullet j}\mid m_{\bullet(-j)},1,c,n)}\\
&=
\frac{\kappa_0(m\mid1,c,n)}
{\xi_0(m\mid1,z,c,n)}
\frac{\kappa_{0\bullet j}(m_{\bullet j}\mid1,c,n)}
{\kappa_{0\bullet j}^{\star}
(m_{\bullet j}\mid m_{\bullet(-j)},1,c,n)} .
\end{split}
\end{equation*}
Since
\begin{equation*}
\kappa_0(m\mid1,c,n)
=
\kappa_{0\bullet j}^{\star}
(m_{\bullet j}\mid m_{\bullet(-j)},1,c,n)
\kappa_{0\bullet(-j)}(m_{\bullet(-j)}\mid1,c,n),
\end{equation*}
we obtain the following identity
\begin{equation*}
\begin{split}
&e_{0\bullet j}(z,m,c,n)
p_0(z,m_{\bullet j}
\mid A=1,M_{\bullet(-j)}=m_{\bullet(-j)},C=c,N=n)\\
&\quad =
\phi_0(z\mid1,c,n)
\kappa_{0\bullet j}(m_{\bullet j}\mid1,c,n).
\end{split}
\end{equation*}
Therefore, we obtain
\begin{equation*}
\begin{split}
&\EE\Big[
e_{0\bullet j}(Z,M,C,N)
\{\eta_{0\bullet j}-\eta_{\bullet j}\}(1,Z,M,C,N)
\mid A=1,M_{\bullet(-j)},C,N\Big]\\
&\quad =
\iint
\{\eta_{0\bullet j}-\eta_{\bullet j}\}
(1,z,m,C,N)
\phi_0(z\mid1,C,N)
\kappa_{0\bullet j}(m_{\bullet j}\mid1,C,N)
\,dz\,dm_{\bullet j}.
\end{split}
\end{equation*}
The part involving $\eta_{0\bullet j}$ is exactly
$r_{0\bullet j}^{(3)}(1,1,M_{\bullet(-j)},C,N)$.  For the part involving
$\eta_{\bullet j}$, compatibility yields
\begin{equation*}
u_{\bullet j}^{(2)}(1,m,c,n)
=
\int
\eta_{\bullet j}(1,z,m,c,n)
\phi_0(z\mid1,c,n)\,dz .
\end{equation*}
Using the compatible definition of $r_{\bullet j}^{(3)}$, together with the true
conditional density of $M_{\bullet j}$ given
$(A=1,M_{\bullet(-j)},C,N)$, gives
\begin{equation*}
\resizebox{\linewidth}{!}{$
\begin{aligned}
&r_{\bullet j}^{(3)}(1,1,m_{\bullet(-j)},c,n)\\
&\quad =
\EE\left[
u_{\bullet j}^{(2)}(1,M,C,N)
\frac{\kappa_{0\bullet j}(M_{\bullet j}\mid1,C,N)}
{\kappa_{0\bullet j}^{\star}
(M_{\bullet j}\mid M_{\bullet(-j)},1,C,N)}
\ \Bigm|\ A=1,M_{\bullet(-j)}=m_{\bullet(-j)},C=c,N=n
\right]\\
&\quad =
\int
u_{\bullet j}^{(2)}(1,m,c,n)
\kappa_{0\bullet j}(m_{\bullet j}\mid1,c,n)
\,dm_{\bullet j}\\
&\quad =
\iint
\eta_{\bullet j}(1,z,m,c,n)
\phi_0(z\mid1,c,n)
\kappa_{0\bullet j}(m_{\bullet j}\mid1,c,n)
\,dz\,dm_{\bullet j}.
\end{aligned}
$}
\end{equation*}
Consequently, we get
\begin{equation*}
\begin{split}
&\EE\Big[
e_{0\bullet j}(Z,M,C,N)
\{\eta_{0\bullet j}-\eta_{\bullet j}\}(1,Z,M,C,N)
\mid A=1,M_{\bullet(-j)},C,N\Big]\\
&\quad =
\{r_{0\bullet j}^{(3)}-r_{\bullet j}^{(3)}\}
(1,1,M_{\bullet(-j)},C,N).
\end{split}
\end{equation*}
Hence the second part of the conditional drift is
\begin{equation*}
\begin{split}
&\EE\Big[
e_{0\bullet j}(\rho_j-\rho_{0j})
\{\eta_{0\bullet j}-\eta_{\bullet j}\}
\mid A=1,C,N\Big]\\
&\quad =
\EE\Big[
(\rho_j-\rho_{0j})
\{r_{0\bullet j}^{(3)}-r_{\bullet j}^{(3)}\}
(1,1,M_{\bullet(-j)},C,N)
\mid A=1,C,N\Big],
\end{split}
\end{equation*}
which is the second product in \eqref{eq:lowdim_drift_tau}.
Averaging over $(C,N)$, multiplying
by $W/N$, and summing over $j$ prove \eqref{eq:lowdim_drift_tau}.
\end{proof}

\subsection{Proof of Theorem~\ref{thm:lowdim_AN}}\label{sec: thm:lowdim_AN proof}

We assume that there is a finite constant $B$ such that, with probability tending to one, the true,
limiting, and fold-specific values of the classifier odds, $d$, $e_{\bullet j}$,
$\eta$, and all auxiliary regressions entering the applicable score are
bounded in absolute value by $B$.

\begin{proof}
We verify the three hypotheses of Lemma~\ref{lem:generic_crossfit} from the
nuisance-level conditions in Assumption~\ref{assm:lowdim_rates}.  For every
fold, condition on the training clusters.  The fold-specific nuisance
functions are then fixed and independent of a generic validation cluster.

Consider $\theta(a,a')$.  In this proof, let
\begin{equation*}
\widehat\rho_r(M,C,N)=\frac{\pi_a}{\pi_{a'}}
\frac{\widehat s^{(-r)}(a'\mid M,C,N)}
{\widehat s^{(-r)}(a\mid M,C,N)}
\end{equation*}
and define $\rho^\star$ and $\rho_0$ analogously.  Boundedness and
\eqref{eq:lowdim_limit_theta} imply
\begin{equation*}
\begin{split}
\|\widehat d^{(-r)}\widehat\rho_r-d^\star\rho^\star\|_{2,P}
\leq B\|\widehat d^{(-r)}-d^\star\|_{2,P}
+B\|\widehat\rho_r-\rho^\star\|_{2,P}=o_p(1)
\end{split}
\end{equation*}
uniformly over folds.  The outcome-residual part satisfies the exact algebraic
identity
\begin{equation*}
\begin{split}
&\widehat d^{(-r)}\widehat\rho_r
\{Y_{\bullet j}-\widehat\eta_{\bullet j}^{(-r)}\}
-d^\star\rho^\star\{Y_{\bullet j}-\eta_{\bullet j}^\star\}\\
&\quad=
\{\widehat d^{(-r)}\widehat\rho_r-d^\star\rho^\star\}
\{Y_{\bullet j}-\eta_{\bullet j}^\star\}
-\widehat d^{(-r)}\widehat\rho_r
\{\widehat\eta_{\bullet j}^{(-r)}-\eta_{\bullet j}^\star\}.
\end{split}
\end{equation*}
The remaining score difference is a bounded linear combination of
\begin{equation*}
\widehat u_{\bullet j}^{(1),(-r)}-u_{\bullet j}^{(1),\star},
\qquad
\widehat u_{\bullet j}^{(2),(-r)}-u_{\bullet j}^{(2),\star},
\qquad
\widehat u_{\bullet j}^{(3),(-r)}-u_{\bullet j}^{(3),\star}.
\end{equation*}
Using $(N^{-1}\sum_j x_j)^2\leq N^{-1}\sum_jx_j^2$, bounded treatment
weights, and bounded nuisance functions, there is a fixed constant $C$ such
that
\begin{equation*}
\resizebox{\linewidth}{!}{$
\begin{aligned}
&|\widehat\psi_{\theta,r}(O)-
\psi_{\theta,\mathrm{ld}}^\star(O)|^2\\
&\quad\leq C\frac{W^2}{N}\sum_{j=1}^N\Bigg[
(1+Y_{\bullet j}^2)
\left\{|\widehat d^{(-r)}\widehat\rho_r-d^\star\rho^\star|^2
+|\widehat\eta_{\bullet j}^{(-r)}-\eta_{\bullet j}^\star|^2\right\}
+\sum_{\ell=1}^3
|\widehat u_{\bullet j}^{(\ell),(-r)}-u_{\bullet j}^{(\ell),\star}|^2
\Bigg].
\end{aligned}
$}
\end{equation*}
Again, by the same truncation argument used in the proof of Theorem~\ref{thm:CF-AN-MR}, together with
the bounded weights, the preceding score bound, and
\eqref{eq:lowdim_limit_theta} provide
\begin{equation}
\|\widehat\psi_{\theta,r}-
\psi_{\theta,\mathrm{ld}}^\star\|_{2,P}=o_p(1)
\label{eq:derived_score_limit_theta}
\end{equation}
uniformly over folds.

Since $h_\theta^\star$ is compatible and satisfies
\eqref{eq:lowdim_MR_theta}, Theorem~\ref{thm:lowdim_drift} gives
$\EE\{\psi_{\theta,\mathrm{ld}}^\star(O)\}=\EE(W)\theta_0(a,a')$.
Applying bias decomposition in Theorem \ref{thm:lowdim_drift} to the fold-specific nuisance functions and
then using the triangle inequality gives
\begin{equation*}
\resizebox{\linewidth}{!}{$
\begin{aligned}
&|\EE\widehat\psi_{\theta,r}-\EE(W)\theta_0(a,a')|\\
&\quad\leq
\EE\Bigg[\frac{W}{N}\sum_{j=1}^N
\EE\Bigg[
|\widehat\rho_r|\,|\widehat d^{(-r)}-d_0|\,
|\eta_{0\bullet j}-\widehat\eta_{\bullet j}^{(-r)}|
+|\widehat\rho_r-\rho_0|\,
|u_{0\bullet j}^{(2)}-\widehat u_{\bullet j}^{(2),(-r)}|
\ \Big|\ A=a,C,N
\Bigg]\Bigg].
\end{aligned}
$}
\end{equation*}
Since $P(A=a\mid C,N)=\pi_a$, for any nonnegative
integrand in the preceding conditional expectation, one has
\begin{equation*}
\EE\{W\EE(\bullet\mid A=a,C,N)\}
=\frac1{\pi_a}\EE\{I(A=a)W(\bullet)\}.
\end{equation*}
Moreover, for any two square-integrable functions evaluated at the
corresponding observed arguments, one has
\begin{equation*}
\EE\{I(A=a)W|fg|\}
\leq \{\EE(Wf^2)\}^{1/2}\{\EE(Wg^2)\}^{1/2}
=\|\sqrt W f\|_{2,P}\|\sqrt W g\|_{2,P}.
\end{equation*}
Since $|\widehat\rho_r|$ is uniformly bounded and
\begin{equation*}
|\widehat\rho_r-\rho_0|
=\frac{\pi_a}{\pi_{a'}}
\left|\frac{\widehat s^{(-r)}(a'\mid M,C,N)}
{\widehat s^{(-r)}(a\mid M,C,N)}
-\frac{s_0(a'\mid M,C,N)}{s_0(a\mid M,C,N)}\right|,
\end{equation*}
the preceding two displays yield
\begin{equation*}
\resizebox{\linewidth}{!}{$
\begin{aligned}
&|\EE\widehat\psi_{\theta,r}-\EE(W)\theta_0(a,a')|\\
&\quad\leq C\Bigg[
\|\sqrt W\{\widehat d^{(-r)}-d_0\}\|_{2,P}
\|\sqrt W\{\widehat\eta^{(-r)}-\eta_0\}\|_{2,P}\\
&\qquad\qquad+
\left\|\sqrt W\left\{
\frac{\widehat s^{(-r)}(a'\mid M,C,N)}
{\widehat s^{(-r)}(a\mid M,C,N)}
-\frac{s_0(a'\mid M,C,N)}{s_0(a\mid M,C,N)}
\right\}\right\|_{2,P}
\|\sqrt W\{\widehat u^{(2),(-r)}-u_0^{(2)}\}\|_{2,P}
\Bigg].
\end{aligned}
$}
\end{equation*}
where $C$ depends only on the fixed treatment probabilities, the uniform
nuisance bound, and $N_{\max}$. 
Condition \eqref{eq:lowdim_rate_theta} makes its
right-hand side $o_p(K^{-1/2})$.  Since the limiting score is centered,
\begin{equation}
|\EE\{\widehat\psi_{\theta,r}-
\psi_{\theta,\mathrm{ld}}^\star\}|=o_p(K^{-1/2})
\label{eq:derived_drift_theta}
\end{equation}
uniformly over folds.  Equations \eqref{eq:derived_score_limit_theta} and
\eqref{eq:derived_drift_theta} verify Lemma~\ref{lem:generic_crossfit}, which
proves \eqref{eq:lowdim_AL_theta} and consistency of the empirical variance.

For $\tau$, let
\begin{equation*}
\widehat\rho_{rj}(M_{\bullet(-j)},C,N)=\frac{\pi}{1-\pi}
\frac{\widehat s_{-j}^{(-r)}(0\mid M_{\bullet(-j)},C,N)}
{\widehat s_{-j}^{(-r)}(1\mid M_{\bullet(-j)},C,N)}
\end{equation*}
and define $\rho_j^\star$ and $\rho_{0j}$ analogously.  The boundedness and
\eqref{eq:lowdim_limit_tau} imply
$\|\widehat e_{\bullet j}^{(-r)}\widehat\rho_{rj}
-e_{\bullet j}^\star\rho_j^\star\|_{2,P}=o_p(1)$ uniformly over folds and
$j\leq N_{\max}$.  The outcome-residual difference satisfies
\begin{equation*}
\begin{split}
&\widehat e_{\bullet j}^{(-r)}\widehat\rho_{rj}
\{Y_{\bullet j}-\widehat\eta_{\bullet j}^{(-r)}\}
-e_{\bullet j}^\star\rho_j^\star
\{Y_{\bullet j}-\eta_{\bullet j}^\star\}\\
&\quad=\{\widehat e_{\bullet j}^{(-r)}\widehat\rho_{rj}
-e_{\bullet j}^\star\rho_j^\star\}
\{Y_{\bullet j}-\eta_{\bullet j}^\star\}
-\widehat e_{\bullet j}^{(-r)}\widehat\rho_{rj}
\{\widehat\eta_{\bullet j}^{(-r)}-\eta_{\bullet j}^\star\}.
\end{split}
\end{equation*}
The augmentation part of the score difference is the bounded linear
combination
\begin{equation*}
\begin{split}
&\frac{A}{\pi}\{(\widehat r_{\bullet j}^{(1),(-r)}-r_{\bullet j}^{(1),\star})
-(\widehat r_{\bullet j}^{(4),(-r)}-r_{\bullet j}^{(4),\star})\}\\
&\quad+\frac{A}{\pi}\{(\widehat r_{\bullet j}^{(2),(-r)}-r_{\bullet j}^{(2),\star})
-(\widehat r_{\bullet j}^{(4),(-r)}-r_{\bullet j}^{(4),\star})\}\\
&\quad+\frac{1-A}{1-\pi}\{(\widehat r_{\bullet j}^{(3),(-r)}-r_{\bullet j}^{(3),\star})
-(\widehat r_{\bullet j}^{(4),(-r)}-r_{\bullet j}^{(4),\star})\}
+(\widehat r_{\bullet j}^{(4),(-r)}-r_{\bullet j}^{(4),\star}).
\end{split}
\end{equation*}
The same averaging and boundedness argument used for $\theta$ gives
\begin{equation*}
\resizebox{\linewidth}{!}{$
\begin{aligned}
&|\widehat\psi_{\tau,r}(O)-
\psi_{\tau,\mathrm{ld}}^\star(O)|^2\\
&\quad\leq C\frac{W^2}{N}\sum_{j=1}^N\Bigg[
(1+Y_{\bullet j}^2)
\left\{|\widehat e_{\bullet j}^{(-r)}\widehat\rho_{rj}
-e_{\bullet j}^\star\rho_j^\star|^2
+|\widehat\eta_{\bullet j}^{(-r)}-\eta_{\bullet j}^\star|^2\right\}
+\sum_{\ell=1}^4
|\widehat r_{\bullet j}^{(\ell),(-r)}-r_{\bullet j}^{(\ell),\star}|^2
\Bigg].
\end{aligned}
$}
\end{equation*}
The truncation argument and \eqref{eq:lowdim_limit_tau} therefore yield
\begin{equation}
\|\widehat\psi_{\tau,r}-
\psi_{\tau,\mathrm{ld}}^\star\|_{2,P}=o_p(1)
\label{eq:derived_score_limit_tau}
\end{equation}
uniformly over folds.  This establishes the role of all four auxiliary
regressions in \eqref{eq:lowdim_limit_tau}.

Compatibility and \eqref{eq:lowdim_MR_tau} give
$\EE\{\psi_{\tau,\mathrm{ld}}^\star(O)\}=\EE(W)\tau_0$.  The exact drift identity,
the triangle inequality, and treatment randomization now give
\begin{equation*}
\resizebox{\linewidth}{!}{$
\begin{aligned}
&|\EE\widehat\psi_{\tau,r}-\EE(W)\tau_0|\\
&\quad\leq\frac1\pi
\EE\Bigg[
\frac{AW}{N}\sum_{j=1}^N
\Bigg\{
|\widehat\rho_{rj}|\,
|\widehat e_{\bullet j}^{(-r)}-e_{0\bullet j}|\,
|\eta_{0\bullet j}-\widehat\eta_{\bullet j}^{(-r)}|
+|\widehat\rho_{rj}-\rho_{0j}|\,
|r_{0\bullet j}^{(3)}-\widehat r_{\bullet j}^{(3),(-r)}|
\Bigg\}
\Bigg].
\end{aligned}
$}
\end{equation*}
Applying the Cauchy--Schwarz inequality displayed above and using
\begin{equation*}
|\widehat\rho_{rj}-\rho_{0j}|
=\frac{\pi}{1-\pi}
\left|\frac{\widehat s_{-j}^{(-r)}(0\mid M_{\bullet(-j)},C,N)}
{\widehat s_{-j}^{(-r)}(1\mid M_{\bullet(-j)},C,N)}
-\frac{s_{0,-j}(0\mid M_{\bullet(-j)},C,N)}
{s_{0,-j}(1\mid M_{\bullet(-j)},C,N)}\right|
\end{equation*}
yield
\begin{equation*}
\resizebox{\linewidth}{!}{$
\begin{aligned}
&|\EE\widehat\psi_{\tau,r}-\EE(W)\tau_0|\\
&\quad\leq C\max_{1\leq j\leq N_{\max}}\Bigg[
\|\sqrt W\{\widehat e_{\bullet j}^{(-r)}-e_{0\bullet j}\}\|_{2,P}
\|\sqrt W\{\widehat\eta_{\bullet j}^{(-r)}-\eta_{0\bullet j}\}\|_{2,P}\\
&\qquad\qquad\qquad\qquad+
\left\|\sqrt W\left\{
\frac{\widehat s_{-j}^{(-r)}(0\mid M_{\bullet(-j)},C,N)}
{\widehat s_{-j}^{(-r)}(1\mid M_{\bullet(-j)},C,N)}
-\frac{s_{0,-j}(0\mid M_{\bullet(-j)},C,N)}
{s_{0,-j}(1\mid M_{\bullet(-j)},C,N)}
\right\}\right\|_{2,P}
\|\sqrt W\{\widehat r_{\bullet j}^{(3),(-r)}-r_{0\bullet j}^{(3)}\}\|_{2,P}
\Bigg].
\end{aligned}
$}
\end{equation*}
Condition \eqref{eq:lowdim_rate_tau} makes this $o_p(K^{-1/2})$ uniformly
over folds. 
Thus Lemma~\ref{lem:generic_crossfit} proves \eqref{eq:lowdim_AL_tau} and the
variance result.

If all nuisance limits are correct,
Lemma~\ref{lem:lowdim_canonical} identifies the limiting influence functions
with the canonical gradients.
\end{proof}

\subsection{Proof of Theorem~\ref{thm:lowdim_practical_drift}}
\label{sec:proof_lowdim_practical_drift}
\begin{proof}
Fix $r$ and condition on the training observations.
The treatment randomization and iterated expectation in
\eqref{eq:lowdim_score_theta} imply
\begin{equation}
\label{eq:practical_mean_theta}
\resizebox{\ifdim\width>\linewidth\linewidth\else\width\fi}{!}{$
\begin{aligned}
\EE\{\widehat\psi_{\theta,r}(O)\}
=\EE\Bigg[\frac{W}{N}\sum_{j=1}^N\Bigg\{&
\EE\Big[\widehat d^{(-r)}(a,Z,M,C,N)\frac{\pi_a}{\pi_{a'}}\frac{\widehat s^{(-r)}(a'\mid M,C,N)}{\widehat s^{(-r)}(a\mid M,C,N)}\{\eta_{0\bullet j}-\widehat\eta_{\bullet j}^{(-r)}\}(a,Z,M,C,N)
\mid A=a,C,N\Big]\\
&+\EE\{\widehat u_{\bullet j}^{(1),(-r)}(a,a',Z,C,N)\mid A=a,C,N\}\\
&+\EE\{\widehat u_{\bullet j}^{(2),(-r)}(a,M,C,N)\mid A=a',C,N\}
-\widehat u_{\bullet j}^{(3),(-r)}(a,a',C,N)\Bigg\}\Bigg].
\end{aligned}
$}
\end{equation}
Likewise, \eqref{eq:lowdim_score_tau} gives
\begin{equation}
\label{eq:practical_mean_tau}
\resizebox{\ifdim\width>\linewidth\linewidth\else\width\fi}{!}{$
\begin{aligned}
\EE\{\widehat\psi_{\tau,r}(O)\}
=\EE\Bigg[\frac{W}{N}\sum_{j=1}^N\Bigg\{&
\EE\Big[\widehat e_{\bullet j}^{(-r)}(Z,M,C,N)\frac{\pi}{1-\pi}\frac{\widehat s_{-j}^{(-r)}(0\mid M_{\bullet(-j)},C,N)}{\widehat s_{-j}^{(-r)}(1\mid M_{\bullet(-j)},C,N)}\{\eta_{0\bullet j}-\widehat\eta_{\bullet j}^{(-r)}\}(1,Z,M,C,N)
\mid A=1,C,N\Big]\\
&+\EE\{\widehat r_{\bullet j}^{(1),(-r)}(1,1,0,Z,C,N)\mid A=1,C,N\}\\
&+\EE\{\widehat r_{\bullet j}^{(2),(-r)}(1,0,M_{\bullet j},C,N)\mid A=1,C,N\}\\
&+\EE\{\widehat r_{\bullet j}^{(3),(-r)}(1,1,M_{\bullet(-j)},C,N)\mid A=0,C,N\}
-2\widehat r_{\bullet j}^{(4),(-r)}(1,1,0,C,N)\Bigg\}\Bigg].
\end{aligned}
$}
\end{equation}
The identities in Proposition~\ref{prop:lowdim_identities} yield
\begin{equation}
\label{eq:practical_true_odds}
\resizebox{\ifdim\width>\linewidth\linewidth\else\width\fi}{!}{$
\begin{aligned}
\kappa_0(m\mid a,C,N)
\frac{\pi_a}{\pi_{a'}}\frac{s_0(a'\mid m,C,N)}{s_0(a\mid m,C,N)}&=\kappa_0(m\mid a',C,N),\\
\kappa_{0\bullet(-j)}(m_{\bullet(-j)}\mid1,C,N)
\frac{\pi}{1-\pi}\frac{s_{0,-j}(0\mid m_{\bullet(-j)},C,N)}{s_{0,-j}(1\mid m_{\bullet(-j)},C,N)}&=\kappa_{0\bullet(-j)}(m_{\bullet(-j)}\mid0,C,N).
\end{aligned}
$}
\end{equation}
Conditioning on $(A,M,C,N)$, cancellation of $\xi_0$ in the definition
of $d_0$ gives
\begin{equation}
\label{eq:practical_true_d_eta}
\resizebox{\ifdim\width>\linewidth\linewidth\else\width\fi}{!}{$
\begin{aligned}
&\EE\big[d_0(a,Z,M,C,N)
\{\eta_{0\bullet j}-\widehat\eta_{\bullet j}^{(-r)}\}(a,Z,M,C,N)
\mid A=a,M=m,C,N\big]\\
&=u_{0\bullet j}^{(2)}(a,m,C,N)
-\int\widehat\eta_{\bullet j}^{(-r)}(a,z,m,C,N)\phi_0(z\mid a,C,N)\,dz.
\end{aligned}
$}
\end{equation}
Conditioning on $(A,M_{\bullet(-j)},C,N)$ and using
\eqref{eq:lowdim_e} similarly gives
\begin{equation}
\label{eq:practical_true_e_eta}
\resizebox{\ifdim\width>\linewidth\linewidth\else\width\fi}{!}{$
\begin{aligned}
&\EE\big[e_{0\bullet j}(Z,M,C,N)
\{\eta_{0\bullet j}-\widehat\eta_{\bullet j}^{(-r)}\}(1,Z,M,C,N)
\mid A=1,M_{\bullet(-j)}=m_{\bullet(-j)},C,N\big]\\
&=r_{0\bullet j}^{(3)}(1,1,m_{\bullet(-j)},C,N)
-\iint\widehat\eta_{\bullet j}^{(-r)}(1,z,m,C,N)
\phi_0(z\mid1,C,N)\kappa_{0\bullet j}(m_{\bullet j}\mid1,C,N)
\,dz\,dm_{\bullet j}.
\end{aligned}
$}
\end{equation}
To verify \eqref{eq:practical_drift_theta}, expand
$\widehat d^{(-r)}=(\widehat d^{(-r)}-d_0)+d_0$
in \eqref{eq:practical_mean_theta}, apply
\eqref{eq:practical_true_d_eta}, and express the expectation given
$A=a'$ by integration against $\kappa_0(m\mid a',C,N)$.
Add and subtract
$\widehat u_{\bullet j}^{(2),(-r)}(a,m,C,N)$ on the right-hand
side of \eqref{eq:practical_true_d_eta}, and then add and subtract
the integral with the true classifier fraction in
\eqref{eq:practical_theta_cancellation}.
Equation~\eqref{eq:practical_true_odds} gives
\begin{equation}
\label{eq:practical_theta_cancellation}
\resizebox{\ifdim\width>\linewidth\linewidth\else\width\fi}{!}{$
\begin{aligned}
&\int\kappa_0(m\mid a,C,N)\frac{\pi_a}{\pi_{a'}}\frac{s_0(a'\mid m,C,N)}{s_0(a\mid m,C,N)}
\{u_{0\bullet j}^{(2)}-\widehat u_{\bullet j}^{(2),(-r)}\}(a,m,C,N)\,dm
+\int\kappa_0(m\mid a',C,N)\widehat u_{\bullet j}^{(2),(-r)}(a,m,C,N)\,dm\\
&=\int\kappa_0(m\mid a',C,N)u_{0\bullet j}^{(2)}(a,m,C,N)\,dm.
\end{aligned}
$}
\end{equation}

By Theorem~\ref{thm:identification}, the expectation of the
right-hand side of \eqref{eq:practical_theta_cancellation}, after
multiplication by $W/N$ and summation over $j$, is
$\nu_0\theta_0(a,a')$.
Substitution of \eqref{eq:practical_true_d_eta} and
\eqref{eq:practical_theta_cancellation} in
\eqref{eq:practical_mean_theta} proves \eqref{eq:practical_drift_theta}.

For \eqref{eq:practical_drift_tau}, expand
$\widehat e_{\bullet j}^{(-r)}=(\widehat e_{\bullet j}^{(-r)}-e_{0\bullet j})+e_{0\bullet j}$
in \eqref{eq:practical_mean_tau}, and apply
\eqref{eq:practical_true_e_eta} and \eqref{eq:practical_true_odds}.
Split the right-hand side of \eqref{eq:practical_true_e_eta}
by adding and subtracting
$\widehat r_{\bullet j}^{(3),(-r)}(1,1,m_{\bullet(-j)},C,N)$.
Split the fitted classifier ratio multiplying
$r_{0\bullet j}^{(3)}-\widehat r_{\bullet j}^{(3),(-r)}$
into its true value and its difference from the true value.
Then
\begin{equation}
\label{eq:practical_tau_cancellation}
\resizebox{\ifdim\width>\linewidth\linewidth\else\width\fi}{!}{$
\begin{aligned}
&\int\kappa_{0\bullet(-j)}(m_{\bullet(-j)}\mid1,C,N)\frac{\pi}{1-\pi}\frac{s_{0,-j}(0\mid m_{\bullet(-j)},C,N)}{s_{0,-j}(1\mid m_{\bullet(-j)},C,N)}
\{r_{0\bullet j}^{(3)}-\widehat r_{\bullet j}^{(3),(-r)}\}(1,1,m_{\bullet(-j)},C,N)\,dm_{\bullet(-j)}\\
&\qquad+\int\kappa_{0\bullet(-j)}(m_{\bullet(-j)}\mid0,C,N)
\widehat r_{\bullet j}^{(3),(-r)}(1,1,m_{\bullet(-j)},C,N)\,dm_{\bullet(-j)}\\
&=\int\kappa_{0\bullet(-j)}(m_{\bullet(-j)}\mid0,C,N)
r_{0\bullet j}^{(3)}(1,1,m_{\bullet(-j)},C,N)\,dm_{\bullet(-j)}.
\end{aligned}
$}
\end{equation}
Theorem~\ref{thm:identification} identifies the expectation of
$(W/N)\sum_{j=1}^N$ applied to the right-hand side of
\eqref{eq:practical_tau_cancellation} with $\nu_0\tau_0$.
Substitution in \eqref{eq:practical_mean_tau} proves
\eqref{eq:practical_drift_tau}.
\end{proof}

\subsection{Proof of Theorem~\ref{thm:lowdim_practical_AN}}
\label{sec:proof_lowdim_practical_AN}
\begin{proof}
First consider $\theta(a,a')$ under \eqref{eq:lowdim_limit_theta}.
Recall that the argument establishing \eqref{eq:derived_score_limit_theta}
in Section~\ref{sec: thm:lowdim_AN proof} uses the component
convergence, the bound $B$, and the second moments in
Assumption~\ref{assm:positivity}.
In particular, the terms multiplying $Y_{\bullet j}$ are uniformly
bounded and converge in $L^2(P)$; truncating $Y_{\bullet j}^2$
as in \eqref{eq:canonical_truncation} gives convergence of their
products with $Y_{\bullet j}$ in $L^2(P)$.
The other terms in \eqref{eq:lowdim_score_theta} are bounded
products of the components in \eqref{eq:lowdim_limit_theta}.
Thus \eqref{eq:derived_score_limit_theta} holds for the fitted
functions considered here. The same reasoning applied to
\eqref{eq:lowdim_limit_tau} gives \eqref{eq:derived_score_limit_tau}.
The limiting scores are square-integrable by the same bounds.

Since $h_\theta^\star$ and $h_\tau^\star$ retain the integral
representations specified in Assumption~\ref{assm:lowdim_rates},
Theorem~\ref{thm:lowdim_drift}, with
\eqref{eq:lowdim_MR_theta} or \eqref{eq:lowdim_MR_tau}, gives
\begin{equation}
\label{eq:practical_limit_centering}
\EE\{\psi_{\theta,\mathrm{ld}}^\star(O)\}
=\nu_0\theta_0(a,a'),\qquad
\EE\{\psi_{\tau,\mathrm{ld}}^\star(O)\}=\nu_0\tau_0,
\end{equation}
under the respective conditions.
The Cauchy--Schwarz inequality and \eqref{eq:derived_score_limit_theta} or
\eqref{eq:derived_score_limit_tau} therefore give
\begin{equation}
\label{eq:practical_mean_convergence}
\max_{1\le r\le R}
\left|\EE\widehat\psi_{\theta,r}-\nu_0\theta_0(a,a')\right|=o_p(1),
\qquad
\max_{1\le r\le R}
\left|\EE\widehat\psi_{\tau,r}-\nu_0\tau_0\right|=o_p(1).
\end{equation}
Apply \eqref{eq:generic_num_decomp} separately with
$\psi^\star=\psi_{\theta,\mathrm{ld}}^\star$ and
$\psi^\star=\psi_{\tau,\mathrm{ld}}^\star$.
The conditional variance argument for \eqref{eq:generic_R1}
gives $R_{1K}=o_p(K^{-1/2})$, and
\eqref{eq:practical_mean_convergence} gives $R_{2K}=o_p(1)$.
The distribution of large numbers and $\mathbb P_K(W)\to_p\nu_0>0$
then prove the stated consistency assertions.

The Cauchy--Schwarz bounds in Section~\ref{sec: thm:lowdim_AN proof}
apply to the products in \eqref{eq:practical_drift_theta} and
\eqref{eq:practical_drift_tau}.
More precisely, the absolute value of the first expectation on the
right-hand side of \eqref{eq:practical_drift_theta}, maximized over
$r$, is bounded by $N_{\max}(1+B)/\pi_{a'}$ times the left-hand
side of \eqref{eq:lowdim_rate_theta}.
For \eqref{eq:practical_drift_tau}, the corresponding bound is
$N_{\max}(1+B)/(1-\pi)$ times the left-hand side of
\eqref{eq:lowdim_rate_tau}.
Under \eqref{eq:lowdim_limit_theta} and \eqref{eq:lowdim_MR_theta},
each product in \eqref{eq:lowdim_rate_theta} has one factor
converging to zero and another bounded in probability.
The analogous assertion follows from \eqref{eq:lowdim_limit_tau}
and \eqref{eq:lowdim_MR_tau} for \eqref{eq:lowdim_rate_tau}.
Consequently, these bounds are $o_p(1)$ under the respective
consistency conditions.
Subtracting the first expectations in
\eqref{eq:practical_drift_theta} and \eqref{eq:practical_drift_tau}
from \eqref{eq:practical_mean_convergence} proves
\eqref{eq:practical_aux_theta} and \eqref{eq:practical_aux_tau}.

If \eqref{eq:lowdim_rate_theta} holds and the left-hand side of
\eqref{eq:practical_aux_theta} is $o_p(K^{-1/2})$,
\eqref{eq:practical_drift_theta} and the same bounds imply
\begin{equation}
\label{eq:practical_drift_rate_theta}
\max_{1\le r\le R}
\left|\EE\{\widehat\psi_{\theta,r}
-\psi_{\theta,\mathrm{ld}}^\star\}\right|=o_p(K^{-1/2}).
\end{equation}
Likewise, \eqref{eq:lowdim_rate_tau} and the $o_p(K^{-1/2})$
version of \eqref{eq:practical_aux_tau} imply
\begin{equation}
\label{eq:practical_drift_rate_tau}
\max_{1\le r\le R}
\left|\EE\{\widehat\psi_{\tau,r}
-\psi_{\tau,\mathrm{ld}}^\star\}\right|=o_p(K^{-1/2}).
\end{equation}
Equations~\eqref{eq:derived_score_limit_theta},
\eqref{eq:practical_limit_centering}, and
\eqref{eq:practical_drift_rate_theta} verify
\eqref{eq:generic_score_conv}--\eqref{eq:generic_drift}
for $\widehat\theta_{\mathrm{ld}}(a,a')$.
Equations~\eqref{eq:derived_score_limit_tau},
\eqref{eq:practical_limit_centering}, and
\eqref{eq:practical_drift_rate_tau} verify them for
$\widehat\tau_{\mathrm{ld}}$.
Lemma~\ref{lem:generic_crossfit} proves
\eqref{eq:lowdim_AL_theta}, \eqref{eq:lowdim_AL_tau},
\eqref{eq:practical_normal_theta}, and \eqref{eq:practical_normal_tau}
under the respective conditions.
Slutsky's theorem proves \eqref{eq:practical_studentized}.
Lemma~\ref{lem:lowdim_canonical} identifies the variances at the
true nuisance limits with those in Theorem~\ref{thm:EIF}.

We next verify the sufficient conditions in
\eqref{eq:practical_centered_rate_theta} and
\eqref{eq:practical_centered_rate_tau}.
For fixed $r$, substitution of \eqref{eq:practical_centered_theta}
into the expectation inside the absolute value in
\eqref{eq:practical_aux_theta} gives
\begin{equation}
\label{eq:practical_centered_identity_theta}
\resizebox{\ifdim\width>\linewidth\linewidth\else\width\fi}{!}{$
\begin{aligned}
\EE\Bigg[\frac WN\sum_{j=1}^N\int&
\{\phi_0-\widehat\phi^{(-r)}\}(z\mid a,C,N)\\
&\times\Bigg[
\widehat u_{\bullet j}^{(1),(-r)}(a,a',z,C,N)
-\int\kappa_0(m\mid a,C,N)
\frac{\pi_a}{\pi_{a'}}
\frac{\widehat s^{(-r)}(a'\mid m,C,N)}
     {\widehat s^{(-r)}(a\mid m,C,N)}
\widehat\eta_{\bullet j}^{(-r)}(a,z,m,C,N)\,dm
\Bigg]dz\Bigg].
\end{aligned}
$}
\end{equation}
By \eqref{eq:practical_true_odds} and \eqref{eq:lowdim_u1}, one has
\begin{equation}\label{eq:practical_centered_bracket_theta}
\begin{split}
&\widehat u_{\bullet j}^{(1),(-r)}(a,a',z,C,N)
-\int\kappa_0(m\mid a,C,N)\frac{\pi_a}{\pi_{a'}}
\frac{\widehat s^{(-r)}(a'\mid m,C,N)}{\widehat s^{(-r)}(a\mid m,C,N)}
\widehat\eta_{\bullet j}^{(-r)}(a,z,m,C,N)\,dm\\
&=\{\widehat u_{\bullet j}^{(1),(-r)}-u_{0\bullet j}^{(1)}\}(a,a',z,C,N)\\
&\quad-\int\kappa_0(m\mid a,C,N)\frac{\pi_a}{\pi_{a'}}
\frac{\widehat s^{(-r)}(a'\mid m,C,N)}{\widehat s^{(-r)}(a\mid m,C,N)}
\{\widehat\eta_{\bullet j}^{(-r)}-\eta_{0\bullet j}\}(a,z,m,C,N)\,dm\\
&\quad-\int\kappa_0(m\mid a,C,N)\frac{\pi_a}{\pi_{a'}}
\Bigg\{\frac{\widehat s^{(-r)}(a'\mid m,C,N)}{\widehat s^{(-r)}(a\mid m,C,N)}
-\frac{s_0(a'\mid m,C,N)}{s_0(a\mid m,C,N)}\Bigg\}
\eta_{0\bullet j}(a,z,m,C,N)\,dm.
\end{split}
\end{equation}
The definition of $d_0$ in \eqref{eq:lowdim_d} yields
\begin{equation}\label{eq:practical_product_measure_theta}
\phi_0(z\mid a,c,n)\kappa_0(m\mid a,c,n)
=d_0(a,z,m,c,n)\phi_0(z\mid a,c,n)\xi_0(m\mid a,z,c,n).
\end{equation}
Write $\phi_0-\widehat\phi^{(-r)}$ as
$\phi_0(1-\widehat\phi^{(-r)}/\phi_0)$ in
\eqref{eq:practical_centered_identity_theta}.
Apply the Cauchy--Schwarz inequality to the product with
$\widehat u_{\bullet j}^{(1),(-r)}-u_{0\bullet j}^{(1)}$ under
$\phi_0(z\mid a,C,N)$.
For the integrals containing
$\widehat\eta_{\bullet j}^{(-r)}-\eta_{0\bullet j}$ and the difference
of the fractions in \eqref{eq:practical_centered_bracket_theta},
\eqref{eq:practical_product_measure_theta} changes the integration
to the distribution of $(Z,M)$ given $(A=a,C,N)$ with multiplier $d_0$.
Bound $d_0$, $\eta_0$, and the fitted classifier odds by $B$, and
use treatment randomization to bound the conditional integrals
by $1/\pi_a$ times the corresponding integrals under $P$.
Since $W$ is a function of $(C,N)$, it supplies a factor $\sqrt W$
to each norm. Summation over $j$ contributes at most $N_{\max}$.
It follows that the left-hand side of \eqref{eq:practical_aux_theta}
is bounded by $N_{\max}(1+B)^3/(\pi_a\pi_{a'})$ times the
left-hand side of \eqref{eq:practical_centered_rate_theta}.

Substitution of \eqref{eq:practical_centered_tau} into the
expectation inside the absolute value in \eqref{eq:practical_aux_tau}
gives
\begin{equation}
\label{eq:practical_centered_identity_tau}
\resizebox{\ifdim\width>\linewidth\linewidth\else\width\fi}{!}{$
\begin{aligned}
\EE\Bigg[\frac WN\sum_{j=1}^N\Bigg\{&
\int\{\phi_0-\widehat\phi^{(-r)}\}(z\mid1,C,N)\\
&\quad\times\Bigg[
\widehat r_{\bullet j}^{(1),(-r)}(1,1,0,z,C,N)
-\int\widehat\eta_{\bullet j}^{(-r)}(1,z,m,C,N)
\kappa_{0\bullet j}(m_{\bullet j}\mid1,C,N)
\kappa_{0\bullet(-j)}(m_{\bullet(-j)}\mid1,C,N)\\
&\hspace{8em}\times\frac{\pi}{1-\pi}
\frac{\widehat s_{-j}^{(-r)}(0\mid m_{\bullet(-j)},C,N)}
     {\widehat s_{-j}^{(-r)}(1\mid m_{\bullet(-j)},C,N)}\,dm
\Bigg]dz\\
&+\int\{\kappa_{0\bullet j}-\widehat\kappa_{\bullet j}^{(-r)}\}
               (m_{\bullet j}\mid1,C,N)\\
&\quad\times\Bigg[
\widehat r_{\bullet j}^{(2),(-r)}(1,0,m_{\bullet j},C,N)
-\iint\widehat\eta_{\bullet j}^{(-r)}(1,z,m,C,N)
\phi_0(z\mid1,C,N)\kappa_{0\bullet(-j)}(m_{\bullet(-j)}\mid1,C,N)\\
&\hspace{8em}\times\frac{\pi}{1-\pi}
\frac{\widehat s_{-j}^{(-r)}(0\mid m_{\bullet(-j)},C,N)}
     {\widehat s_{-j}^{(-r)}(1\mid m_{\bullet(-j)},C,N)}
\,dz\,dm_{\bullet(-j)}\Bigg]dm_{\bullet j}\\
&+\iint\widehat\eta_{\bullet j}^{(-r)}(1,z,m,C,N)
\{\widehat\phi^{(-r)}-\phi_0\}(z\mid1,C,N)
\{\widehat\kappa_{\bullet j}^{(-r)}-\kappa_{0\bullet j}\}
          (m_{\bullet j}\mid1,C,N)\\
&\qquad\times\kappa_{0\bullet(-j)}(m_{\bullet(-j)}\mid1,C,N)
\frac{\pi}{1-\pi}
\frac{\widehat s_{-j}^{(-r)}(0\mid m_{\bullet(-j)},C,N)}
     {\widehat s_{-j}^{(-r)}(1\mid m_{\bullet(-j)},C,N)}\,dz\,dm
\Bigg\}\Bigg].
\end{aligned}
$}
\end{equation}
Indeed, the contributions of $\widehat r^{(1),(-r)}$ and
$\widehat r^{(2),(-r)}$ follow directly from the equality for
$2\widehat r^{(4),(-r)}$ in \eqref{eq:practical_centered_tau}.
The equality for $\widehat r^{(3),(-r)}$ and the identity
\begin{equation}
\label{eq:practical_density_product_expansion}
\begin{aligned}
&\widehat\phi^{(-r)}(z\mid1,c,n)
 \widehat\kappa_{\bullet j}^{(-r)}(m_{\bullet j}\mid1,c,n)
-\phi_0(z\mid1,c,n)\kappa_{0\bullet j}(m_{\bullet j}\mid1,c,n)\\
&=\{\widehat\phi^{(-r)}-\phi_0\}(z\mid1,c,n)
       \kappa_{0\bullet j}(m_{\bullet j}\mid1,c,n)\\
&\quad+\phi_0(z\mid1,c,n)
       \{\widehat\kappa_{\bullet j}^{(-r)}-\kappa_{0\bullet j}\}
            (m_{\bullet j}\mid1,c,n)\\
&\quad+\{\widehat\phi^{(-r)}-\phi_0\}(z\mid1,c,n)
       \{\widehat\kappa_{\bullet j}^{(-r)}-\kappa_{0\bullet j}\}
            (m_{\bullet j}\mid1,c,n)
\end{aligned}
\end{equation}
give the remaining integrals in
\eqref{eq:practical_centered_identity_tau}.

By \eqref{eq:practical_true_odds} and
\eqref{eq:lowdim_r1}--\eqref{eq:lowdim_r2}, replacing
$\widehat\eta^{(-r)}$ and $\widehat s_{-j}^{(-r)}$ by
$\eta_0$ and $s_{0,-j}$ in the integrals inside the two brackets
in \eqref{eq:practical_centered_identity_tau} gives
$r_{0\bullet j}^{(1)}(1,1,0,z,C,N)$ and
$r_{0\bullet j}^{(2)}(1,0,m_{\bullet j},C,N)$, respectively.
Adding and subtracting these integrals, and then adding and
subtracting the integrals with $\eta_0$ and
$\widehat s_{-j}^{(-r)}$, expresses the brackets as sums involving
$\widehat r^{(1),(-r)}-r_0^{(1)}$ or
$\widehat r^{(2),(-r)}-r_0^{(2)}$,
$\widehat\eta^{(-r)}-\eta_0$, and
\[
\frac{\widehat s_{-j}^{(-r)}(0\mid m_{\bullet(-j)},C,N)}
     {\widehat s_{-j}^{(-r)}(1\mid m_{\bullet(-j)},C,N)}
-\frac{s_{0,-j}(0\mid m_{\bullet(-j)},C,N)}
      {s_{0,-j}(1\mid m_{\bullet(-j)},C,N)}.
\]
The support condition in Assumption~\ref{assm:positivity} for every
$(a,a')$ gives mutual absolute continuity of
$\kappa_0(\bullet\mid0,c,n)$ and $\kappa_0(\bullet\mid1,c,n)$,
and hence of their $M_{\bullet(-j)}$ marginals.
Its condition for $\tau$ therefore gives domination of
$\phi_0(\bullet\mid1,c,n)\kappa_{0\bullet j}(\bullet\mid1,c,n)
\kappa_{0\bullet(-j)}(\bullet\mid1,c,n)$ by the distribution of
$(Z,M)$ given $(A=1,C=c,N=n)$.
Consequently, the definition of $e_{0\bullet j}$ in \eqref{eq:lowdim_e}
gives the almost-everywhere identity
\begin{equation}
\label{eq:practical_product_measure_tau}
\begin{aligned}
&\phi_0(z\mid1,c,n)\kappa_{0\bullet j}(m_{\bullet j}\mid1,c,n)
 \kappa_{0\bullet(-j)}(m_{\bullet(-j)}\mid1,c,n)\\
&\qquad=e_{0\bullet j}(z,m,c,n)
 \phi_0(z\mid1,c,n)\xi_0(m\mid1,z,c,n).
\end{aligned}
\end{equation}
Use \eqref{eq:practical_product_measure_tau} and
$\EE\{W\EE(\bullet\mid A=1,C,N)\}=\EE\{AW(\bullet)\}/\pi$
to apply the Cauchy--Schwarz inequality under $P$.
The products with $\widehat r^{(1),(-r)}-r_0^{(1)}$ and
$\widehat r^{(2),(-r)}-r_0^{(2)}$ give the first two products
in \eqref{eq:practical_centered_rate_tau}.
The products with $\widehat\eta^{(-r)}-\eta_0$ and the difference
of fractions give its product of the two sums of norms.
The last integral in \eqref{eq:practical_centered_identity_tau}
gives its last product, using the bounds on
$\widehat\eta^{(-r)}$, $e_{0\bullet j}$, and the fitted classifier odds.
Thus the left-hand side of \eqref{eq:practical_aux_tau}
is bounded by $N_{\max}(1+B)^3/\{\pi(1-\pi)\}$ times
the left-hand side of \eqref{eq:practical_centered_rate_tau}.
These inequalities prove the stated sufficient conditions at
both $o_p(1)$ and $o_p(K^{-1/2})$ rates.
\end{proof}

\section{Nuisance and auxiliary function estimation}
\label{sec:nuisance-estimation}

Recall that the canonical nuisance parameters are $(\eta,\phi,\kappa,\xi)$. The ratios $\omega^{(1)}$ and $\omega_{\bullet j}^{(2)}$ and the integrated outcome regressions $u_{\bullet j}^{(1)}$--$u_{\bullet j}^{(3)}$ and $r_{\bullet j}^{(1)}$--$r_{\bullet j}^{(4)}$ are auxiliary functions obtained from these parameters, as defined in Section~\ref{sec:eif}.
The conditional mean $\eta_{\bullet j}$ can be estimated using a regression method appropriate for $Y_{\bullet j}$, for example, a generalized linear model or a flexible regression method such as Super Learner, with predictors $(A,Z,M,C,N)$.
Alternatively, within each treatment group $A=a$, regress
$Y_{\bullet j}$ on $(Z,M,C,N)$ to estimate
$\eta_{\bullet j}(a,z,m,c,n)$ separately for $a=0,1$.

To estimate $\kappa$, first estimate the conditional distribution
of each $M_{\bullet j}$ given $(A,C,N)$.
For example, following \citet{cheng2026semiparametric}, specify
normal conditional distributions, estimate their means by
regression, and estimate their variances from the training residuals.
Specify a Gaussian copula for the conditional dependence among
$M_{\bullet1},\ldots,M_{\bullet N}$.
Estimate its correlation parameters by maximizing the copula
likelihood with the fitted marginal distribution functions held
fixed, as described in their Appendix A.1.
Evaluate $\kappa$ by multiplying the fitted
$\kappa_{\bullet j}$ over $j=1,\ldots,N$ and multiplying this product
by the fitted copula density evaluated at the marginal
distribution functions.

To estimate $\xi$, fit the conditional distributions of
$M_{\bullet j}$ given $(A,Z,C,N)$ and a Gaussian copula for their
conditional dependence motivated by \citet{cheng2026semiparametric}.
Similarly, to estimate $\phi$, fit the conditional distributions of the
coordinates of $Z$ given $(A,C,N)$ and a Gaussian copula for
their conditional dependence.
For each of $\xi$ and $\phi$, estimate the marginal means,
residual variances, and correlation parameters for the copula separately,
and calculate the joint density by the same multiplication
of marginal densities and the specified copula density.

We obtain $\kappa_{\bullet j}$ and $\kappa_{\bullet(-j)}$ used for
$\tau$ by integrating the fitted $\kappa$ over
$m_{\bullet(-j)}$ and $m_{\bullet j}$, respectively.
We compute the auxiliary functions by substituting the
fitted $\eta_{\bullet j}$, $\phi$, $\kappa$, $\xi$,
$\kappa_{\bullet j}$, and $\kappa_{\bullet(-j)}$
into their defining ratios and integrals in Section~\ref{sec:eif}.

For the low-dimensional estimator, we estimate $s$ by regressing $A$ on $(M,C,N)$ and $s_{-j}$ by regressing $A$ on $(M_{\bullet(-j)},C,N)$. 
We estimate $d$ using the augmented method proposed by \citet{rudolph2024practical}, with the two classifiers and their predictors given in Section~\ref{sec:augmented_classification}. The densities $\kappa_{\bullet j}$ and $\kappa_{\bullet j}^{\star}$ have the scalar response $M_{\bullet j}$, conditioned on $(A,C,N)$ and $(M_{\bullet(-j)},A,C,N)$, respectively; they can be fitted through conditional mean and residual-distribution models or obtained from the same fitted $\kappa$. Substitution into \eqref{eq:lowdim_e} gives $e_{\bullet j}$. The conditional expectations in \eqref{eq:lowdim_u1}--\eqref{eq:lowdim_u3} and Proposition~\ref{prop:lowdim_identities} specify the responses and predictors for the auxiliary regressions.
These regressions estimate $u^{(1)},u^{(2)},u^{(3)}$ and
$r^{(1)},r^{(2)},r^{(3)},r^{(4)}$ from the same fitted $\eta$.
Theorem~\ref{thm:lowdim_practical_AN} gives sufficient conditions on
their fitted values through \eqref{eq:lowdim_rate_theta},
\eqref{eq:practical_aux_theta}, \eqref{eq:lowdim_rate_tau}, and
\eqref{eq:practical_aux_tau}.
For example, Table~\ref{tab:sim-nuisance-fitting} records the particular fits and covariate restrictions used in our simulations.

\section{Additional numerical details}
\label{sec:additional-numerical-details}

\subsection{Data-generating model}
\label{sec:sim-data-generating-model}

Denote
\(q_i=(N_i-15)/5\) and
\(\bar X_i=N_i^{-1}\sum_{j=1}^{N_i}X_{ij}\).  
The baseline variables and
treatment were generated as
\[
\begin{aligned}
N_i&\sim\operatorname{Unif}\{10,\ldots,20\},
&V_i\mid N_i&\sim N(0.20q_i,1),\\
X_{ij}&=0.55V_i+0.15q_i+\varepsilon^X_{ij},
&A_i&\sim\operatorname{Bernoulli}(0.5),
\end{aligned}
\]
where the \(\varepsilon^X_{ij}\)'s were independent standard normal errors.
For an \(N\)-vector, let
\(\Sigma_N(\sigma,\rho)=\sigma^2\{(1-\rho)I_N+\rho J_N\}\),
where \(J_N\) is the matrix of ones.  The post-treatment confounder was
generated from
\[
\begin{split}
\mu_Z(a,C_i)
&=-0.45+0.35q_i+0.65V_i+0.55\bar X_i
  +a(0.152+0.072q_i),\\
\boldsymbol Z_i
&=\mu_Z(A_i,C_i)\boldsymbol 1_{N_i}+\boldsymbol E_i^Z,
\qquad
\boldsymbol E_i^Z\sim N\{\boldsymbol0,\Sigma_{N_i}(0.85,0.06)\}.
\end{split}
\]
The marginal mediator mean was
\(\mu_M(a,C_i)=-0.30-0.30q_i+0.65V_i-0.55\bar X_i+0.192a\),
and the mediator was generated through the compatible conditional
representation
\[
\begin{aligned}
\boldsymbol M_i
&=\mu_M(A_i,C_i)\boldsymbol 1_{N_i}
+0.10\{\boldsymbol Z_i-\mu_Z(A_i,C_i)\boldsymbol 1_{N_i}\}
+\boldsymbol E_i^M,\\
\boldsymbol E_i^M
&\sim N\{\boldsymbol0,\Sigma_{N_i}(0.90,0.06)\},
\end{aligned}
\]
with \(\boldsymbol E_i^M\) independent of \(\boldsymbol E_i^Z\).
Consequently, both \(M_i\mid A_i,C_i\) and
\(M_i\mid Z_i,A_i,C_i\) are exactly exchangeable Gaussian.

Let \(\bar Z_{i,-j}\) and \(\bar M_{i,-j}\) denote leave-one-out peer
means.  Conditional on all observed arguments, the outcome mean under arm
\(a\) was
\[
\begin{aligned}
\mu_{Y,ij}(a)
={}&0.25+0.70a+0.18q_i+0.22V_i+0.15X_{ij}+0.10\bar X_i
 +0.12aV_i-0.10a\bar X_i\\
&+1.80\big[ (0.50+0.12V_i+0.12a)Z_{ij}
 +(0.30+0.10a)\bar Z_{i,-j}\\
&\hspace{0.54in} +(1.10+0.14V_i+0.20a)M_{ij}
 +(0.40+0.12a)\bar M_{i,-j}\\
&\hspace{0.54in} +0.06Z_{ij}^2+0.10M_{ij}^2
 +0.25M_{ij}\bar M_{i,-j}+(0.70-0.15a)Z_{ij}M_{ij}\big].
\end{aligned}
\]
We set
\(\boldsymbol Y_i=\boldsymbol\mu_Y(A_i)+\boldsymbol E_i^Y\), where
\(\boldsymbol E_i^Y\sim N\{\boldsymbol0,\Sigma_{N_i}(1.20,0.08)\}\)
and was independent of the preceding errors. 
We have the cluster-average \((W_i=1)\) true values
\(0.468018\), \(0.330551\), and \(0.192446\) for IDE, IIME, and ISME,
respectively.  The individual-average results and true values are reported in
Section~\ref{sec:sim-individual-results}.

\subsection{Stabilization}
\label{sec:sim-stabilization}

Recall that $s$ and $s_{-j}$ are the treatment probabilities defined in Section~\ref{sec:lowdim_reparam}, whereas $\ell_a$ and $\ell_a^M$ are the probabilities of a reference copy in the augmented sample defined in Section~\ref{sec:augmented_classification}. 
Throughout the study, fitted values of $s$, $s_{-j}$, $\ell_a$,
and $\ell_a^M$ were truncated to $[0.025,0.975]$.
For the canonical estimators, fitted values of $\omega^{(1)}$
and $\omega_{\bullet j}^{(2)}$ in Theorem~\ref{thm:EIF}
were truncated to $[0.05,20]$.
For the low-dimensional estimators, fitted values of $d$,
the ratio $\kappa_{\bullet j}/\kappa_{\bullet j}^{\star}$
in \eqref{eq:lowdim_e}, and their product $e_{\bullet j}$
were truncated to $[0.05,20]$.
The fitted ratios
$s(a'\mid m,c,n)/s(a\mid m,c,n)$ and
$s_{-j}(0\mid m_{\bullet(-j)},c,n)/
s_{-j}(1\mid m_{\bullet(-j)},c,n)$
were also truncated to $[0.05,20]$.
After combining these quantities according to
\eqref{eq:lowdim_omega1} and \eqref{eq:lowdim_omega2},
the resulting estimates of $\omega^{(1)}$ and
$\omega_{\bullet j}^{(2)}$ were truncated to the same interval.

We adapted the outcome-regression adjustment in Appendix A.4 of \citet{cheng2026semiparametric}. Within each validation fold, separately for each $\theta(a,a')$, $\tau$, and choice of $W$, we added a constant to the fitted outcome regression. This constant was the weighted mean of the outcome residuals, using the coefficient of the outcome residual in Theorem~\ref{thm:EIF} as its weight, including $W_i/N_i$, the treatment indicator, and the inverse randomization probability. The weighted residuals after adjustment therefore summed to zero. We added the same constant to the associated auxiliary regressions: their differences in the EIF were unchanged, while the final $u_{\bullet j}^{(3)}$ or $r_{\bullet j}^{(4)}$ increased by that constant. All other nuisance fits remained fixed at their training-fold values. In our setting, the weights also account for $Z$, and the adjustment is made separately for $W_i=1$ and $W_i=N_i$. The low-dimensional estimator uses the same adjustment with the weights represented as in Proposition~\ref{prop:lowdim_identities}.

Recall the estimated influence functions $\widehat D_{\theta,\mathrm{can},i}$ and $\widehat D_{\tau,\mathrm{can},i}$ defined after Theorem~\ref{thm:CF-AN-MR}, and their low-dimensional counterparts defined after Theorem~\ref{thm:lowdim_AN}. For each method, we evaluated these influence functions using the adjusted fits. For IDE, we subtracted the influence function for $\theta(0,0)$ from that for $\theta(1,0)$ within each cluster. For IIME, we subtracted that for $\theta(1,0)$ from that for $\tau$; for ISME, we subtracted that for $\tau$ from that for $\theta(1,1)$. Taking these differences before squaring retains the covariance between the estimates. The squared standard error was $K^{-1}$ times $\mathbb P_K$ applied to the squared difference. Each reported 95\% interval was centered at the corresponding estimated effect, with half-width equal to its standard error multiplied by $\sqrt{K/(K-3)}\,t_{K-3,0.975}$. The G-computation estimator used the 2.5th and 97.5th percentiles of 200 cluster-bootstrap estimates.

\subsection{Nuisance fitting and theoretical guarantees}
\label{sec:sim-additional-scenarios}

Table~\ref{tab:sim-nuisance-fitting} reports the exact working models used in the final simulations.

For the low-dimensional estimators, we computed
$\widehat u^{(1)}$--$\widehat u^{(3)}$ and
$\widehat r^{(1)}$--$\widehat r^{(4)}$ using exact Gaussian moments
of the scenario-specific parametric outcome regression under the
training-fold fits of $\phi$ and $\kappa$.
We used the full working models for $\phi$ and $\xi$ in every
low-dimensional scenario and obtained $\widehat\kappa$ by
integrating $\widehat\xi$ against $\widehat\phi$.
The applicable convergence conditions for these estimated auxiliary
functions are given in Theorem~\ref{thm:lowdim_practical_AN}.

{\scriptsize
\setlength{\tabcolsep}{3pt}
\begin{longtable}{@{}>{\raggedright\arraybackslash}p{0.18\textwidth}>{\raggedright\arraybackslash}p{0.38\textwidth}>{\raggedright\arraybackslash}p{0.38\textwidth}@{}}
\caption{Nuisance estimation and misspecification scenarios in the final simulations.}
\label{tab:sim-nuisance-fitting}\\
\toprule
Block & Correct construction & Misspecification \\
\midrule
\endfirsthead
\toprule
Block & Correct construction & Misspecification \\
\midrule
\endhead
Canonical \(\eta\) & Arm-specific Gaussian regression on
\((q,V,X,\bar X,Z,\bar Z_{-j},M,\bar M_{-j})\), with
\((VZ,VM,Z^2,M^2,M\bar M_{-j},ZM)\).  Parametric fitting used least squares; the implementation
labeled flexible used Super Learner with
\texttt{SL.library} set to \texttt{c("SL.glm","SL.ranger")}. & Keep
\((q,Z,\bar Z_{-j},M,\bar M_{-j})\), with
\((Z^2,M^2,M\bar M_{-j},ZM)\), omitting
\((V,X,\bar X)\) and every baseline interaction. \\
Canonical \(\phi\) & Exchangeable Gaussian \(Z\) process with mean regressors
\((A,q,V,\bar X,Aq)\); fit the mean by least squares or the same Super Learner
library and estimate residual variance and correlation in each training fold.
& Use only \((A,q)\) in the mean and re-estimate the
exchangeable residual covariance. \\
Canonical \(\xi\) & Conditional exchangeable Gaussian mediator distribution with mean
regressors \((Z,A,q,V,\bar X,Aq)\), fitted parametrically or by the same Super
Learner library. & Use \((Z,A,q)\), including the
defining conditioning variable \(Z\), and re-estimate the residual covariance. \\
Canonical \(\kappa\) & For the parametric EIF, derive the marginal mediator
mean, variance, and exchangeable correlation from the full-model fits of \(\phi\) and \(\xi\).  For the flexible EIF,
fit the marginal mean directly with the same Super Learner library. & Fit a reduced marginal exchangeable
Gaussian mediator distribution with mean regressors \((A,q)\). \\
Canonical partial means & Exact Gaussian moments for the parametric estimator. & Integrate the corresponding reduced
outcome or process distribution. \\
Parametric g-computation & Use parametric \(\eta,\phi,\kappa\), exact Gaussian
moment integration, and a 200-resample cluster percentile bootstrap. & Use the same reduced models indexed by the
scenario. \\
Reference distribution \(\widetilde\lambda\) & Fit the arm-specific mediator reference
mean on raw \((N,V,\bar X)\).  Generate ten independent reference vectors per
training cluster using either fitted Gaussian residuals or centered empirical
residual resampling. & Both Gaussian and residual-bootstrap references were considered under each nuisance scenario. The Gaussian reference satisfies the support requirement in Section~\ref{sec:augmented_classification}, while the unsmoothed RB reference was included as an empirical comparison. \\
Low-dimensional \(d\) & Use the augmented method proposed by
\citet{rudolph2024practical}.  Full and mediator-only classifiers contain the
exact linear, quadratic, sum, and exchangeable-Gaussian sufficient statistics;
fit grouped glmnet with \texttt{lambda.min}, a 300-tree RF, or
Super Learner with \texttt{c("SL.glm","SL.ranger")} (the latter's forest has
500 trees). & Refit both augmented
classifiers after retaining \(N\) and the defining \(Z\)- and \(M\)-summaries
but deleting \((V,\bar X)\) and all features involving either baseline. \\
Low-dimensional \(s,s_{-j}\) & Fit the exact exchangeable-Gaussian
discriminant basis using Super Learner with
\texttt{c("SL.glm","SL.ranger")} and cluster-grouped internal folds. & Use the appropriate full- or
leave-one-out mediator summary and \(N\), omit \((V,\bar X)\), and refit the
same ensemble. \\
Low-dimensional \(e_{\bullet j}\) & Evaluate
\(\kappa_j/\kappa_j^\star\) analytically from the compatible fitted
\(\kappa\) and set \(e_{\bullet j}=d\kappa_j/\kappa_j^\star\). & Use the
restricted-classifier estimate of \(d\) with the same fitted scalar
\(\kappa_j/\kappa_j^\star\). \\
\bottomrule
\end{longtable}
}

Tables~\ref{tab:canonical-scenarios} and \ref{tab:lowdim-scenarios}
distinguish two different claims: consistency and asymptotic linearity.  P denotes point consistency of
the effect estimator; AL denotes asymptotic linearity with the influence
function used for variance estimation.
Under a misspecification scenario, a multiple-robust P branch can hold
while the product of one fixed nuisance error and one root-\(K\) estimation error remains first order.  We therefore assert AL for the cross-fitted EIF estimators only under all-correct nuisance estimation.  A zero
entry means that the corresponding theorem does not provide statistical guarantee for the label ``1."

\begin{center}
\begin{minipage}{\textwidth}
\centering
\captionof{table}{Canonical nuisance scenarios and separate theoretical guarantees.}
\label{tab:canonical-scenarios}
\scriptsize
\resizebox{\linewidth}{!}{%
\begin{tabular}{@{}lccccccccl@{}}
\toprule
Scenario & \(\eta\) & \(\phi\) & \(\kappa\) & \(\xi\) &
EIF IDE-P & EIF IIME-P & EIF ISME-P & EIF AL & g-comp P/AL \\
\midrule
C-1111 &1&1&1&1&1&1&1&1&1/1\\
C-1110 &1&1&1&0&1&1&1&0&1/1\\
C-1101 &1&1&0&1&1&0&0&0&0/0\\
C-1100 &1&1&0&0&1&0&0&0&0/0\\
C-1011 &1&0&1&1&1&1&1&0&0/0\\
C-1010 &1&0&1&0&1&1&1&0&0/0\\
C-1001 &1&0&0&1&0&0&0&0&0/0\\
C-1000 &1&0&0&0&0&0&0&0&0/0\\
C-0111 &0&1&1&1&1&1&1&0&0/0\\
C-0110 &0&1&1&0&0&0&0&0&0/0\\
C-0101 &0&1&0&1&0&0&0&0&0/0\\
C-0100 &0&1&0&0&0&0&0&0&0/0\\
C-0011 &0&0&1&1&1&1&1&0&0/0\\
C-0010 &0&0&1&0&0&0&0&0&0/0\\
C-0001 &0&0&0&1&0&0&0&0&0/0\\
C-0000 &0&0&0&0&0&0&0&0&0/0\\
\bottomrule
\end{tabular}}
\par\smallskip
\parbox{0.96\textwidth}{\footnotesize}
\end{minipage}
\end{center}

\begin{center}
\begin{minipage}{\textwidth}
\centering
\captionof{table}{Compatibility-preserving low-dimensional scenarios and separate theoretical guarantees.}
\label{tab:lowdim-scenarios}
\scriptsize
\resizebox{\linewidth}{!}{%
\begin{tabular}{@{}lccccccccccl@{}}
\toprule
Scenario & \(\eta_\theta\) & \(d\) & \(s\) &
\(\eta_\tau\) & \(e\) & \(s_{-j}\) & IDE-P & IIME-P & ISME-P & AL & Active point-consistency route \\
\midrule
LD-AC &1&1&1&1&1&1&1&1&1&1&all nuisances correct\\
LD-T2 &1&0&0&1&1&1&1&1&1&0&\(\eta_\theta\); all \(\tau\) nuisances\\
LD-T3 &0&1&1&1&1&1&1&1&1&0&\((d,s)\); all \(\tau\) nuisances\\
LD-T4 &0&1&0&1&1&1&0&0&1&0&same-arm \(\theta\) through \(d\)\\
LD-T5 &0&0&1&1&1&1&0&0&0&0&no complete \(\theta\) route\\
LD-T6 &0&0&0&1&1&1&0&0&0&0&no complete \(\theta\) route\\
LD-U2 &1&1&1&1&0&0&1&1&1&0&\(\eta_\tau\); all \(\theta\) nuisances\\
LD-U3 &1&1&1&0&1&1&1&1&1&0&\((e,s_{-j})\); all \(\theta\) nuisances\\
LD-U4 &1&1&1&0&1&0&1&0&0&0&IDE only; incomplete \(\tau\) route\\
LD-U5 &1&1&1&0&0&1&1&0&0&0&IDE only; incomplete \(\tau\) route\\
LD-U6 &1&1&1&0&0&0&1&0&0&0&IDE only; no \(\tau\) route\\
\bottomrule
\end{tabular}}
\par\smallskip
\parbox{0.96\textwidth}{\footnotesize}
\end{minipage}
\end{center}

\subsection{Additional simulation results}
\label{sec:sim-additional-results}

\subsubsection{Additional cluster-average results}
\label{sec:sim-additional-cluster-results}

Figures~\ref{fig:sim-app-can-IDE-C}--\ref{fig:sim-app-ld-ISME-C}
report bias and coverage for all cluster-average scenarios and methods,
including the IIME results omitted from Section \ref{sec:simulation}.

The additional canonical scenarios show that simultaneous misspecification of two nuisances need not introduce persistent bias when a sufficient
nuisance combination remains correct
(Figures~\ref{fig:sim-app-can-IDE-C}--\ref{fig:sim-app-can-ISME-C}).
Under C-0011, both EIF implementations had small biases at the largest $K$
for all three effects, consistent with correct $(\kappa,\xi)$.
C-1010 showed a similar pattern through correct $(\eta,\kappa)$,
although its IDE intervals were more conservative than under C-1011.
The g-computation showed appreciable IDE bias in both scenarios.
The scenario C-1100, like C-1101 in Section~\ref{sec:sim-canonical}, showed small EIF bias for IDE
but persistent biases for IIME and ISME, reflecting the different point-consistency requirements for the direct and mediation effects.

The IIME results in Figure~\ref{fig:sim-app-can-IIME-C} generally
followed the point-estimation patterns seen for ISME, but their coverage
patterns were not identical. Under C-1100 and C-1101, both EIF
implementations underestimated IIME while overestimating ISME.
Nevertheless, IIME coverage remained closer to nominal than ISME
coverage at larger $K$, despite the lingering IIME bias.
Under C-0101, C-0100, C-0001, and C-0000, the mediation-effect biases were more pronounced than under C-1100 and C-1101, and IIME coverage was poorer. 

Other scenarios without a point-consistency guarantee further separate
small observed bias from robustness.
Under C-0110 and C-0010, the EIF estimators had relatively small IDE
biases at larger $K$, but their IIME and ISME biases persisted;
IDE intervals were conservative in both scenarios.
Under C-1001 and C-1000, appreciable biases remained for all three
effects, with little separation between the two EIF implementations
at the largest $K$. In these two scenarios, the g-computation estimator had smaller absolute IIME bias than either EIF implementation, illustrating that
the EIF estimators did not uniformly improve point estimation outside their sufficient nuisance conditions.
As in the results shown in Section~\ref{sec:sim-canonical}, increasing $K$ did not generally remove
bias caused by misspecification.

Figures~\ref{fig:sim-app-ld-IDE-C}--\ref{fig:sim-app-ld-ISME-C}
extend the low-dimensional comparisons to the LD-U scenarios.
IDE results were unchanged across LD-AC and LD-U2--LD-U6 for each
estimator since the nuisances entering IDE were unchanged.
Under LD-U2 and LD-U3, IIME and ISME estimates remained close to their
targets, consistent with their point-consistency guarantees.
Their coverage nevertheless differed: LD-U2 tended to be more
conservative, especially for the SL implementations, whereas LD-U3 coverage was closer to nominal at larger $K$.
Under LD-U4 and LD-U6, negative IIME biases and positive ISME biases continued as $K$ increased for most methods, even though coverage
was often near or above nominal.
LD-U5 generally had smaller mediation-effect biases than LD-U4,
but, like LD-U4 and LD-U6, does not possess point-consistency guarantee for these effects.

The complete six-estimator comparison shows more variation in coverage
than in bias. Under LD-T2, LD-T5, and LD-T6, the SL implementations
were generally more conservative than {\tt glmnet} for all three effects;
RF also became conservative in several of these scenarios as $K$
increased. These patterns extend the SL results described in Section~\ref{sec:sim-lowdim}. The two reference distributions gave similar results
for a given classifier under LD-AC and many misspecification scenarios.
However, this similarity was less uniform for the mediation effects under misspecification. At the largest $K$, the SL implementations had
smaller absolute IIME and ISME biases under LD-U6 than {\tt glmnet} and RF,
whereas they had larger absolute biases under LD-T5.
The G and RB versions of SL also separated more visibly for the mediation effects under LD-U5. 
Thus, no classifier or reference distribution was uniformly preferable across the misspecified scenarios.

Under all-correct nuisance estimation, the canonical and low-dimensional
EIF estimators had similar IDE and IIME RMSE, whereas the canonical
estimators had smaller RMSE for the ISME estimand.
At the largest $K$, low-dimensional confidence intervals were wider
than the parametric canonical EIF intervals, with the largest
relative difference for ISME. 
Within the low-dimensional methods,
the SL implementations generally had the widest IDE and ISME
intervals. 
The low-dimensional ISME coverage was closer to nominal
at larger $K$, while its IDE coverage was somewhat more conservative.

\clearpage
\begin{landscape}
\noindent\begin{minipage}{\linewidth}
\centering
\includegraphics[width=0.96\linewidth,height=0.85\textheight,keepaspectratio]{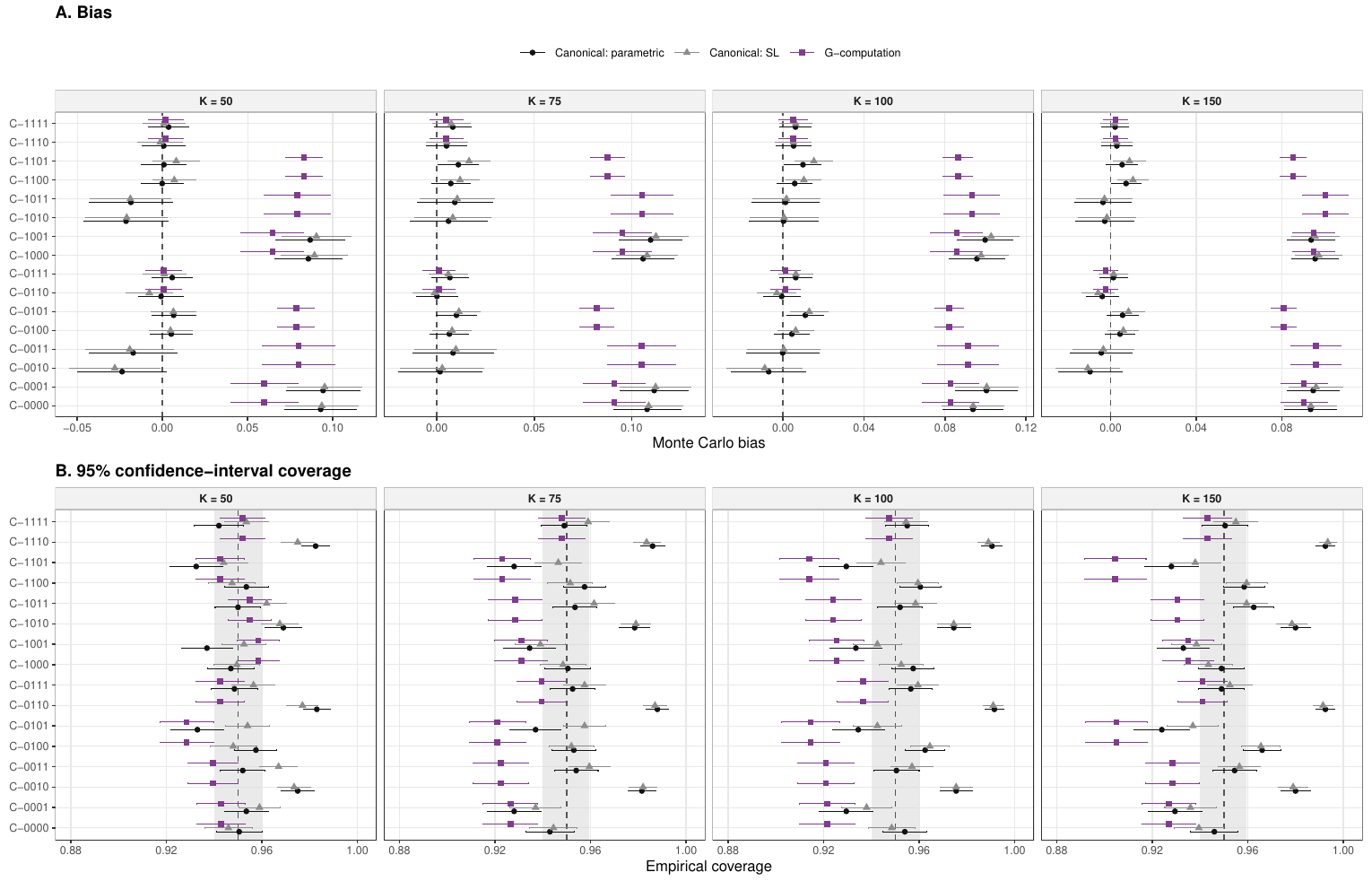}
\captionof{figure}{Complete canonical-study cluster-average IDE results for
all 16 nuisance scenarios, three estimators, and four values of \(K\).}
\label{fig:sim-app-can-IDE-C}
\end{minipage}
\end{landscape}

\clearpage
\begin{landscape}
\noindent\begin{minipage}{\linewidth}
\centering
\includegraphics[width=0.96\linewidth,height=0.85\textheight,keepaspectratio]{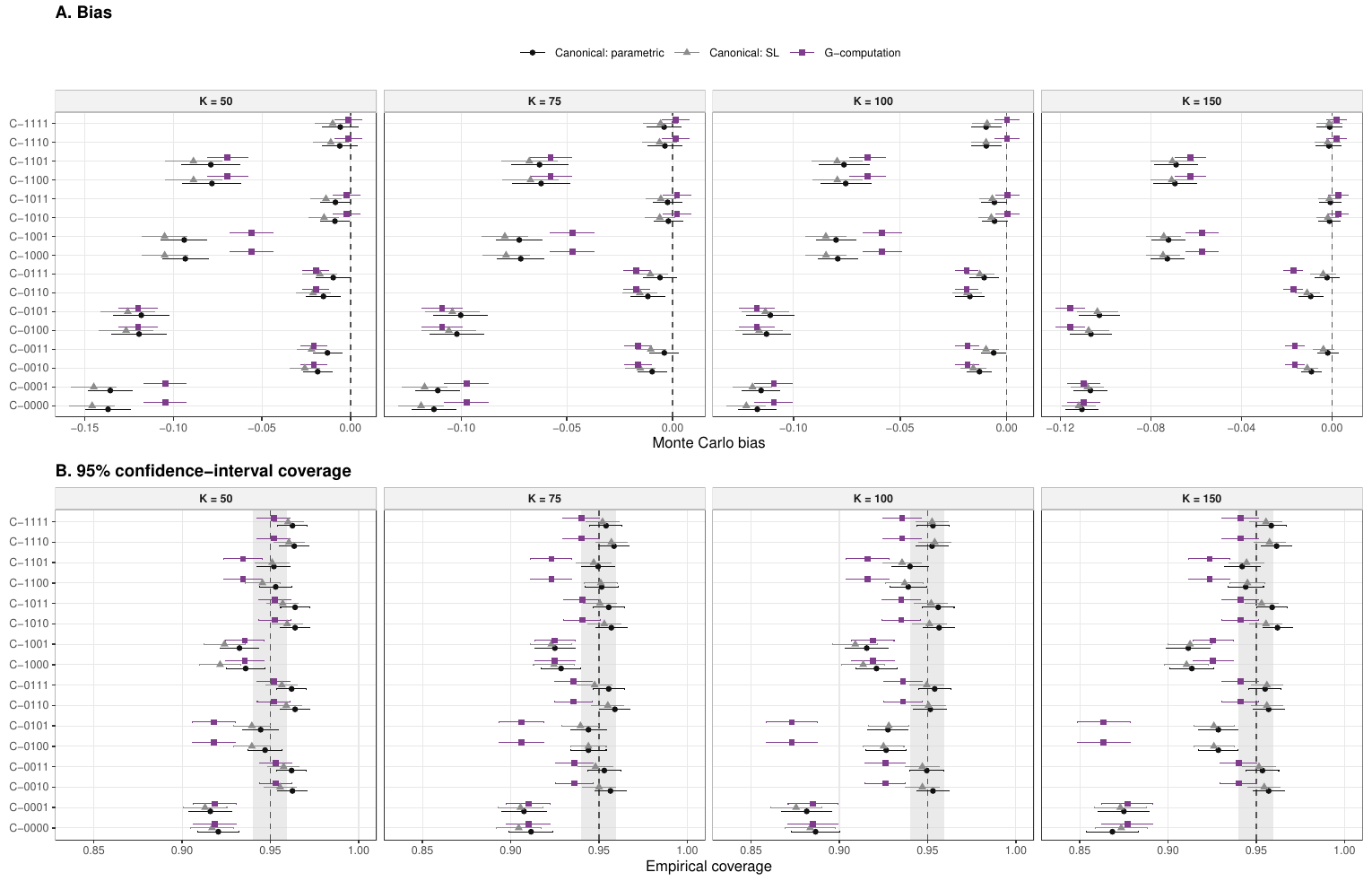}
\captionof{figure}{Complete canonical-study cluster-average IIME results.}
\label{fig:sim-app-can-IIME-C}
\end{minipage}
\end{landscape}

\clearpage
\begin{landscape}
\noindent\begin{minipage}{\linewidth}
\centering
\includegraphics[width=0.96\linewidth,height=0.85\textheight,keepaspectratio]{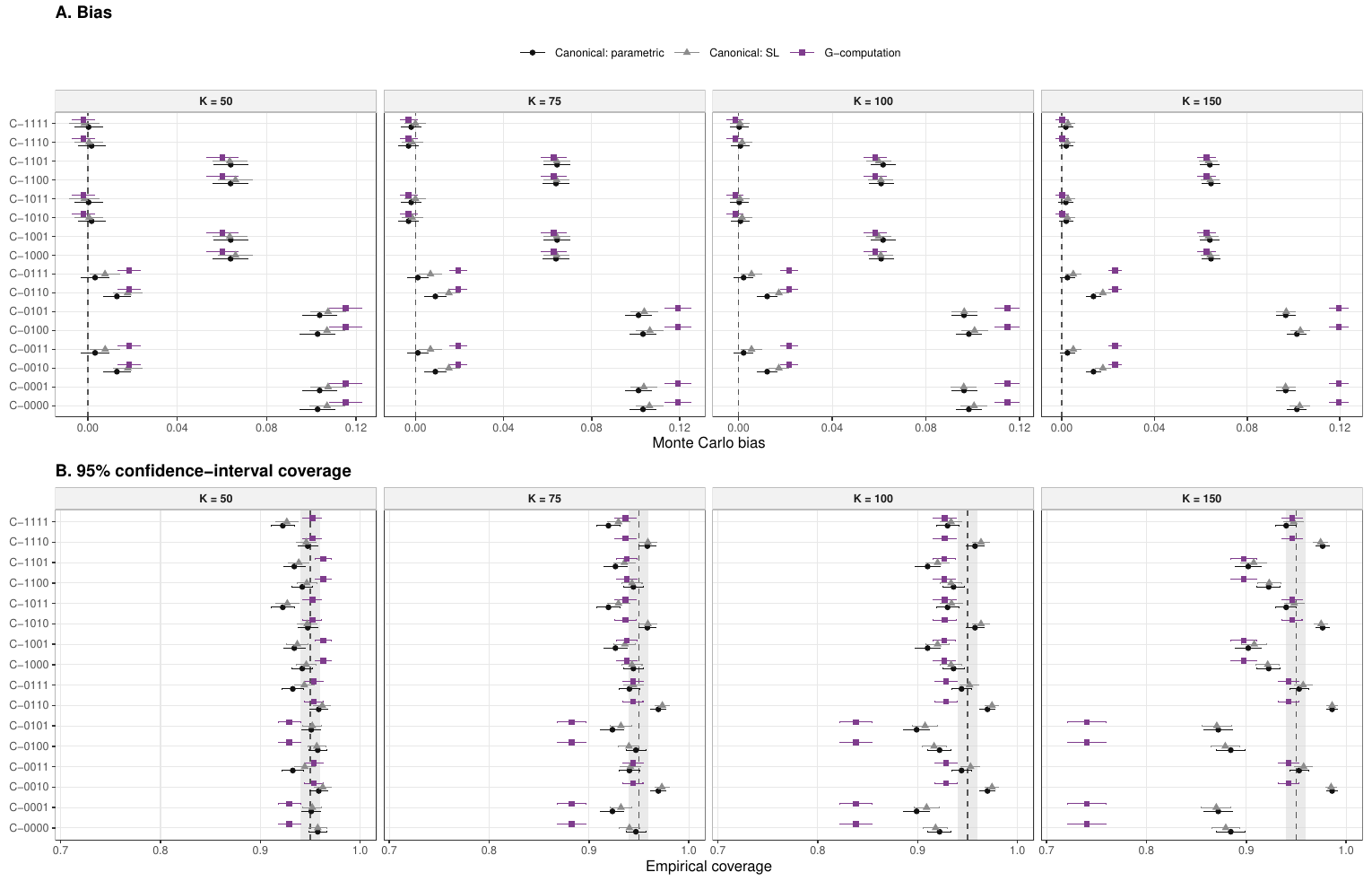}
\captionof{figure}{Complete canonical-study cluster-average ISME results.}
\label{fig:sim-app-can-ISME-C}
\end{minipage}
\end{landscape}

\clearpage
\begin{landscape}
\noindent\begin{minipage}{\linewidth}
\centering
\includegraphics[width=0.96\linewidth,height=0.85\textheight,keepaspectratio]{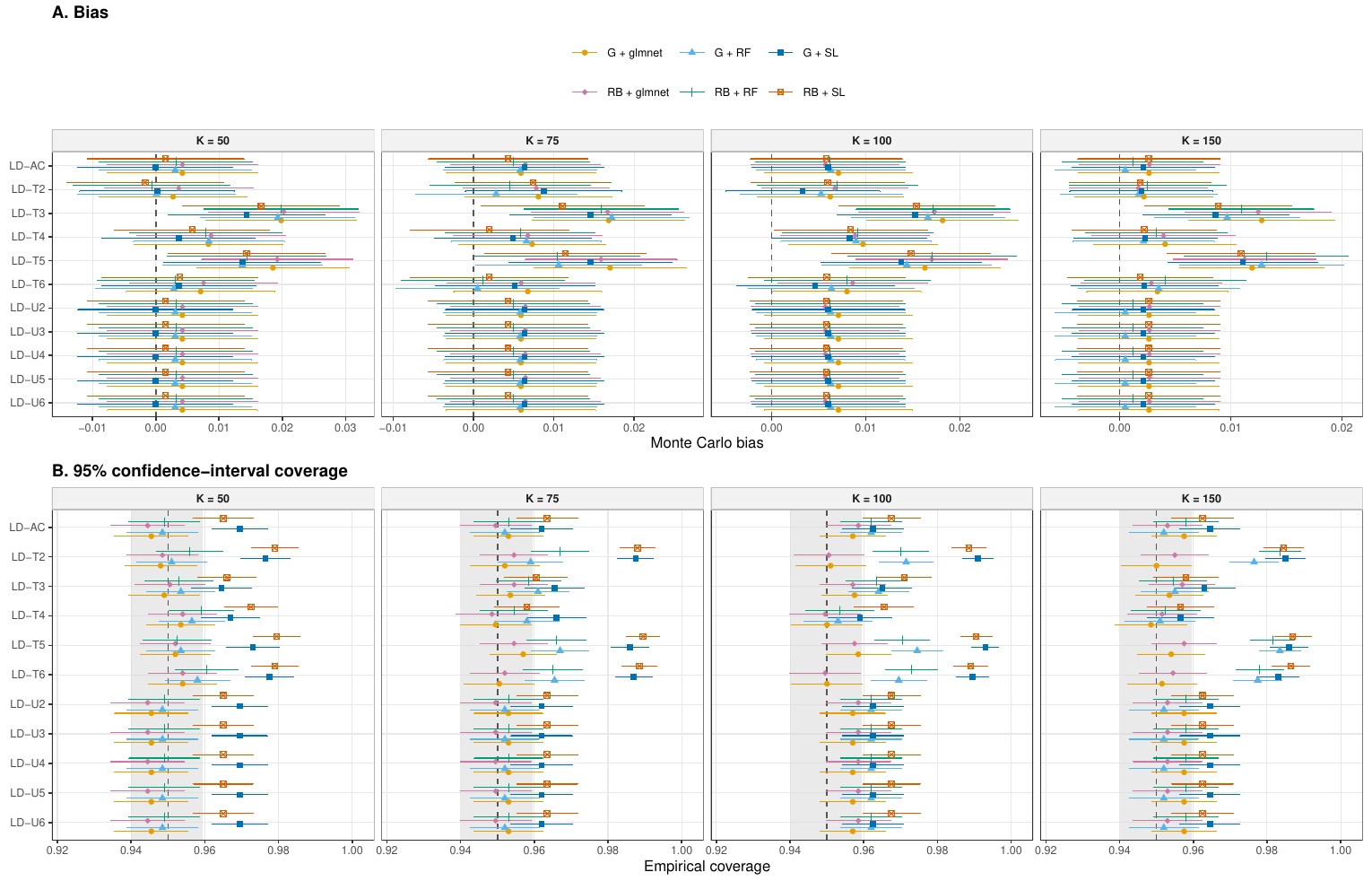}
\captionof{figure}{Complete low-dimensional-study cluster-average IDE results
for all 11 scenarios, six reference-distribution--classifier pairs, and four values of
\(K\).}
\label{fig:sim-app-ld-IDE-C}
\end{minipage}
\end{landscape}

\clearpage
\begin{landscape}
\noindent\begin{minipage}{\linewidth}
\centering
\includegraphics[width=0.96\linewidth,height=0.85\textheight,keepaspectratio]{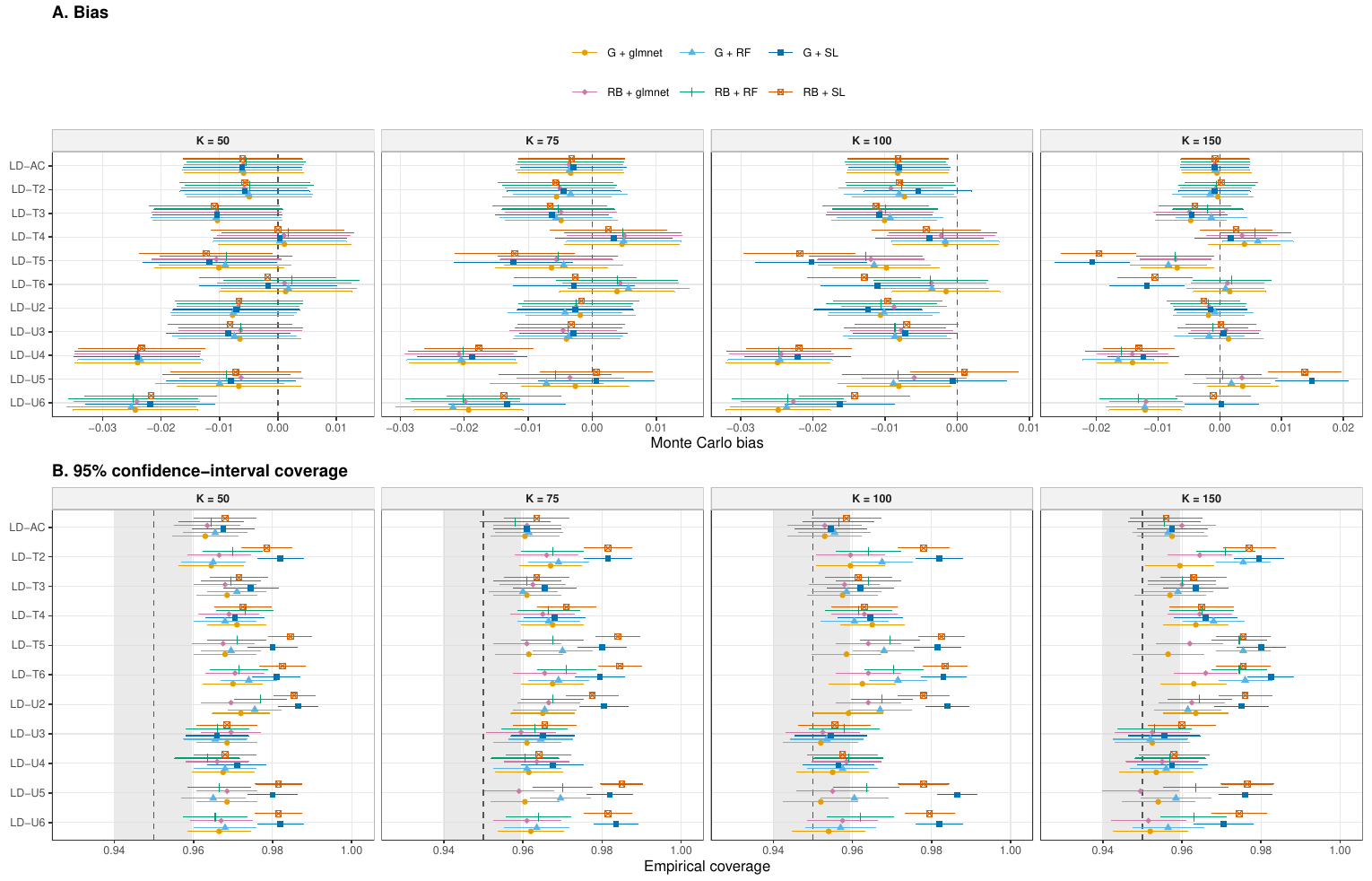}
\captionof{figure}{Complete low-dimensional-study cluster-average IIME results.}
\label{fig:sim-app-ld-IIME-C}
\end{minipage}
\end{landscape}

\clearpage
\begin{landscape}
\noindent\begin{minipage}{\linewidth}
\centering
\includegraphics[width=0.96\linewidth,height=0.85\textheight,keepaspectratio]{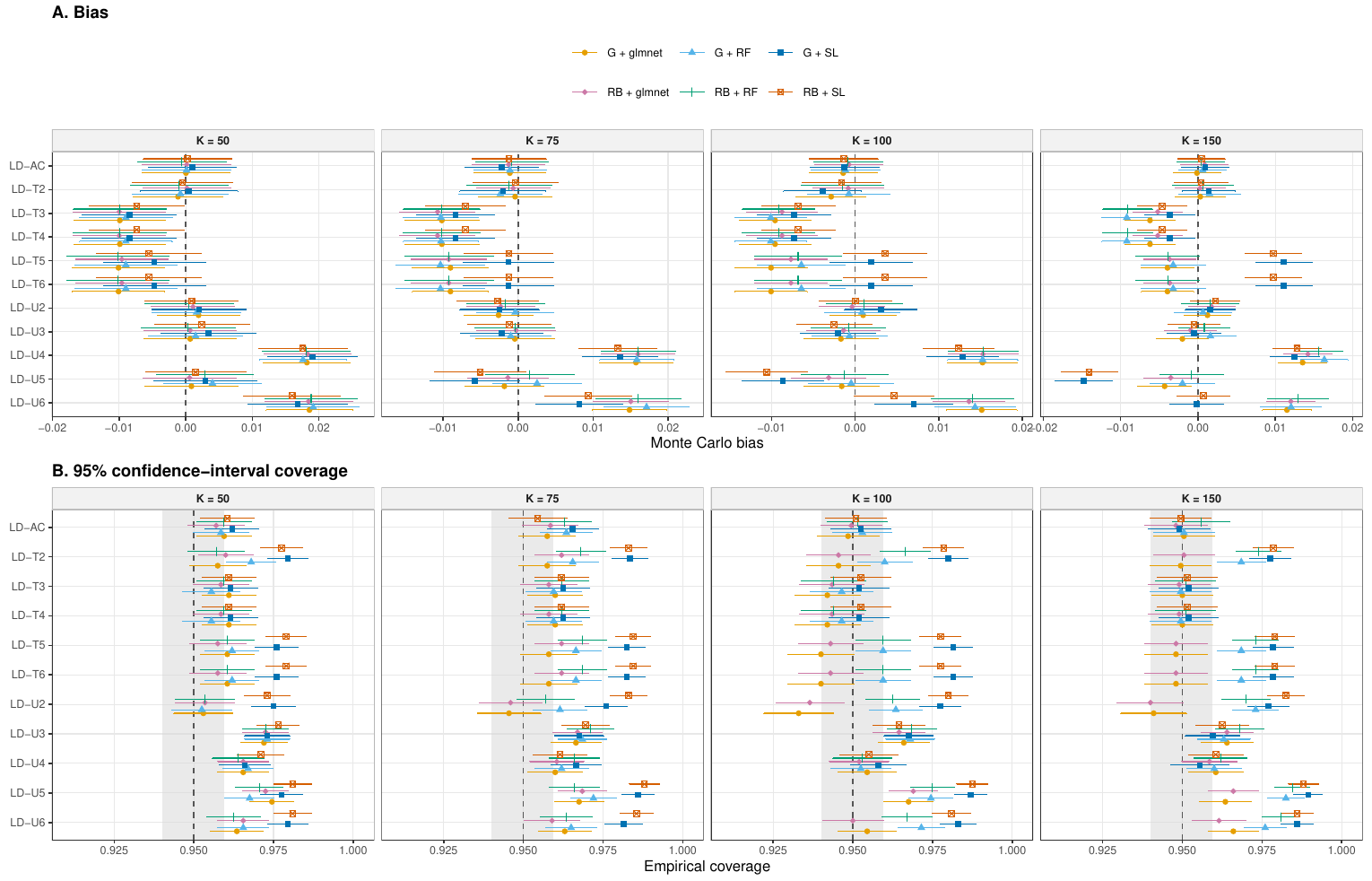}
\captionof{figure}{Complete low-dimensional-study cluster-average ISME results.}
\label{fig:sim-app-ld-ISME-C}
\end{minipage}
\end{landscape}

% \clearpage
% \begin{landscape}
% \centering
% \begin{minipage}{\linewidth}
% \centering
% \includegraphics[width=0.88\linewidth]{final_results/report/figures_reorganized/appendix_all_correct_cluster.pdf}
% \captionof{figure}{All-correct comparison of the two stabilized canonical EIF
% estimators and six low-dimensional EIF estimators for cluster-average IDE,
% IIME, and ISME.  Panels A--C report RMSE, empirical 95\% coverage, and mean
% 95\% confidence-interval width.  G-computation is omitted because it uses a
% percentile-bootstrap interval.}
% \label{fig:sim-app-all-correct-C}
% \vspace{0.5em}

% {\small
% \noindent\textbf{Complete cluster-average tables.}
% Each numerical entry in Tables~\ref{tab:sim-can-IDE-C}--%
% \ref{tab:sim-ld-ISME-C} gives bias, RMSE, empirical 95\% coverage, and mean
% estimated SE divided by empirical SD, in that order.  A final value below one
% indicates variance underestimation.  The P/AL/CI column separately gives the
% theoretical point-consistency, reported-influence-function asymptotic-linearity,
% and interval-validity status.  G and RB denote Gaussian and residual-bootstrap
% reference distributions.\par}
% \end{minipage}
% \end{landscape}

% \input{final_results/report/simulation_tables_cluster.tex}

\subsubsection{Individual-average results}
\label{sec:sim-individual-results}

For $W_i=N_i$, one can easily compute the true IDE, IIME, and
ISME values of $0.489481$, $0.342965$, and $0.188840$, respectively, based on the data geneating process given in Section~\ref{sec:sim-data-generating-model}.
Note that both the point estimator and the cluster influence function accounted
for estimation of $E(N_i)$.
Figures~\ref{fig:sim-app-can-IDE-I}--\ref{fig:sim-app-ld-ISME-I}
report the complete individual-average results.
The bias and coverage patterns were broadly similar to those under
cluster-average weighting; changing the target population did not
materially change the comparisons among methods or scenarios.

For the canonical estimators, C-0011 and C-1010 again yielded small
biases at larger $K$, whereas C-1100 and C-1101 retained small IDE
biases but lingering mediation-effect biases.
The latter two scenarios again underestimated IIME and overestimated
ISME, with IIME coverage closer to nominal than ISME coverage.
The conservative IDE coverage under C-1110, C-0110, and C-0010
also persisted. Under C-0001 and C-0000, substantial mediation-effect
biases and poor coverage remained as $K$ increased, as in the
cluster-average results.

The low-dimensional estimators likewise reproduced the cluster-average
patterns. IDE remained close to its target across the LD-U scenarios,
whereas LD-U4 and LD-U6 produced negative IIME biases and positive
ISME biases for most methods.
The relatively conservative SL intervals under LD-T2, LD-T5, and
LD-T6 were also evident under individual-average weighting.
Under LD-AC, differences between G and RB remained small, and
coverage was close to nominal at larger $K$.
The all-correct interval-width ordering was also similar across
the two weighting schemes: low-dimensional intervals were wider
than the parametric canonical EIF intervals for IDE and ISME,
especially for the SL implementations, with little difference for IIME.

\clearpage
\begin{landscape}
\noindent\begin{minipage}{\linewidth}
\centering
\includegraphics[width=0.96\linewidth,height=0.85\textheight,keepaspectratio]{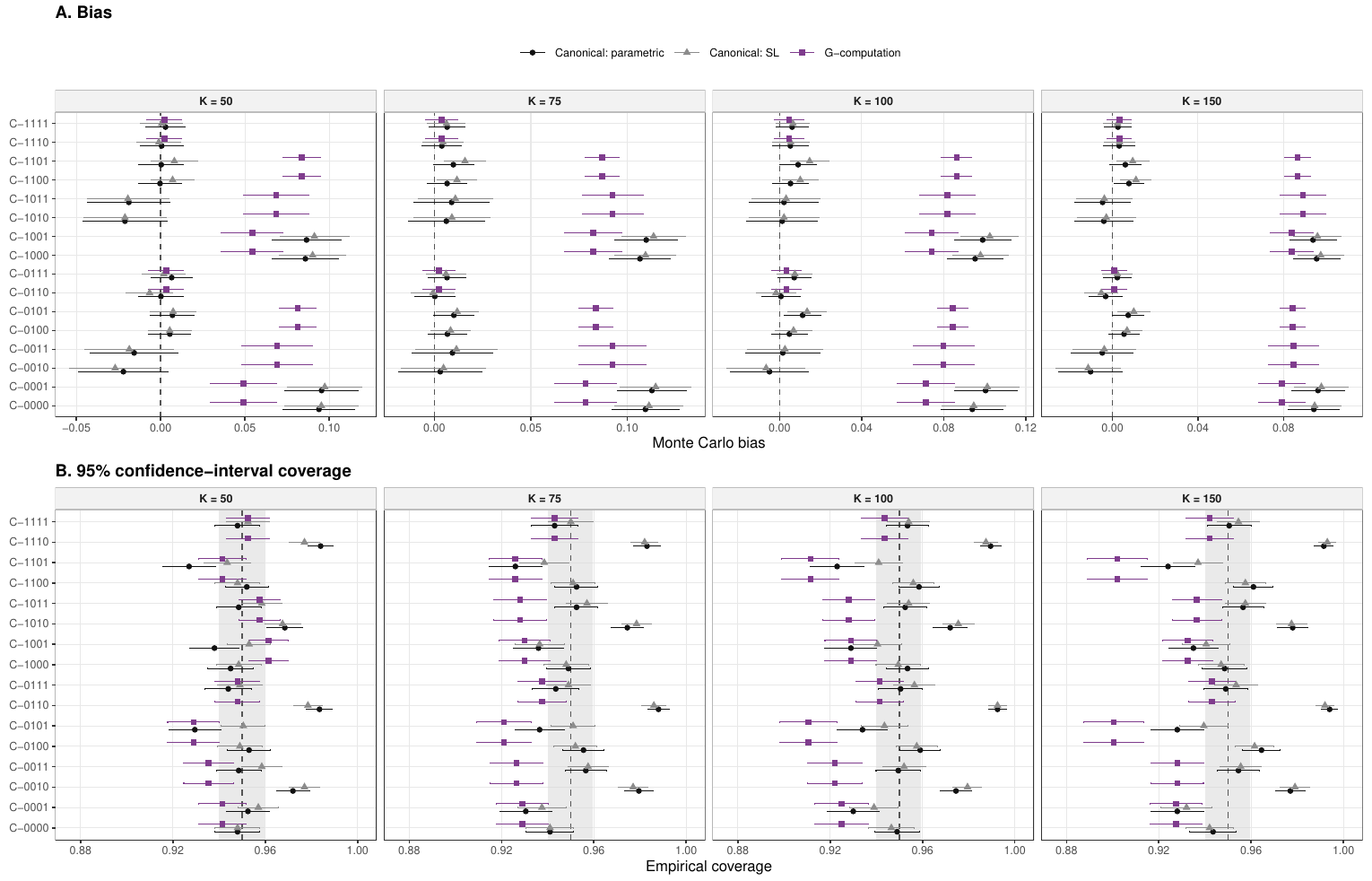}
\captionof{figure}{Complete canonical-study individual-average IDE results for
all 16 nuisance scenarios, three estimators, and four values of \(K\).}
\label{fig:sim-app-can-IDE-I}
\end{minipage}
\end{landscape}

\clearpage
\begin{landscape}
\noindent\begin{minipage}{\linewidth}
\centering
\includegraphics[width=0.96\linewidth,height=0.85\textheight,keepaspectratio]{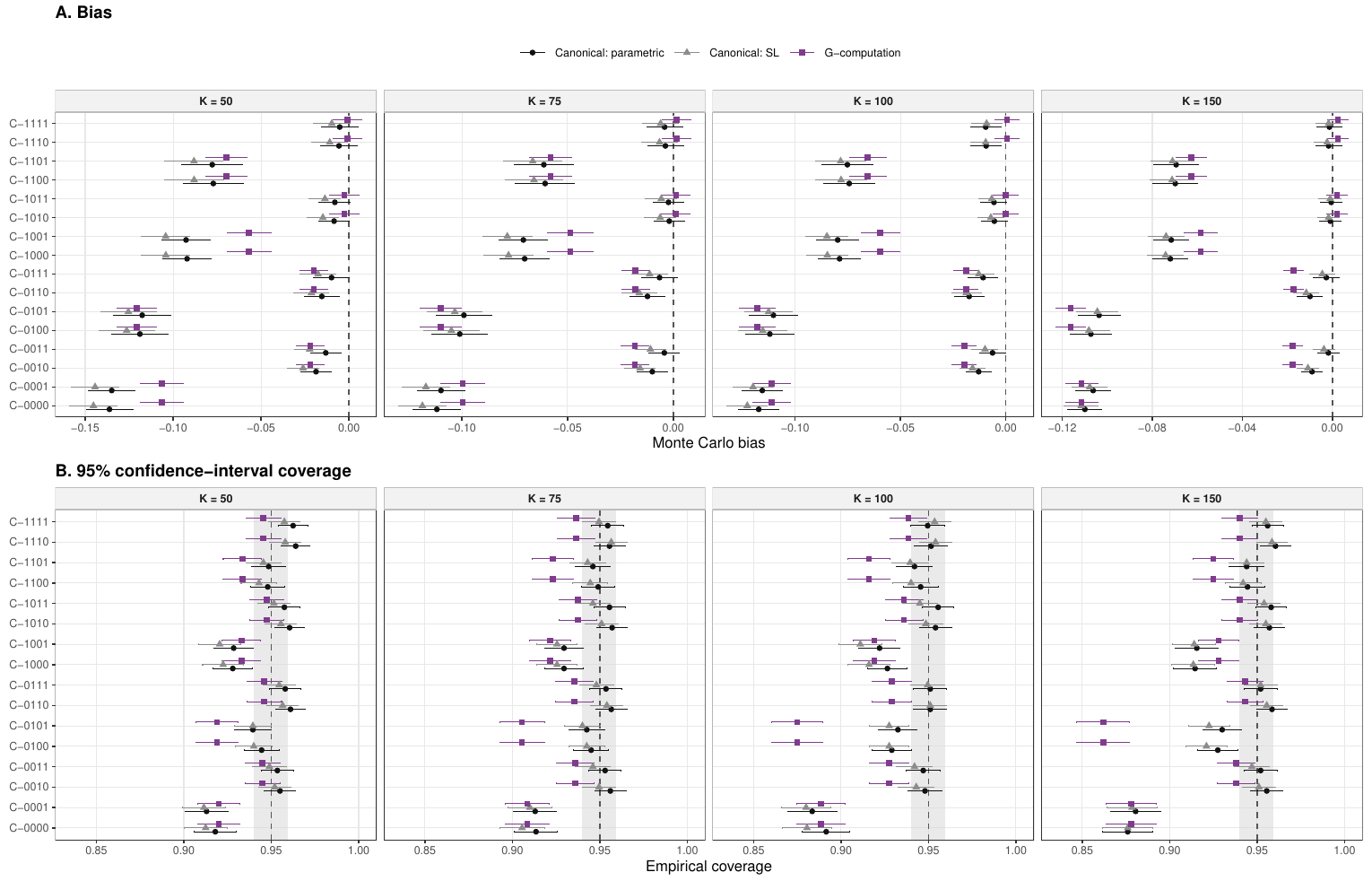}
\captionof{figure}{Complete canonical-study individual-average IIME results.}
\label{fig:sim-app-can-IIME-I}
\end{minipage}
\end{landscape}

\clearpage
\begin{landscape}
\noindent\begin{minipage}{\linewidth}
\centering
\includegraphics[width=0.96\linewidth,height=0.85\textheight,keepaspectratio]{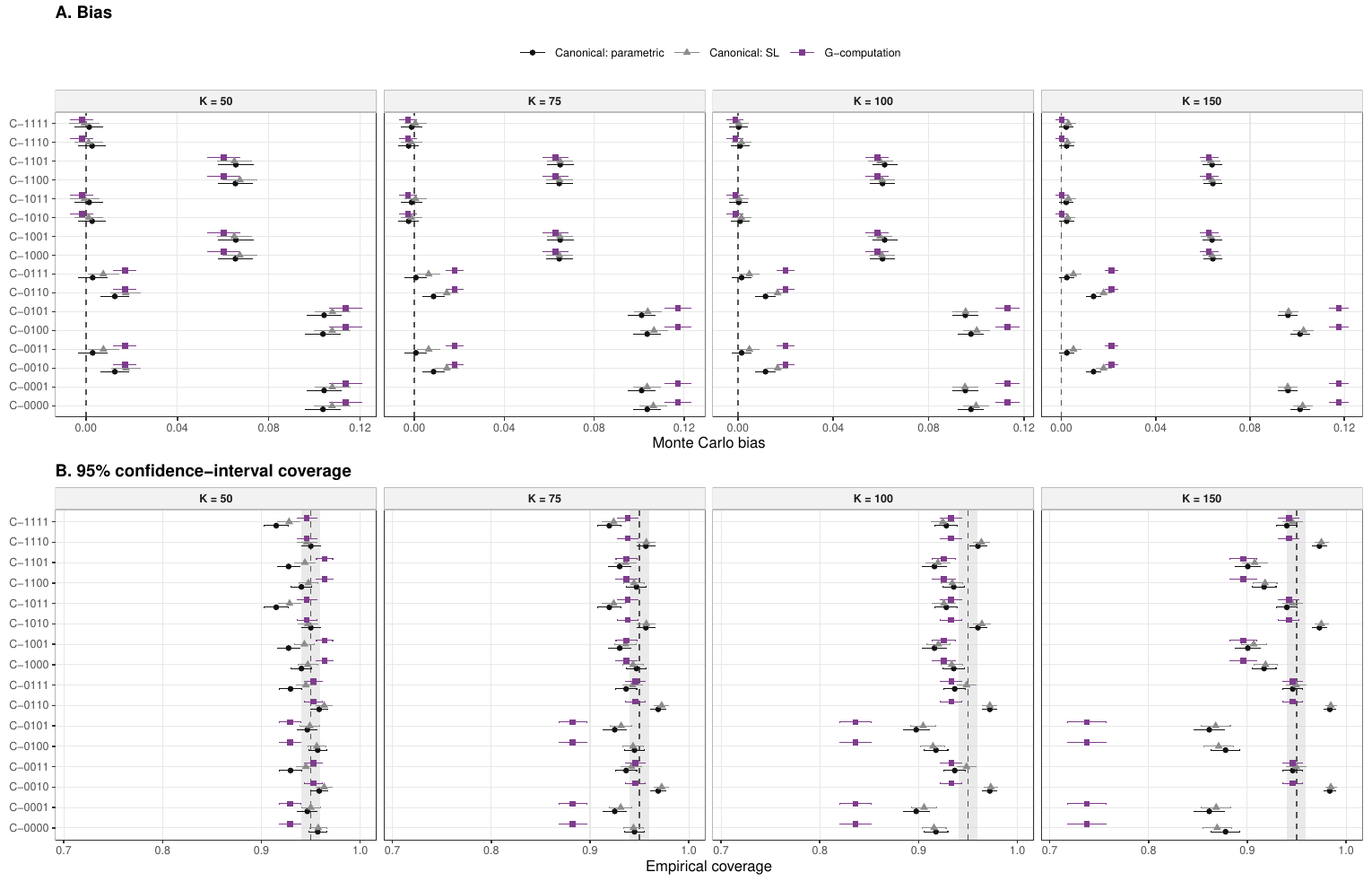}
\captionof{figure}{Complete canonical-study individual-average ISME results.}
\label{fig:sim-app-can-ISME-I}
\end{minipage}
\end{landscape}

\clearpage
\begin{landscape}
\noindent\begin{minipage}{\linewidth}
\centering
\includegraphics[width=0.96\linewidth,height=0.85\textheight,keepaspectratio]{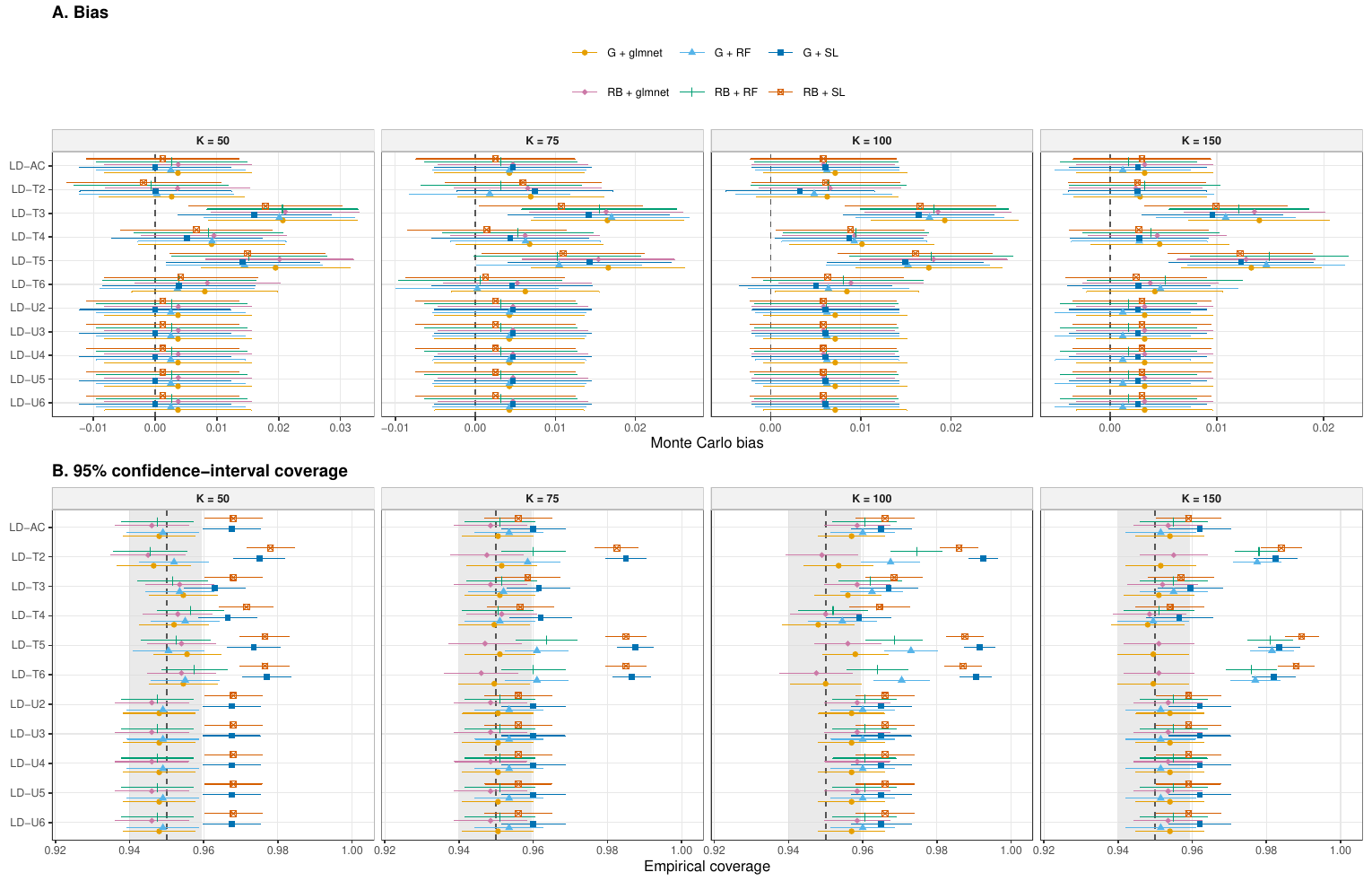}
\captionof{figure}{Complete low-dimensional-study individual-average IDE
results for all 11 scenarios, six reference-distribution--classifier pairs, and four
values of \(K\).}
\label{fig:sim-app-ld-IDE-I}
\end{minipage}
\end{landscape}

\clearpage
\begin{landscape}
\noindent\begin{minipage}{\linewidth}
\centering
\includegraphics[width=0.96\linewidth,height=0.85\textheight,keepaspectratio]{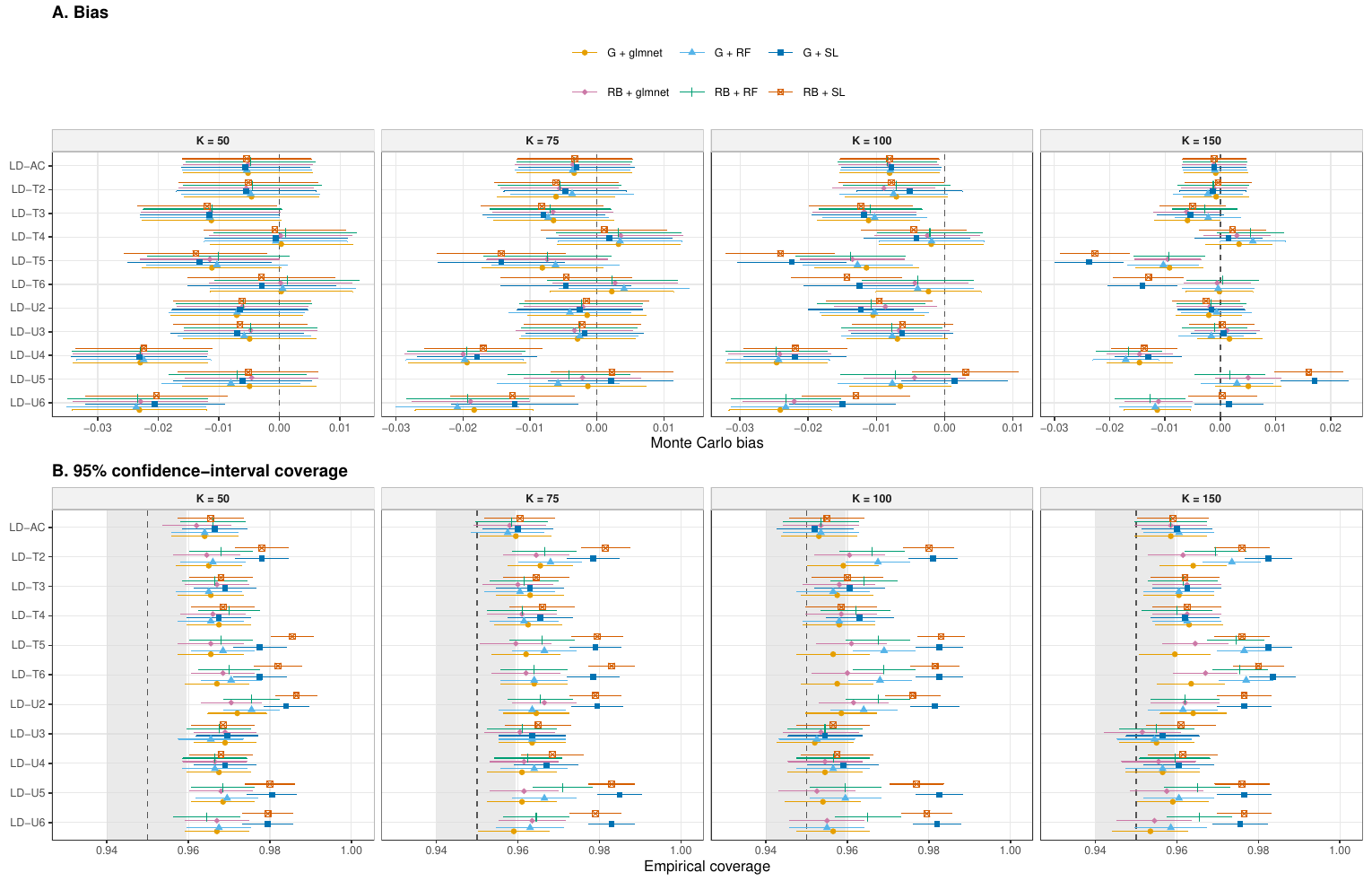}
\captionof{figure}{Complete low-dimensional-study individual-average IIME results.}
\label{fig:sim-app-ld-IIME-I}
\end{minipage}
\end{landscape}

\clearpage
\begin{landscape}
\noindent\begin{minipage}{\linewidth}
\centering
\includegraphics[width=0.96\linewidth,height=0.85\textheight,keepaspectratio]{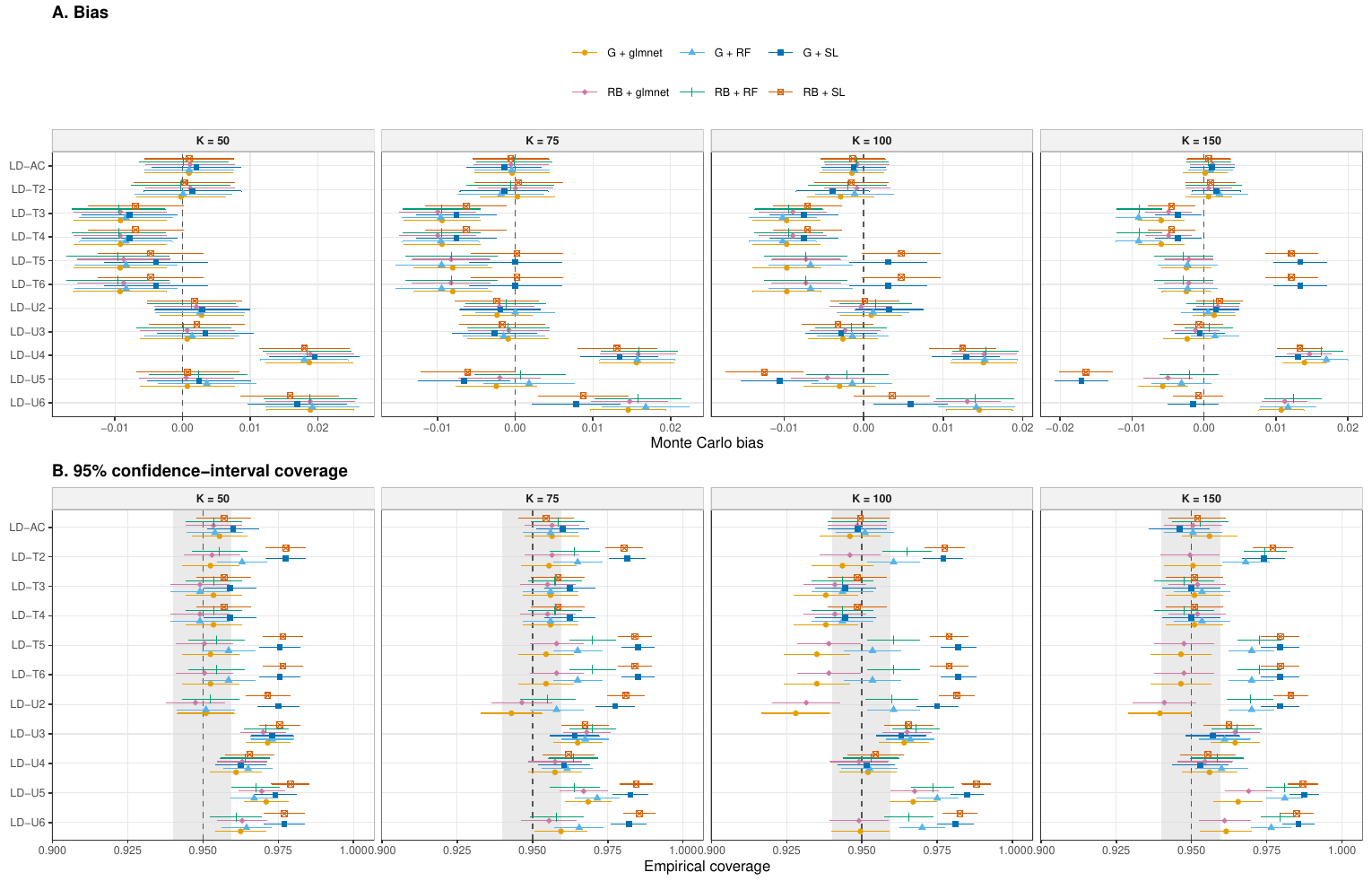}
\captionof{figure}{Complete low-dimensional-study individual-average ISME results.}
\label{fig:sim-app-ld-ISME-I}
\end{minipage}
\end{landscape}

% \clearpage
% \begin{landscape}
% \centering
% \begin{minipage}{\linewidth}
% \centering
% \includegraphics[width=0.88\linewidth]{final_results/report/figures_reorganized/appendix_all_correct_individual.pdf}
% \captionof{figure}{All-correct comparison of the two stabilized canonical EIF
% estimators and six low-dimensional EIF estimators for individual-average IDE,
% IIME, and ISME.  Panels A--C report RMSE, empirical 95\% coverage, and mean
% 95\% confidence-interval width.}
% \label{fig:sim-app-all-correct-I}
% \vspace{0.5em}

% {\small
% \noindent\textbf{Complete individual-average tables.}
% Each numerical entry and each P/AL/CI code has the same definition as in the
% cluster-average tables.\par}
% \end{minipage}
% \end{landscape}

% \input{final_results/report/simulation_tables_individual.tex}

\clearpage
\section{PPACT trial mediation effects}\label{sec:ppact-comparison}

In this section, we provide an additional table to Section~\ref{sec:ppact-application} for the comparison between the natural effects and the interventional effects.

\par\medskip
\begingroup
\renewenvironment{table}[1][]{\par\noindent\begin{minipage}{\linewidth}\captionsetup{type=table}}{\end{minipage}\par}
\begin{table}[t]
\centering
\caption{Comparison of natural mediation effects and interventional mediation effects in the PPACT trial.}
\label{tab:ppact-comparison}
\begin{tabular}{lcc}
\toprule
Effect
& Interventional effects
& Natural effects \\
& (canonical EIF estimator)
& (canonical EIF estimator) \\
\midrule
Total effect
& $-0.543\;(-0.868,\,-0.219)$
& $-0.619\;(-0.919,\,-0.320)$ \\

Indirect effect
& $-0.380\;(-0.577,\,-0.184)$
& $-0.342\;(-0.579,\,-0.106)$ \\

Direct effect
& $-0.163\;(-0.426,\,0.101)$
& $-0.277\;(-0.502,\,-0.052)$ \\

Individual mediation effect
& $-0.255\;(-0.373,\,-0.138)$
& $-0.243\;(-0.424,\,-0.061)$ \\

Spillover mediation effect
& $-0.125\;(-0.291,\,0.041)$
& $-0.100\;(-0.326,\,0.126)$ \\
\bottomrule
\end{tabular}

\vspace{3pt}
\begin{minipage}{0.94\textwidth}
\footnotesize
\textit{Note:} Natural effects were estimated using the canonical EIF estimator, also denoted the EIF$_1$ estimator, of \citet{cheng2026semiparametric}. The associated nuisance parameters were estimated using the same library of machine learning algorithms employed in the interventional effect analysis.
\end{minipage}
\end{table}
\endgroup

% PRESERVED APPENDIX END

\end{document}